\documentclass[pre,twocolumn,superscriptaddress,amsmath,amssymb,longbibliography]{revtex4-2}
\usepackage[utf8]{inputenc}
\usepackage[T1]{fontenc}
\usepackage{mathtools}
\usepackage{hyperref}
\usepackage{booktabs}
\usepackage{bm}
\usepackage{xcolor}
\usepackage{graphicx}
\usepackage{tikz}
\usetikzlibrary{arrows.meta,positioning,shapes.geometric}
\usepackage[english]{babel}

\newcommand{\Coll}{\mathcal{C}}      
\newcommand{\W}{\mathcal{W}}
\newcommand{\D}{\mathcal{D}}
\newcommand{\Lg}{\mathcal{L}}

\newcommand{\Da}{\mathrm{Da}}

\newcommand{\pder}[2]{\frac{\partial #1}{\partial #2}}

\newtheorem{remark}{Remark}
\let\oldremark\remark
\renewcommand{\remark}{\oldremark\normalfont}

\begin{document}

\title{The Statistical physics of unsaturated soil water:\\ kinetic theory and non commutative pore water dynamics}

\author{Riccardo Rigon}
\email{riccardo.rigon@unitn.it}
\affiliation{Centro Agricoltura Alimenti Ambiente (C3A), University of Trento, 38098 San Michele all'Adige (TN), Italy}
\affiliation{Dipartimento di Ingegneria Civile Ambientale e Meccanica, University of Trento, 30123 Mesiano, Trento, Italy}

\date{September 3, 2026}

\begin{abstract}
We develop a statistical-mechanical theory of water in unsaturated soil whose outcome is a
continuum field equation for the pore-occupancy $g(r,\mathbf{x},t)\in[0,1]$, the fraction of
pores of radius $r$ that are water-filled at position $\mathbf{x}$ and time $t$.  The theory is
built by passing through three scales.  At the \emph{microscale}, single inter-pore water
transfers are set by Hagen--Poiseuille rates $\kappa(r)$ and a driving potential $\Phi(r,r')$, the
difference of the pore-class chemical potentials, taken here in its capillary--gravitational form
but open to adsorptive, osmotic, or thermal refinement without change to the theory.  Averaging over a representative volume gives a \emph{mesoscale} master
equation whose gain and loss terms relax the occupancy toward the equilibrium step
$g_{\rm eq}=H(r^*\!-r)$ fixed by chemical-potential minimization.  Contracting the volume to a
point promotes $g$ to a field and yields the \emph{continuum} balance
$\partial_t g + \nabla\!\cdot\mathbf{F} = \mathcal{C}[g] - \mathcal{E} - \mathcal{T}$, with
$\mathcal{C}[g]$ the redistribution operator and $\nabla\!\cdot\mathbf{F}$ the transport
divergence, the central equation of the theory, of which everything else is a limit, a moment,
or a boundary resolution.

The kinetic equation is an Onsager gradient flow: it descends the Gibbs free energy with a
state-dependent mobility, with $d\mathcal{F}/dt\le0$ a theorem for the isothermal, unforced
system and mass conservation as its zeroth moment.  A single dimensionless group, the
pore-resolved Damk\"ohler number $\Da(r,\mathbf{x})=\tau_{\rm redis}(r)/\tau_{\rm forcing}(\mathbf{x})$,
organizes the behavior and unifies phenomenologies long modelled separately.  A Chapman--Enskog
reduction identifies Richards' equation as the quasi-static ($\Da\to0$) limit, with the matric
potential $\psi$ and the hydraulic conductivity $K(\theta)$ emerging only there, $K$ vanishing
below the percolation threshold; capillary-bundle and critical-path conductivity models are its
diagonal and spectral limits, and dynamic capillary-pressure models its one-moment
approximation.  Hysteresis is the holonomy of a forcing bundle over the space of boundary
histories, a geometric phase rather than per-pore bistability, with a falsifiable loop-area law
$\mathcal{H}\propto I^2$ that no rate-independent model reproduces.  Preferential flow is what
the same equation does where $\Da\gtrsim1$, so the long-standing dichotomy between
Richards-equation and preferential flow becomes a continuous, $\Da$-controlled crossover within
one framework.  Out of the quasi-static limit $g(r)$ is the irreducible state variable and no
scalar potential characterizes the system.  All inputs, $f(r)$, $C(r,r')$, $\bar L$,
$\bar\tau$, are geometric properties of the pore network, measurable from micro-CT images and
calibrated against no macroscopic retention or conductivity data.
\end{abstract}


\maketitle

\section{Introduction}
\label{sec:intro}

The dynamics of unsaturated porous media is inherently irreversible \cite{Alonso1990}. Under realistic forcing, the system does not evolve through a sequence of equilibrium states, but through a succession of non-equilibrium configurations. Capillary hysteresis is the most familiar manifestation of this irreversibility: identical macroscopic states such as the same water and energy content can correspond to different internal configurations depending on whether the system was reached by wetting or drying, slowly or rapidly~\cite{Haines1930,Mualem1974}. 

Another long-standing difficulty in unsaturated flow is the coexistence of two seemingly distinct behaviors. When wetting is moderate or absent, water redistribution is slow, capillary-controlled, and close to local equilibrium; flow is dominated by matric forces and is reasonably described by the classical continuum model known as Richards equation \cite{Richardson1922,Richards1931,Tubini2022}. In contrast, during intense or rapid wetting events, water preferentially occupies larger, well-connected pores, bypassing finer pores that would be favored by equilibrium thermodynamics.  This dual behavior has been emphasized in the seminal critiques of Richards' equation by Germann and Beven~\cite{Beven1982,Germann1985,Germann2014}, who argued that preferential flow and bypass mechanisms violate the assumptions underlying single-domain formulations.

These observations have motivated the development of dual-domain and dual-permeability extensions of Richards' equation, in which the pore space is partitioned into interacting fast and slow domains~\cite{Gerke1993}. While successful in many applications, such approaches introduce additional state variables and exchange terms phenomenologically, without explaining how the two flow regimes emerge from a common physical description.
A complementary, Lagrangian line of work represents soil water as an ensemble of particles of unequal mobility~\cite{ZeheJackisch2016}.  That approach abandons the classical assumption that all the water at a point moves with a single velocity, but the fraction of pores of each size that is water-filled is still not treated as a variable with its own dynamics.  Promoting that pore-class occupancy to a dynamical field is the step taken here.

From a physical perspective, the key issue is that macroscopic variables such as water content are coarse-grained descriptors. Many distinct pore-scale configurations correspond to the same macroscopic state, yet differ in connectivity, accessibility, transport efficiency, and energy content. When forcing is slow compared to internal relaxation processes, such as film flow, vapor transport, and interfacial rearrangement, the system remains close to equilibrium and capillary ordering dominates~\cite{HassanizadehGray1993}. When forcing is rapid, relaxation cannot keep pace, and water occupies geometrically accessible pathways rather than thermodynamically optimal ones. The same porous medium therefore exhibits qualitatively different behavior depending on the relative timescales of forcing and relaxation. These dynamics generated  the phenomenon known as hydraulic hysteresis \cite{Poulovassilis1962,Topp1971-zr} which  is not explained by Richards equation and usually treated with ad hoc methods \cite{Kool1987-le}.

These considerations motivate a shift in perspective, in which three elements play complementary roles.  Thermodynamics defines the equilibrium targets.  Network topology constrains which microscopic rearrangements are geometrically admissible.  Relaxation dynamics governs the rate at which the system can move through configuration space.  When relaxation is fast relative to forcing, the system tracks equilibrium and classical matric flow emerges; when forcing is fast, relaxation is incomplete, flow bypasses the smallest pores, and preferential pathways appear.  The transition between the two behaviors is continuous and dynamically controlled, rather than imposed by the model structure.

The theory developed here explicitly tracks this irreversible evolution through non-equilibrium states and naturally unifies these behaviors within a single formalism. Although motivated by water flow in porous media, the framework is not specific to capillarity. It provides a general description of systems in which local thermodynamic equilibria exist, but macroscopic behavior emerges from constrained, history-dependent transitions between them. Similar structures arise in plastic deformation, magnetic hysteresis, and glassy dynamics, where irreversibility results from the interplay between thermodynamic driving, geometric constraints, and finite relaxation times~\cite{Preisach1935,Bertotti1998,Bouchaud1998}. Phenomenological understanding, together with over a century of measurements \cite{Haines1930,Richards1931,Davidson1969}, points to two distinct scales: an internal one, whose dynamics is governed by the topology and geometry of the soil medium \cite{Vogel2010-wi}, and a larger one dominated by transport.  The natural mathematical representation of such a two-scale system is a fiber bundle \cite{Schutz1980-uu}. 

The remainder of this paper formalizes this perspective using tools from statistical physics and explores its implications for transport and upscaling in porous media. Hysteresis, memory, rate dependence, and preferential flow emerge naturally as consequences of irreversible non-equilibrium evolution, rather than as ad hoc modifications of equilibrium theory.

Although we speak of soil water throughout, nothing in the construction is
specific to water or to soil.  The fluid enters the theory through three
parameters only: the surface tension and contact angle $\gamma\cos\theta_c$ (the
capillary head), the dynamic viscosity $\mu$ (the Hagen--Poiseuille rate), and
the density $\rho_w$ (the gravitational head and the pressure-to-head
conversion); the medium enters only through the geometric inputs $f(r)$,
$C(r,r')$, $\bar L$, $\bar\tau$.  The kinetic equation, its thermodynamic
structure, and the Chapman--Enskog reduction therefore hold verbatim for any
Newtonian wetting liquid displacing a passive gas phase in any rigid porous
medium, oil or brine in reservoir rock, cryogenic liquids in regolith, melt in
partially molten aggregates, upon substituting $(\gamma\cos\theta_c,\mu,\rho)$;
only the adsorptive and osmotic refinements of $\mu_w$ discussed in
Sec.~\ref{sec:filling} carry water-specific physics. 

\section{The cylindrical pore network}
\label{sec:network}

\subsection{Network representation}
\label{sec:geometry}

We represent the pore space by a network of communicating cylinders.  Each pore $i$ is characterized by a quadruplet:
\begin{equation}
\text{Pore}_i = \{r_i,\, L_i,\, \tau_i,\, n_i\}
\end{equation}
where $r_i$ is the radius, $L_i$ the length along the centerline, $\tau_i = L_i/d_{\mathrm{straight},i} \geq 1$ the tortuosity, with $d_{\mathrm{straight},i}$ the straight-line (Euclidean) distance between the two endpoints of the pore centerline, so that $\tau_i=1$ for a straight pore, and $n_i$ the coordination number (number of connected neighbors).  The pore volume is $V_i = \pi r_i^2 L_i$.  The network is:
\begin{equation}
\text{Network} = \{\text{Pore}_i,\; E(i,j)\}
\end{equation}
where $E(i,j) \in \{0,1\}$ is the adjacency matrix.


We work with statistics averaged over a Representative Elementary Volume (REV, \cite{Bachmat1987-cv}), the smallest volume over which the pore-space geometry is statistically stationary: the pore-size distribution $f(r)$ (normalized to $\int f\,dr = 1$), the mean length $\bar{L}$, the mean tortuosity $\bar{\tau}$, and the connectivity matrix $C(r,r')$, the probability that a pore of radius $r$ connects to a pore of radius~$r'$.
The connectivity matrix $C(r,r')$ is the probability that a pore of radius $r$ and a pore of radius $r'$, drawn uniformly at random from the pore network, are directly connected.
 It is a property of the pore geometry, not of the filling state.  Its physical content:
\begin{itemize}
\item \emph{If $C(r,r')$ is broad} (each pore connects to pores of many sizes): conductivity varies smoothly with saturation, and the network is relatively insensitive to the size of the drained pores.
\item \emph{If $C(r,r')$ is narrow} (like-sized pores preferentially connect): the network has strong correlated structure, percolation transitions are sharp, and heterogeneity penalties are large.
\item \emph{If $C(r,r') \propto f(r')$} (Mualem's random-pairing hypothesis~\cite{Mualem1976}): every pore connects equally to all others weighted by abundance, yielding the classical Mualem formula as a limiting case (Sec.~\ref{sec:K_derived}).
\end{itemize}
$C(r,r')$ is measurable from micro-CT images by pore-network extraction (see Sec.~\ref{sec:GL} for the extraction pipeline).  It does not require fitting; it is a structural input.

\subsection{Hagen--Poiseuille rate constant}
\label{sec:HP}

Flow through a fully filled cylinder of radius $r$ is governed by the Hagen--Poiseuille (HP) equation.  The rate constant (reciprocal time per unit driving force) for steady flow through a filled cylinder embedded in the network is:
\begin{equation}
\kappa_{\rm HP}(r) = \frac{r^2}{8\mu\, \bar{L}^2\, \bar{\tau}^2} \qquad [\mathrm{Pa}^{-1}\,\mathrm{s}^{-1}]
\label{eq:kappa}
\end{equation}
where $\mu$ [$\mathrm{Pa\cdot s}$] is the dynamic viscosity.  The $r^2$ scaling means that large pores equilibrate fast; small pores equilibrate slowly.  Physically, $\kappa_{\rm HP}$ is a rate constant per unit driving force: given a pressure difference across one pore length, it returns an inverse time.  During active filling the rate is modified by Lucas--Washburn meniscus invasion, giving a state-dependent generalization $\kappa(r,g)$ developed in Sec.~\ref{sec:LW_rate} once the filling fraction has been introduced.

In all transport formulas below, the effective pore conductance is $\kappa_{\rm eff}(r,g) = \kappa(r,g)\,\Xi(r)$, where $\Xi(r) \leq 1$ collects geometric corrections for cross-sectional shape, tortuosity, roughness, constriction, and film flow (Sec.~\ref{sec:K_derived}).  For notational simplicity we write $\kappa(r,g)$ throughout, absorbing $\Xi$ implicitly except where the distinction matters.

\subsection{Inter-pore driving potential}
\label{sec:Phi}

The driving potential $\Phi(r,r',\hat{\bm e})$ for flow along the network edge that joins a
pore of radius $r'$ (donor) to a pore of radius $r$ (receiver) is, in full generality, the
difference of the water chemical potentials of the two pore classes expressed as a head,
$\rho_w\mathrm{g}\,\Phi = \mu_w(r') - \mu_w(r)$, with $\mu_w$ the complete potential of
Eq.~\eqref{eq:mu_main}, capillary, adsorptive, osmotic, gravitational, and any further term one
cares to include.  Nothing in the construction that follows depends on \emph{which} of
these terms $\Phi$ carries: the transition rates, the gain--loss operators, the gradient-flow
and holonomy structure, and the Chapman--Enskog reduction all use $\Phi$ only through its
values.  Refining or extending the physics of the driving potential therefore changes the inputs
to the theory, not the theory itself.  For concreteness, and because it already exposes
the capillary/gravity competition that governs most of the phenomenology, in the following we
retain only the capillary and gravitational terms.  These enter through geometrically distinct
channels: capillarity depends on the pore radii and is isotropic at the pore scale, whereas
gravity is a fixed vector $-\mathrm{g}\,\hat{\bm z}$ in physical space and acts on an edge only
through its projection onto the edge orientation $\hat{\bm e}$ (the unit vector pointing from the
donor to the receiver).  Hence
\begin{equation}
\begin{aligned}
\Phi(r,r',\hat{\bm e}) ={}& \underbrace{\frac{2\gamma\cos\theta_c}{\rho_w \mathrm{g}}\!\left(\frac{1}{r} - \frac{1}{r'}\right)}_{\text{capillary}} \\[2pt]
&- \underbrace{(\hat{\bm e}\!\cdot\!\hat{\bm z})\,\bar{L}}_{\text{gravity}}
\end{aligned}
\label{eq:Phi}
\end{equation}
where $\gamma$ [$\mathrm{N/m}$] is the surface tension, $\theta_c$ the contact angle, $\rho_w$ [$\mathrm{kg/m^3}$] the water density, $\hat{\bm z}$ the upward vertical, and $\bar{L}$ the pore length.  The gravity term deserves a word on the length it carries.  The capillary term is a genuine difference of two Young--Laplace \emph{heads} and needs no length inserted by hand; gravity, by contrast, is a body force $-\rho_w\mathrm{g}\,\hat{\bm z}$, and to appear as a head in $\Phi$ it must be integrated over the elevation drop between the two pores.  That drop is $\Delta z = (\hat{\bm e}\!\cdot\!\hat{\bm z})\,\bar{L}$: here $\bar{L}$ is the length of the edge joining the two pore centres, so $(\hat{\bm e}\!\cdot\!\hat{\bm z})\,\bar{L}$ is its vertical projection.\footnote{This edge length is the same $\bar{L}$ that sets the mean pore length in the network statistics, and, crucially for the coarse-graining, the same inter-pore spacing $\ell\sim\bar{L}$ used in the gradient expansion of Sec.~\ref{sec:spatial} and Appendix~\ref{app:single_parameter}.  It is \emph{not} the viscous length appearing squared in $\kappa_{\rm HP}\propto r^2/(\mu\bar{L}^2\bar{\tau}^2)$ [Eq.~\eqref{eq:kappa}], which measures dissipative resistance along the pore; the two coincide numerically in a statistically homogeneous network but play distinct roles.  Writing the gravity term this way is what lets the Chapman--Enskog limit (Sec.~\ref{sec:CE}) turn $\Delta z\to\ell\,\hat{\bm e}\!\cdot\!\nabla z=\ell\,\hat{\bm e}\!\cdot\!\hat{\bm z}$ into the clean $+\hat{\bm z}$ of Richards' $q=-K(\nabla\psi+\hat{\bm z})$, with no length left over to explain away.}  Throughout, $\mathrm{g}$ (upright roman) denotes the gravitational acceleration [$\mathrm{m/s^2}$]. 
$\Phi$ has dimensions of length (hydraulic head).  Reversing the edge exchanges donor and receiver, $r\leftrightarrow r'$ and $\hat{\bm e}\to-\hat{\bm e}$, under which both terms change sign, so $\Phi$ is antisymmetric in the proper sense $\Phi(r,r',\hat{\bm e}) = -\Phi(r',r,-\hat{\bm e})$.    The dimensional bookkeeping deserves a comment, because it fixes the units of the whole kinetic construction: $\Phi$ is a head [m], so $\rho_w\mathrm{g}\,\Phi$ [Pa] is the pressure difference that actually drives the flow, and $\kappa$ [$\mathrm{Pa^{-1}\,s^{-1}}$] is the Hagen--Poiseuille rate per unit driving pressure (Eq.~\eqref{eq:kappa}).  Their product $\kappa\cdot\rho_w\mathrm{g}\cdot\Phi$ therefore has dimensions of $\mathrm{s}^{-1}$: it is the inter-pore exchange rate, the elementary frequency out of which the transition rates of Sec.~\ref{sec:rates} are built.  A detailed derivation of Eq.~\eqref{eq:Phi} is in Appendix \ref{sec:driving_potential}.

The capillary term is positive when $r < r'$ (water moves from large to small pores, toward higher suction), reflecting the thermodynamic preference for filling small pores first.
The gravitational term adds a head $-(\hat{\bm e}\!\cdot\!\hat{\bm z})\,\bar{L}$ over one pore length, positive when the receiver lies below the donor ($\hat{\bm e}\!\cdot\!\hat{\bm z}<0$, flow assisted downward).  Averaging over an isotropic edge ensemble gives $\langle\hat{\bm e}\!\cdot\!\hat{\bm z}\rangle = 0$, so the gravitational drive vanishes in the (isotropic) intra-REV redistribution and survives only at the inter-REV scale, as in Sec.~\ref{sec:connection}.  The orientation average cancels the net \emph{drive}, but the gravitational \emph{potential} still enters the equilibrium configuration pore by pore (Eq.~\eqref{eq:geq_smeared}): it broadens the sharp radius step whenever the REV spans an appreciable elevation range.  The residual net projection $\alpha \equiv -\langle\hat{\bm e}\!\cdot\!\hat{\bm z}\rangle = \cos\zeta$ measures the anisotropy of structured soils (Appendix~\ref{sec:angular}).
The competition between capillarity and gravity defines the capillary length:
\begin{equation}
\ell_c = \frac{2\gamma\cos\theta_c}{\rho_w \mathrm{g}\, \alpha\, \bar{L}}
\label{eq:ell_c}
\end{equation}
For $r \gg \ell_c$, gravity dominates; for $r \ll \ell_c$, capillarity dominates.

\section{Configuration space and equilibrium}
\label{sec:config}

\subsection{The filling distribution}
\label{sec:filling}

The microscopic state of the pore network is specified by the filling fraction:
\begin{equation}
g: \mathbb{R}^+ \to [0,1], \quad r \mapsto g(r,\mathbf{x},t)
\label{eq:g_def}
\end{equation}
giving the fraction of pores of radius $r$ at macroscopic position $\mathbf{x}$ that are water-filled at time $t$.

The state space is $\mathcal{S} = \{g: \mathbb{R}^+ \to [0,1] \,|\, g \text{ measurable}\}$.
The macroscopic water content is a derived quantity:
\begin{equation}
\theta[g] = \phi \int_0^\infty g(r)\, f(r)\,dr
\label{eq:theta}
\end{equation}
where $\phi$ is porosity.

Many distinct $g \in \mathcal{S}$ map to the same $\theta$.  This degeneracy is the source of history dependence: the macroscopic water content does not determine the microscopic configuration.

\subsection{State-dependent rate constant: Lucas--Washburn to Hagen--Poiseuille transition}
\label{sec:LW_rate}

With $g(r)$ now defined, we can specify how the rate constant depends on the filling state.  The Hagen--Poiseuille rate $\kappa_{\rm HP}(r)$ [Eq.~\eqref{eq:kappa}] describes steady flow through a fully filled cylinder.  But pores being invaded are not fully filled: a meniscus advances under capillary suction according to the Lucas--Washburn (LW) equation~\cite{Washburn1921}:
\begin{equation}
\ell\,\frac{d\ell}{dt} = \frac{r^2}{8\mu}\,\Delta P
\label{eq:LW_ode}
\end{equation}
where $\ell(t)$ is the position of the meniscus along the pore axis and $\Delta P$ [Pa] is the pressure difference across the invading water column (the capillary suction $2\gamma\cos\theta_c/r$ plus any imposed head), giving $\ell(t) \propto \sqrt{t}$: self-decelerating invasion.  The instantaneous volume rate $dV/dt \propto 1/\ell$ is fast at the start and slow near completion.

The kinetic equation operates on $g(r)$, an ensemble average over many pores at all invasion stages.  When $g(r)$ is small, most pores being invaded are at early stages ($\ell \ll \bar{L}$, fast rate).  When $g(r) \to 1$, the few remaining pores are nearly complete ($\ell \approx \bar{L}$, HP rate).  The ensemble-averaged meniscus position is $\bar{\ell}(g) = g\,\bar{L}$, giving the \textbf{state-dependent rate constant}:
\begin{equation}
\kappa(r,g) = \kappa_{\rm HP}(r)\cdot\frac{1}{g}
  = \frac{r^2}{8\mu\,\bar{L}^2\,\bar{\tau}^2}
  \cdot \frac{1}{g}
\qquad (g > 0)
\label{eq:kappa_LW}
\end{equation}
This diverges as $g \to 0$ (first pores fill almost instantaneously) and reduces to $\kappa_{\rm HP}$ at $g = 1$ (fully filled, steady Poiseuille).

\textbf{Regularization.}  The divergence at $g \to 0$ is bounded by the finite single-pore LW filling time, obtained by integrating Eq.~\eqref{eq:LW_ode} from $\ell=0$ to $\ell=\bar L$,
\begin{equation}
\tau_{\rm LW}(r) = \frac{4\mu\,\bar{L}^2}{r^2\,\Delta P}\, .
\label{eq:tau_LW}
\end{equation}
A physically consistent cutoff is:
\begin{equation}
\kappa(r,g) = \kappa_{\rm HP}(r)\cdot
  \frac{1}{\max(g,\, g_{\min})}
\label{eq:kappa_reg}
\end{equation}
where $g_{\min} = 1/n_p$ ($n_p$ = number of pores of class $r$ per REV), the resolution limit of the mean-field description.

\textbf{Known limits.}  At $g = 1$: $\kappa = \kappa_{\rm HP}$ (steady throughflow).  At $g = 1/2$: $\kappa = 2\,\kappa_{\rm HP}$ (time-averaged LW rate; derived in the companion paper~\cite{Rigon2026CE}).  The factor of two follows directly from the $1/g$ law: during Lucas--Washburn invasion the meniscus sits at depth $\ell$ in a pore of length $\bar L$, so the filled fraction of a class is $g \simeq \ell/\bar L$, while the instantaneous LW rate scales as $\kappa_{\rm HP}\,\bar L/\ell = \kappa_{\rm HP}/g$ (the driving head is fixed but the viscous column shortens as $\ell\to 0$).  Averaging the instantaneous rate over the ensemble of pores uniformly distributed in filling stage reproduces $\kappa(r,g)=\kappa_{\rm HP}/g$, which evaluates to $2\,\kappa_{\rm HP}$ at $g=1/2$.

\textbf{Two roles of $\kappa$.}  The state-dependent $\kappa(r,g)$ governs the kinetic terms (gain--loss, Sec.~\ref{sec:GL}): how fast $g(r)$ evolves.  The HP rate $\kappa_{\rm HP}(r)$ governs the conductivity formula (Sec.~\ref{sec:K_derived}): how much flows through fully filled pores at quasi-static conditions.  In the CE limit ($g \approx g_{\rm eq}$, contributing pores at $g \approx 1$), the two merge.  Outside this limit, the LW enhancement accelerates equilibration but does not change the equilibrium conductivity.

\subsection{Equilibrium distribution}
\label{sec:geq}

For a system at water content $\theta$, the unconstrained minimum (ignoring accessibility) places water in the pores of lowest chemical potential \cite{Millington1961-yf}.  We derive this result from the free energy.

\begin{figure*}[t]
\centering
\includegraphics[width=\textwidth]{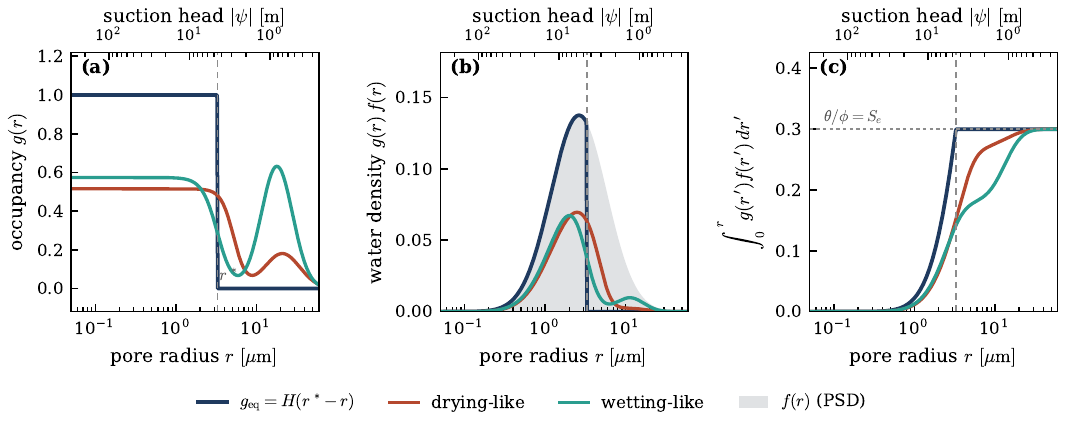}
\caption{\textbf{The filling fraction $g(r)$ as the irreducible state variable.}
(a)~Occupancy $g(r)$ for a Kosugi pore-size distribution (median $5\,\mu$m, $\sigma=0.8$) at fixed water content $\theta/\phi=S_e=0.3$: the \emph{reference} (single-elevation) equilibrium is the Young--Laplace step $g_{\rm eq}^{\rm ref}=H(r^*\!-r)$ (navy, $r^*\!=3.3\,\mu$m; in a gravity field the true equilibrium is the smeared form, Eq.~\eqref{eq:geq_smeared}), while two prototypical non-equilibrium states, ``drying-like'' (incomplete drainage, rust) and ``wetting-like'' (macropore activation with incomplete fine-pore filling, teal), spread occupancy smoothly across pore classes.
(b)~The corresponding water density $g(r)\,f(r)$ over the PSD $f(r)$ (grey); all three states enclose the same area, i.e.\ the same $\theta$.
(c)~The cumulative water $\int_0^r g\,f\,dr'$, which generalizes the retention CDF: the equilibrium curve coincides with the classical CDF up to $r^*$ and then plateaus at $S_e$, whereas the non-equilibrium curves reach the plateau more slowly because part of their water resides in pores larger than $r^*$.
The non-equilibrium profiles are illustrative states constrained only by mass conservation $\int g\,f\,dr=S_e$; they are not solutions of the redistribution operator (cf.\ Fig.~\ref{fig:g_relaxation}).}
\label{fig:g_state}
\end{figure*}

\textbf{Variational argument.}  The Gibbs free energy of the pore water is (Sec.~\ref{sec:thermo}, Eq.~\eqref{eq:Gibbs_main}):
\begin{equation}
\mathcal{F}[g] = \int_0^\infty \mu_w(r)\, g(r)\, d\nu
\end{equation}
with $d\nu = \phi\,f(r)\,dr$.  Here $\mu_w(r)$ is the chemical potential of the water held in pores of radius $r$, the work required to add a unit volume of water to that pore class, whose leading, capillary-dominated form is $\mu_w(r) = \mu_w^0 - 2\gamma\cos\theta_c/r + \ldots$, with $\mu_w^0$ the chemical potential of bulk free water; the complete expression, including the adsorptive, osmotic and gravitational terms, is given in Eq.~\eqref{eq:mu_main} of Sec.~\ref{sec:thermo}.  \begin{remark}[Volume measure]
\label{rem:measure}
This is
a natural choice because $g(r)$ is a filling
fraction and $d\nu$ is the volume of pore
class~$r$, so $\delta\mathcal{F}/\delta g$ is the
chemical potential per unit volume of water, the
thermodynamic force conjugate to~$g$.
\end{remark}
Minimizing $\mathcal{F}$ at fixed $\theta = \phi\int g\,f\,dr$ by a Lagrange multiplier $\lambda$, it is:
\begin{equation}
\frac{\delta\mathcal{F}}{\delta g}\bigg|_{\rm eq} = \mu_w(r) = \lambda \quad \forall r \text{ where } 0 < g < 1
\label{eq:KKT}
\end{equation}
Equation~\eqref{eq:KKT} says, in words, that wherever the occupancy is strictly between its bounds, all such pore classes must sit at the same chemical potential $\lambda$: otherwise, moving a little water from a class with higher $\mu_w$ to one with lower $\mu_w$ would decrease $\mathcal{F}$ while conserving $\theta$, and $g$ would not be a minimizer.  Since $\mu_w(r) = \mu_w^0 - 2\gamma\cos\theta_c/r$ is a strictly monotone decreasing function of $r$, the condition $\mu_w(r) = \lambda$ can be satisfied at most at a single radius $r^*$: partial filling ($0<g<1$) is therefore possible at most at that single, marginal radius, and no interval of $r$ can have $0 < g_{\rm eq}(r) < 1$.  Every other class must sit at a bound, and the constrained-optimization (Karush--Kuhn--Tucker, KKT) complementarity conditions decide which one: $g = 1$ where $\mu_w(r) < \lambda$ (pores that bind water more strongly than the marginal class fill completely) and $g = 0$ where $\mu_w(r) > \lambda$ (pores that bind it more weakly empty completely).  Combining the two statements, the unique minimizer is the step function:
\begin{equation}
g_{\mathrm{eq}}^{\rm ref}(r,\theta) = H(r^*(\theta) - r)
\label{eq:geq}
\end{equation}
where $r^*(\theta)$ is the frontier radius determined by $\phi\int_0^{r^*} f\,dr = \theta$ (as Fig. \ref{fig:g_state} a).  All pores smaller than $r^*$ are filled (lower chemical potential, stronger capillary binding); all larger pores are empty.  The strict monotonicity of $\mu_w(r)$ is what rules out intermediate occupancies $0 < g < 1$: at equilibrium, a pore of radius $r$ is either entirely full or entirely empty, a consequence of thermodynamics, not a separate physical postulate.

\textbf{The step is only the reference equilibrium.}  The sharp step in $r$ presumes that $\mu_w$ depends on radius alone.  The full chemical potential of Eq.~\eqref{eq:mu_main} also carries elevation and composition, $\mu_w=\mu_w(r,\mathbf{x})$, so the stationarity $\mu_w=\lambda$ fixes a threshold in the \emph{total} potential,
\begin{equation}
g_{\rm eq}^{\rm ref}(r,\mathbf{x}) = H\!\big(r^*(\mathbf{x})-r\big),\qquad \mu_w\!\big(r^*(\mathbf{x}),\mathbf{x}\big)=\lambda,
\label{eq:geq_total}
\end{equation}
with $r^*(\mathbf{x})$ fed by the capillary, gravitational, osmotic, and adsorptive heads.  Equation~\eqref{eq:geq_total} is only the \emph{reference} equilibrium, the singular single-elevation ($\rho_w\mathrm{g}\,\ell\to0$) idealization, and in a gravity field it is never the true equilibrium of a finite REV.  Within an averaging volume of vertical extent $\ell$ the pores of a class~$r$ are spread over elevations $z$ with distribution $p_r(z)$, and equilibrium $\mu_{\rm cap}(r)+\rho_w\mathrm{g}\,z=\lambda$ is met pore by pore, so the class-averaged occupancy is a \emph{smeared} step,
\begin{equation}
g_{\rm eq}(r,\mathbf{x}) = \int H\!\big(\lambda-\mu_{\rm cap}(r)-\rho_w\mathrm{g}\,z\big)\,p_r(z)\,dz .
\label{eq:geq_smeared}
\end{equation}
The smearing is \emph{two-sided}: classes just above $r^*$ keep their low-lying members filled, large pores holding a little water, while classes just below $r^*$ lose their high-lying members, small pores left unfilled, precisely the partial occupancies the bare step forbids.  Concretely, a large pore low in the profile stays filled while a smaller pore higher up has drained whenever $\rho_w\mathrm{g}\,\Delta z\gtrsim 2\gamma\cos\theta_c\,(1/r-1/R)$, each pore holding water within its Jurin rise $h_r=2\gamma\cos\theta_c/(\rho_w\mathrm{g}\,r)$.  The width $\sim(dr^*/d\mu_{\rm cap})\,\rho_w\mathrm{g}\,\ell$ closes only as $\ell\to0$, so $H$ is recovered exactly in the limit where gravity has no room to act.  This gravitational smearing is physically distinct from the entropic smoothing $\psi_T$ of Eq.~\eqref{eq:geq_FD}: the latter is a symmetric, scalar \emph{fluctuation} width, the former a generally asymmetric, \emph{deterministic} width carrying the length $\ell$ and the direction $\hat{\bm z}$.  The equilibrium retention curve therefore has at least three distinguishable sources of width, the pore-size spread $f(r)$, the configurational temperature $\psi_T$, and the gravitational $\rho_w\mathrm{g}\,\ell$, the last rendering it weakly non-local (mildly sample-height dependent), in line with the long-recognized sample-size sensitivity of pressure-plate retention data.

\emph{The equilibrium thus lies off the one-parameter family $H(r^*-r)$.}  No regime, not even quasi-static equilibrium, admits a single scalar ($\theta$, $r^*$, or $\psi$) as a complete description: the retention curve was the last place the classical scalar picture was meant to be exact, and gravity removes even that.  The filling fraction $g(r,\mathbf{x})$ is the irreducible state variable \emph{at} equilibrium as much as away from it, so the equilibrium collapse recovers the gravity-inclusive \emph{hydrostatic} retention physics ($r^*(z)=2\gamma\cos\theta_c/[\rho_w\mathrm{g}(z-z_{\rm wt})]$, Jurin's law pore by pore), not the bare Brooks--Corey/Kosugi step.  The reference~\eqref{eq:geq_total} remains an excellent stand-in for fine media, where the large-pore Jurin rise $h_R=2\gamma\cos\theta_c/(\rho_w\mathrm{g}\,R)$ is metres, and becomes a genuine effect only in sands and gravels, where $h_R$ is centimetres.  Of the non-capillary terms only gravity smears the step structurally: \emph{osmotic} $(-\nu RT c)$ is radius-blind, relocating $r^*(\mathbf{x})$ but leaving the step sharp (its new content is dynamical, coupled solute transport, osmotic flow $\propto\nabla c$, reordering pores only if $c$ becomes radius-correlated), and \emph{adsorptive} $(\Pi)$ renormalizes the threshold $\mu_{\rm cap}\to\mu_{\rm cap}+\Pi$, with a position-dependent tail only at the dry end through the ambient-humidity gradient. 

Note that at equilibrium, there is a one-to-one mapping between $\theta$, the critical radius $r^*$, and the matric potential can be obtained as 
\begin{equation}
\psi = -2\frac{\gamma\cos\theta_c}{\rho_w \mathrm{g}\, r^*}
\label{eq:YL}
\end{equation}
 This mapping is the basis of the Richards equation \cite{Tubini2022}.  Out of equilibrium, when $g \neq g_{\rm eq}$, different pore classes are at different local potentials and no single scalar $\psi$ characterizes the system.  The filling fraction $g(r)$ is then the irreducible state variable (Fig.~\ref{fig:g_state}).

When the system is not at equilibrium, redistribution proceeds by the same principle: water moves toward pores of lower chemical potential, whether those are smaller pores (capillary gradient dominates) or lower-elevation pores (gravitational potential dominates).  What is commonly called ``drainage'' is simply redistribution away from the current configuration toward the chemical potential minimum, the same physics as ``wetting,'' driven by the same potential $\Phi(r,r')$.  Evapotranspiration is likewise driven to minimize the chemical potential even if more complicated dynamics are implied \cite{Lehmann2008,DAmatoRigon2025}.


\subsection{Accessibility and percolation}
\label{sec:accessibility}

The \textbf{accessibility function} $a(r)$ gives the fraction of pores of radius $r$ that are connected to the external boundary (air phase for drainage, water source for wetting) through a continuous path of already invading pores.

For water connectivity, the relevant quantity is:
\begin{equation}
a_w(r) \approx \mathcal{P}\!\left[\int_0^r g(r')\, f(r')\, dr'\right]
\label{eq:aw}
\end{equation}
where $\mathcal{P}[\phi_w]$ is the percolation function, the probability that a site belongs to the spanning cluster given the occupied fraction $S_e$.


This introduces a critical water content $\theta_c = \phi\, S_c$ below which the water network fragments.  The accessibility function $a_w(r)$ vanishes for pore classes that are not connected to the spanning cluster, producing the topological cutoff that defines field capacity~\cite{Hunt2017,Hunt2001}.

When the water phase does not span the REV ($\theta<\theta_c$), no class belongs to a spanning cluster, so $a_w(r)=0$ for every $r$, which weights each class by $a_w(r)$ inside the integral (Sec.~\ref{sec:K_derived}), vanishes identically.  
A  complete treatment requires tracking the topology of the 
water-occupied subset $\{r : g(r) > 0\}$ as a function of~$\theta$.  
The natural tool is the Euler characteristic $M_2$, the third 
Minkowski functional of the water-filled 
phase~\cite{Vogel2010-wi}, which counts connected components minus loops 
and captures the fragmentation transition. 

This fragmentation has a dynamical reading that reconciles the percolation cutoff with the equilibrium target.  Below $\theta_c$ the free energy still possesses its minimizer $g_{\rm eq}(r,\mathbf{x})$, but the only operator that relaxes $g$ toward it, the liquid-network redistribution $\mathcal{C}$, is gated by $a_w(r)$ and switches off once the wetting phase ceases to span the REV.  The state is then \emph{arrested} short of equilibrium, $g\neq g_{\rm eq}$, held not by an energy barrier but by the loss of a connected relaxation pathway: a \emph{topological} metastability.  Thermodynamic equilibrium ($g_{\rm eq}$, set by $\mathcal{F}$) and reachability (set by $a_w$ and percolation) are carried by different objects and decouple exactly here, which is why the percolation cutoff does not contradict the variational construction of $g_{\rm eq}$, it constrains whether $g_{\rm eq}$ can be \emph{reached}, not whether it \emph{exists}.  The same obstruction underlies residual saturation, the gravity-stranded water of Eq.~\eqref{eq:geq_smeared}, and the wetting--drying holonomy: in each case $g_{\rm eq}$ is well defined but the dynamics cannot carry $g$ to it.  The present liquid-network model leaves $g$ frozen below $\theta_c$; a real soil still creeps toward $g_{\rm eq}$ through the slow film and vapour channels bracketed out here, so ``metastable'' is to be read as arrested with respect to the modelled dynamics, the dry-end residual relaxation deferred to that channel.

\section{The kinetic equation}
\label{sec:kinetic}

The filling fraction evolves according to the master equation representing the water budget for pores of radius $r$:
\begin{equation}
\pder{g(r,t)}{t} = \mathcal{G}(r) - \Lg(r) - \mathcal{E}(r) - \mathcal{T}(r)
\label{eq:master}
\end{equation}
where $\mathcal{G}$ is the gain (imbibition from other pore classes), $\Lg$ is the loss (redistribution to other classes), $\mathcal{E}$ is evaporation, and $\mathcal{T}$ is root uptake.  Precipitation does not appear as a separate term: in the spatially explicit formulation (Sec.~\ref{sec:spatial}), it enters as the inter-fiber exchange. 

The elementary process underlying $\mathcal{G}$ and $\Lg$ is a single pore-class-pair redistribution event: a filled pore of radius~$r'$ (donor) transfers water to an empty pore of radius $r < r'$ (receiver) when the chemical-potential difference $\Phi(r,r') = [\mu_w(r')-\mu_w(r)]/(\rho_w \mathrm{g}) > 0$ favours it,
The transfer proceeds at rate $\kappa(r)\,C(r,r')\,\Phi(r,r')$ per unit donor fraction.  The occupancy gates, $g(r')$ for a filled donor, $[1-g(r)]$ for an empty receiver, enter through the Stosszahlansatz~\cite{Boltzmann1872} and are detailed with Eqs.~\eqref{eq:G}--\eqref{eq:L} below.

The global counterpart is obtained by integrating over all pore sizes using the volume measure $d\nu = \phi\,f(r)\,dr$, recovering the water content $\theta$ (Eq.~\ref{eq:theta}), which represents the water mass budget conservation.

\subsection{Construction of the transition rates}
\label{sec:rates}


The kinetic equation is a master equation for the REV-averaged filling fractions.  Its construction follows four logical steps:

\textbf{Step 1: Single pore classes-pair physics.}  Consider one filled donor pore of radius $r'$ connected to one empty receiver pore of radius $r$.  The driving force is the chemical potential difference $\mu_w(r') - \mu_w(r) = \rho_w \mathrm{g}\,\Phi(r,r')$ (the inter-pore potential, Eq.~\eqref{eq:Phi}).  For viscous (Stokes) flow through the connecting throat, Hagen--Poiseuille gives the flow rate per unit pressure difference as $\kappa(r)$ [Pa$^{-1}$s$^{-1}$].  The transition rate from the filled state of $r'$ to the filled state of $r$ is:
\begin{equation}
w(r' \to r) = \kappa(r)\,C(r,r')\,\Phi(r,r') \quad [\mathrm{s}^{-1}]
\label{eq:transition_rate}
\end{equation}
This is  Hagen--Poiseuille flow with $C(r,r')$ encoding whether the two classes are connected.

\textbf{Step 2: REV averaging (Stosszahlansatz).}  Within the REV, $g(r)$ is the fraction of pores of radius $r$ that are water-filled.  For pore-pair exchange, the relevant quantities are: fraction of donor class $r'$ that is filled ($g(r')$), and fraction of receiver class $r$ that is empty ($1-g(r)$).  These are treated as statistically independent, the porous-media analog of the molecular chaos hypothesis in Boltzmann theory~\cite{Boltzmann1872,Cercignani1988}, so that the probability of the donor is full and the receiver is empty:
\begin{equation}
\Pr({\rm donor=full, receiver=empty})
  = g(r')\cdot[1-g(r)]
\label{eq:chaos}
\end{equation}
This is exact for uncorrelated filling; corrections for spatial correlations enter at $O(\xi/\ell_{\rm REV})$, where $\xi$ is the correlation length of the filling pattern (Sec.~\ref{sec:GL}).

\textbf{Step 3: Summing over donors and PSD weighting.}  This step builds the gain term explicated in Sec.~\ref{sec:GL}, where each factor is read off term by term.

\textbf{Step 4: Uniqueness of the gain--loss structure.}  The bilinear form $g(r')[1-g(r)]$ is a form consistent with: (a)~no gain into a full pore ($g = 1 \Rightarrow \mathcal{G} = 0$); (b)~no loss from an empty pore ($g = 0 \Rightarrow \Lg = 0$); (c)~detailed balance at the equilibrium $g_{\rm eq}$ (Sec.~\ref{sec:geq}); (d)~antisymmetry of the driving force $\Phi(r,r') = -\Phi(r',r)$.  Any rate that satisfies (a)--(d) and is linear in both occupancies must have this form.


\subsection{Gain and loss terms}
\label{sec:GL}

Step 3 above builds the gain term describing the water arriving at pores of radius $r$ from pores of radius $r'$:
\begin{equation}
\mathcal{G}(r) = \kappa(r,g)\,[1-g(r)] \!\int_0^\infty\! C(r,r')\,\rho_w\mathrm{g}\,\big[\Phi(r,r')\big]_+ g(r') f(r')\, dr'
\label{eq:G}
\end{equation}
Reading right to left: $f(r')\,dr'$ is the volume fraction of potential donors; $g(r')$ is the fraction of those that are filled (available water); $[\Phi(r,r')]_+=\max\{\Phi(r,r'),0\}$ is the driving force [Eq.~\eqref{eq:Phi}] rectified to its favourable direction, so that a transfer runs only downhill in chemical potential (the opposite direction is the loss term below);  $C(r,r')$ is the probability that the two pore classes are connected; $[1-g(r)]$ is the fraction of receiving pores that are empty (receiving capacity); $\kappa(r,g)$ is the state-dependent rate constant [Eq.~\eqref{eq:kappa_LW}], which incorporates Lucas--Washburn invasion dynamics for the ensemble of pores at various stages of filling.  As already noted after Eq.~\eqref{eq:Phi}, the product $\kappa(r,g) \cdot \rho_w \mathrm{g} \cdot \Phi$ has dimensions of $\mathrm{s}^{-1}$: it is the rate at which the ensemble of empty cylinders fills, given connected filled donors, accounting for the self-decelerating meniscus advancement.

Note that the factor $1/g$ in $\kappa(r,g)$ multiplied by $[1-g]$ produces $[1-g]/g$ in the gain term, which diverges as $g \to 0$.  The first few pores of a class fill very rapidly (meniscus at $\ell \ll \bar{L}$, LW rate $\gg$ HP rate).  The regularization~\eqref{eq:kappa_reg} bounds this at $\kappa_{\max} = \kappa_{\rm HP}/g_{\min}$.


%

\textbf{Domain of validity near the percolation threshold.}
The mean-field factorization governs the redistribution
rate; the percolation quantity $a_w(r)$
(Sec.~\ref{sec:accessibility}) enters as
static topological inputs computed independently from the
pore network $C(r,r')$ and $f(r)$.  Near the percolation
threshold $\theta_c$, the conductivity cutoff
and field capacity (both carried by $a_w \to 0$ for
disconnected classes) remain valid because they follow from
network topology, not from the rate equation. 

The factorization~\eqref{eq:chaos} is a mean-field closure, not a theorem: it treats the donor and receiver occupancies as statistically independent, valid while the water-phase correlation length $\xi$ stays small compared with the REV.  Its failure is itself physical.  As $\theta\to\theta_c^{+}$, $\xi$ grows, neighbouring occupancies correlate, and the invasion organizes into a few spanning channels rather than a uniform advance, this correlated, channelized regime \emph{is} preferential flow and fingering.  Preferential flow should therefore not be regarded as an independent constitutive process.  Rather, it emerges from the breakdown of the molecular-chaos assumption, represented by the leading configurational-correlation correction, $\langle g(r)\,g(r')\rangle-g(r)\,g(r')$, neglected by the closure approximation.  In porous media, this correction plays the same role as the pair-correlation terms neglected by the Stosszahlansatz in the kinetic theory of dilute gases.  The mean-field theory thus carries its own validity flag, accurate where flow is diffuse, degrading exactly where flow becomes preferential, and the correlated regime is the natural object of a future cluster (BBGKY-type) expansion. 

Percolation \cite{Hunt2001} enters $\mathcal{G}$ and $\Lg$ through the interplay between 
the static pore geometry and the dynamic water-phase topology.  


The accessibility function $a_w(r)$ 
(Sec.~\ref{sec:accessibility}) is a partial encoding of this topological 
information, projected onto pore-radius space; the full Euler characteristic $M_2$ (the third Minkowski functional introduced in Sec.~\ref{sec:accessibility})
contains richer structure, including the correlation length~$\xi$ that 
determines the spatial range of water-phase coupling.  

The loss term has the conjugate structure:
\begin{equation}
\Lg(r) = \kappa(r,g)\, g(r) \!\int_0^\infty\! C(r,r')\,\rho_w\mathrm{g}\big[\Phi(r',r)\big]_+[1{-}g(r')] f(r')\, dr'
\label{eq:L}
\end{equation}
Reading right to left: $f(r')$ are potential receivers; $[1-g(r')]$ is the acceptance probability (fraction empty); $\Phi(r',r)$ is the driving force (now from donor $r$ to receiver $r'$); $C(r,r')$ the connection probability; $g(r)$ the donor probability (fraction filled); $\kappa(r,g)$ the state-dependent rate [Eq.~\eqref{eq:kappa_LW}].  For the loss term, $\kappa(r,g)$ uses the current occupation of the donor class: fully saturated donors ($g = 1$) donate at the HP rate; partially filled donors ($g < 1$) donate faster per unit occupied volume because their menisci are at intermediate positions where the LW flow is faster.  The same factorization applies: $g(r)\cdot[1-g(r')]$ is the joint probability that a pore of class $r$ has water and a connected pore of class $r'$ has room, assumed independent.

\textbf{The closure is a mean-field exclusion process.}
The pair weight $g(r')[1-g(r)]$ is the transition rule of a lattice gas with
exclusion: a class receives only if it has vacancy ($1-g$) and donates only if it
is occupied ($g$).  Three properties follow.  (i)~The flux vanishes as $g\to1$ and
as $g\to0$, so the box $0\le g\le1$ is invariant for the occupancy-independent
Hagen--Poiseuille rate; with the Lucas--Washburn enhancement $\kappa\propto1/g$ the
loss term does not switch off as $g\to0$ and the bound is enforced by the
regularization~\eqref{eq:kappa_reg}.  (ii)~The
mobility factor $g(1-g)$ is the reciprocal of the curvature of the mixing entropy
of Sec.~\ref{sec:thermo}, so the kinetics is the gradient flow of a free energy
that includes that entropy.  (iii)~The mean-field truncation degrades only where
pair correlations become long-ranged ($\xi\to\infty$ near the percolation
threshold), as the leading term of a cluster expansion in $\xi/L_{\rm REV}$; this
coincides with the breakdown of the REV concept and of the Chapman--Enskog
reduction (Sec.~\ref{sec:breakdown}), not before~\cite{KipnisLandim1999}.



Drying mechanisms produce distinct pore emptying sequences.  
\begin{itemize}
\item \emph{Capillary drainage} (the laboratory idealization): Gas invades from the boundary, constrained by network topology.  Emptying proceeds from largest accessible pores to smallest.  The accessibility $a(r)$ is controlled by gas-phase percolation.
\item \emph{Stage~I evaporation} (atmosphere-limited, Sec.~\ref{sec:forcing_I}): Water at the surface, preferentially in smaller pores, where it redistributed after infiltration, is removed by atmospheric demand.  The liquid network replenishes the surface through capillary flow.  Pore emptying is controlled by the atmospheric boundary condition, not by the air-phase topology \cite{Or2013-pf}.
\item \emph{Stage~II evaporation} (diffusion-limited): The liquid network disconnects from the surface.  Water vaporizes within the soil, preferentially in larger pores (higher vapor pressure at lower capillary pressure), then diffuses upward as vapor.  This creates a pore emptying sequence opposite to capillary drainage: large pores empty by vaporization even when the gas phase has no hydraulic access to them.
\item \emph{Root uptake} (plant-mediated): Roots extract water through the root--soil interface, accessing pores that are hydraulically connected to the root network.  This can remove water from pores that are inaccessible to both the gas-phase boundary and the surface evaporation front, producing yet another emptying sequence.
\end{itemize}
These mechanisms collectively tend to leave water in more strongly-held configurations (more negative $\psi$) compared to the non-selective filling during wetting.  The result: at the same water content $\theta$, the drying path has water in smaller pores (lower energy, more negative $\psi$) while the wetting path has water in larger pores (higher energy, less negative $\psi$).  Measurements consistently confirm this: $\psi_{\rm wet}(\theta) > \psi_{\rm dry}(\theta)$ (less negative on wetting)~\cite{Poulovassilis1962}, a direct consequence of the non-equilibrium nature of the wetting process and the mechanism-dependent selectivity of drying.

The trapping described by $a(r) < 1$ is a topological phenomenon (percolation connectivity of the gas phase), not the ``ink-bottle effect'' of the classical literature, which invokes a local geometric mechanism (large bodies behind narrow necks).  Lattice Boltzmann simulations~\cite{Hosseini2024} demonstrate that hysteresis persists even when the classical ink-bottle mechanism is explicitly eliminated, confirming that the topological asymmetry, not the local large-pore/small-pore geometry, is the primary source.

There is only one equilibrium configuration $g_{\rm eq}$ at each $\theta$ (Eq.~\ref{eq:geq}).  What differs between wetting and drying is not the target but the path to the target: during drying, the single-cluster expansion constraint prevents the system from reaching $g_{\rm eq}$; during wetting, multi-site nucleation approaches $g_{\rm eq}$ from a different direction with different intermediate configurations.  Hysteresis is a topological and kinematic phenomenon, not a thermodynamic one.

%

\subsection{Thermodynamic structure}
\label{sec:thermo}

The free-energy functional and the kinetic equation are not independent: the kinetic equation is the gradient flow of the free energy.  

The chemical potential of water in a pore of radius~$r$,
at uniform temperature and atmospheric pressure, is:
\begin{equation}
\mu_w(r) = \mu_w^0
  - \frac{2\gamma\cos\theta_c}{r}
  + \Pi(h(r))
  - \nu R T\, c(r)
  + \rho_w \mathrm{g}\, z
\label{eq:mu_main}
\end{equation}
which generalizes the simplified capillary-plus-gravity potential used throughout the paper~\cite{Tuller1999}.  $\mu_w^0$ is the chemical potential of bulk free
water, and the remaining terms encode from left to right capillary,
adsorptive, osmotic, and gravitational contributions.

\textbf{Free-energy functional.}  The \textbf{Gibbs free energy} of the pore water is:
\begin{equation}
\mathcal{F}[g] = \underbrace{
  \phi \int_0^\infty
  \mu_w(r)\, g(r)\,  f(r)\,dr
  }_{\text{capillary contribution}}
  + \underbrace{
  M_1 \int_0^{h_{\rm max}} \Pi(h)\, dh
  }_{\text{adsorptive contribution}}
\label{eq:Gibbs_main}
\end{equation}
where $M_1$ [$\mathrm{m^2/m^3}$] is the specific surface area of the pore walls, i.e.\ the pore--solid interfacial area per unit REV volume (its role is discussed below), while the capillary part uses the volume measure $d\nu = \phi\,f(r)\,dr$.

\textbf{Configurational entropy and convexity.}  Because $g(r)$ is an occupancy variable bounded between 0 and 1, the free energy includes, in addition to the energetic contributions introduced above, a combinatorial (mixing) entropy.  This term represents the configurational entropy associated with distributing the occupied pore fraction among pores belonging to class $r$ within the REV~\cite{KipnisLandim1999}.  Writing it explicitly,
\begin{multline}
\mathcal{F}[g] = \phi\!\int_0^\infty\!\! \Big[\mu_w(r)\,g
  + \psi_T\big(g\ln g + (1-g)\ln(1-g)\big)\Big] f\,dr \\
  + \mathcal{F}_{\rm ads},
\label{eq:F_entropy}
\end{multline}
where $\psi_T>0$ is a configurational potential scale (an effective temperature
expressed in pressure units, set by pore-scale disorder and thermal/Haines-jump
fluctuations~\cite{Haines1930}) and $\mathcal{F}_{\rm ads}$ is the adsorptive term of
Eq.~\eqref{eq:Gibbs_main}.  The bracketed term $g\ln g+(1-g)\ln(1-g)$ is the configurational (mixing) entropy of a two-state occupancy: within class~$r$ a fraction $g$ of the pores is filled and $1-g$ empty, and by Stirling's formula $-[g\ln g+(1-g)\ln(1-g)]$ counts, per pore, the number of ways of distributing the filled fraction among the class-$r$ pores (the counting is carried out in Appendix~\ref{app:entropy}, which is more than a technical aside: it makes explicit the mean-field closure linking the microstructure to the mesoscale, since the exchangeability of pores within a class is exactly the Stosszahlansatz whose failure near percolation generates the correlation corrections of Sec.~\ref{sec:GL}).  It vanishes at the packed limits $g\in\{0,1\}$, is maximal at $g=1/2$, and its curvature $1/[g(1-g)]$ sets the entropic stiffness that makes $\mathcal{F}$ strictly convex.  Operationally, $\psi_T$ is fixed by the width of the equilibrium retention transition, equivalently, by the magnitude of within-class occupancy fluctuations, so a sharper measured retention step implies a smaller $\psi_T$; its quantitative calibration against retention data or pore-network fluctuation statistics is left to future work.  The functional derivative is then
\begin{equation}
\frac{\delta\mathcal{F}}{\delta g(r)} = \mu_w(r)
  + \psi_T \ln\!\frac{g(r)}{1-g(r)} ,
\label{eq:dF_entropy}
\end{equation}
and stationarity $\delta\mathcal{F}/\delta g = \lambda$ at fixed $\theta$ gives a
Fermi--Dirac occupancy,
\begin{equation}
g_{\rm eq}(r) = \frac{1}{1+\exp\!\big[(\mu_w(r)-\lambda)/\psi_T\big]} .
\label{eq:geq_FD}
\end{equation}
As $\psi_T\to 0$ (the athermal limit used throughout the
rest of the paper), Eq.~\eqref{eq:geq_FD} sharpens to the step
$g_{\rm eq}=H(r^*\!-r)$ of Eq.~\eqref{eq:geq}, with $r^*$ fixed by
$\mu_w(r^*)=\lambda$; all results below are recovered exactly in this limit.
For finite $\psi_T$ the equilibrium retention curve is a smooth sigmoid rather
than a step, so part of the observed smoothness of $\theta(\psi)$ is due to
occupancy fluctuations rather than purely a consequence of the width of $f(r)$.
The entropy also settles the well-posedness of the variational structure: the
integrand of Eq.~\eqref{eq:F_entropy} is strictly convex in $g$ for
$\psi_T>0$ (its second variation is $\psi_T/[g(1-g)]>0$), so the equilibrium
\eqref{eq:geq_FD} is the unique global minimizer. 

\textbf{Role of the specific surface area $M_1$.}
The chemical potential $\mu_w$ is an intensive
quantity: the disjoining pressure~$\Pi(h)$ depends on
the film thickness~$h(r)$ alone, not on how much surface
is present.  What changes with $M_1$ is the
amount of adsorbed water.  The total adsorbed
water content in the REV is
$\theta_{\rm ads} = M_1 \cdot h_{\rm eq}(\mu_w)$,
where $h_{\rm eq}(\mu_w)$ is the equilibrium film
thickness obtained by inverting $\Pi(h) = \mu_w^0 - \mu_w$.
Since $\mu_w$ is intensive and $M_1$ is extensive
(surface per unit REV volume), $\theta_{\rm ads}$
scales linearly with~$M_1$.  Clayey soils with
$M_1 \sim 10^7$--$10^8$~m$^2$/m$^3$ therefore retain
far more adsorbed water than sandy soils with
$M_1 \sim 10^4$--$10^5$~m$^2$/m$^3$, not because
the binding energy per molecule is different, but
because there are more molecules bound per unit REV
volume.  The crossover between capillary-dominated
and adsorption-dominated regimes occurs at the smallest
capillary pore radius~$r_{\rm min}$: above
$\theta_{\rm cap,min} = \phi\int_0^{r_{\rm min}} f\,dr$,
capillarity dominates; below it, adsorption dominates.
A full treatment of the adsorptive regime requires further developments, e.g ~\cite{Luo2022-mu}.

\textbf{The driving potential as a free-energy gradient.}  The inter-pore driving potential~$\Phi$ is the difference of functional derivatives:
\begin{equation}
\rho_w \mathrm{g}\;\Phi(r,r')
  = \frac{\delta\mathcal{F}}{\delta g(r')}
  - \frac{\delta\mathcal{F}}{\delta g(r)}
  = \mu_w(r') - \mu_w(r)
\label{eq:Phi_thermo}
\end{equation}
This identifies $\Phi$ as the thermodynamic force conjugate to the water flux between pore classes $r$ and~$r'$.  The antisymmetry
$\Phi(r,r') = -\Phi(r',r)$ follows immediately.

\textbf{The kinetic equation as a gradient flow.}  Substituting \eqref{eq:Phi_thermo} into the gain--loss terms:
\begin{equation}
\pder{g}{t} = -\int_0^\infty M(r,r')\,
  \left[\frac{\delta\mathcal{F}}{\delta g(r)}
  -\frac{\delta\mathcal{F}}{\delta g(r')}\right]\,f(r')\,dr'
\label{eq:gradient_flow}
\end{equation}
where the intra-REV redistribution mobility is
\begin{equation}
M(r,r') = \kappa_s(r,r')\,C(r,r')\,\sqrt{g(r)\,[1-g(r)]\,g(r')\,[1-g(r')]} ,
\label{eq:M}
\end{equation}
symmetric in $r\leftrightarrow r'$ and non-negative, each factor is
non-negative ($\kappa_s\ge0$, $C\ge0$, $0\le g\le1$), so $M\ge0$ pointwise and the
Rayleigh dissipation below is non-negative.
The \emph{geometric} mean of the two occupancy factors, rather than their
product, is what Arrhenius (heat-bath) exclusion rates deliver on linearization: by an \emph{Arrhenius} (or heat-bath) rate we mean a transfer rate of the form $w_{r'\to r}\propto\kappa_s\,e^{-[\mu_w(r)-\mu_w(r')]/2\psi_T}$, the standard choice for a lattice gas exchanging with a thermal bath \cite{Kawasaki1966,Huang2009}: the rate depends on the driving through the Boltzmann factor of half the chemical-potential jump, so that forward and backward rates satisfy detailed balance with respect to $\mathcal{F}$ automatically.  Expanding this rate about equilibrium, where $\mu_w=\mu_0+\psi_T\ln[g/(1-g)]$, the occupancy factors of donor and receiver enter as $\sqrt{g(1-g)\,g'(1-g')}$, which is the mobility of Eq.~\eqref{eq:M} (Table~\ref{tab:symbols}); the double product would gate each pair twice, once as exclusion, once inside
the intrinsic conductance, and, as the companion paper
shows~\cite{Rigon2026CE}, would hand the slow end of the relaxation spectrum to
the edges of the active band instead of to the frontier $r^*$.  Here
\begin{equation}
\kappa_s(r,r') = \frac{2\,\kappa(r)\,\kappa(r')}{\kappa(r)+\kappa(r')}
\label{eq:kappa_sym}
\end{equation}
is the symmetric pair conductance, the harmonic mean of the two single-pore conductances, i.e.\ the bottleneck-limited series resistance of the edge, which restores the reciprocity $M(r,r')=M(r',r)$ required by the gradient-flow structure (replacing the single-pore $\kappa(r)$ of the directed rate~\eqref{eq:transition_rate} would break it).
Equation~\eqref{eq:gradient_flow} should be read as follows.  The free energy $\mathcal{F}$ assigns to every configuration $g$ a scalar ``height'', and its functional derivative $\delta\mathcal{F}/\delta g(r')=\mu_w(r')$ is the slope of that landscape in the direction of pore class $r'$: it measures how much $\mathcal{F}$ would decrease if a little water were removed from class $r'$.  The difference structure is essential: for symmetric $M$ it is what makes the flow conserve the water content $\int g\,f\,dr$ exactly, pair by pair.  The kinetic equation moves $g$ downhill: each class $r$ changes at a rate obtained by summing the slope differences against all the classes $r'$ it can exchange water with, weighted by the mobility $M(r,r')$ of Eq.~\eqref{eq:M}, which measures how easily that exchange can actually occur, it vanishes whenever either class is completely full or completely empty (nothing to move, or no room to receive) or the two classes are not connected ($C=0$).  Equation~\eqref{eq:gradient_flow} is the equilibrium-consistent completion of the gain--loss dynamics of Eqs.~\eqref{eq:G}--\eqref{eq:L}: the two agree in their driving (the slope of $\mathcal{F}$, which reduces to the capillary head when the mixing entropy is dropped), in their gates, and in their equilibrium, and differ only in how the gates enter the mobility, the geometric mean of Eq.~\eqref{eq:M} being the Arrhenius linearization of the rectified rates.  Its merit is that the thermodynamic driving and the kinetic accessibility (the mobility) appear as separate factors.
This is the Onsager gradient flow~\cite{Onsager1931a,Onsager1931b,Jordan1998,Mielke2025}: the kinetic equation descends the free-energy landscape $\mathcal{F}$ at a rate set by the state-dependent mobility $M$.  The Rayleighian, the Onsager principle from which Eq.~\eqref{eq:gradient_flow} follows, and its further consequences (Onsager reciprocity, the Wasserstein/JKO metric structure, and the GENERIC extension to non-isothermal forcing) are given in Appendix~\ref{app:variational}.

\begin{figure*}[t]
\centering
\includegraphics[width=\textwidth]{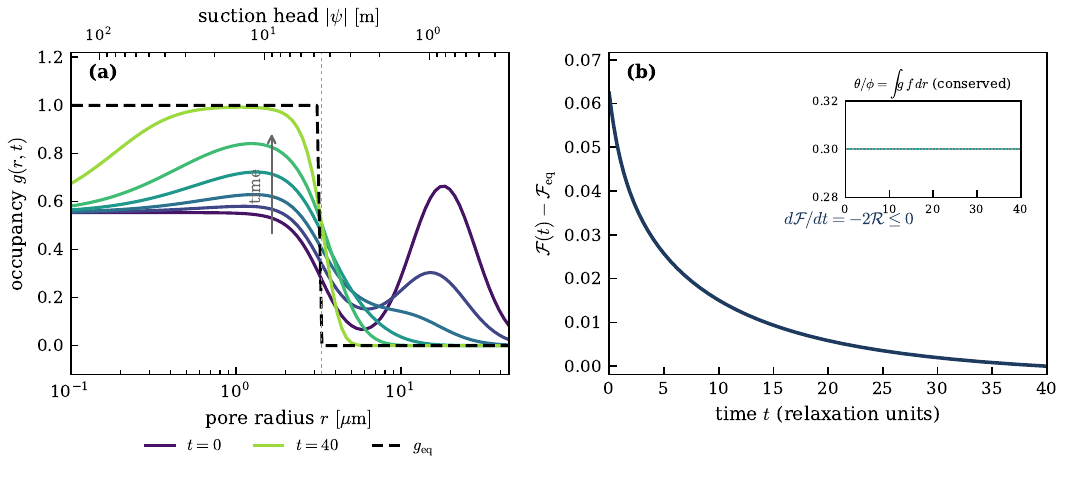}
\caption{\textbf{Relaxation of the kinetic equation and the $H$-theorem.}
Numerical integration of the mass-conserving redistribution dynamics $\partial_t g = \mathcal{G}-\Lg$ [Eqs.~\eqref{eq:G}--\eqref{eq:gradient_flow}] on $90$ pore classes (Kosugi PSD, capillary driving potential, symmetric pair conductance $\kappa_s$ of Eq.~\eqref{eq:kappa_sym}, like-size-biased connectivity $C$), starting from the ``wetting-like'' state of Fig.~\ref{fig:g_state}.
(a)~The occupancy $g(r,t)$ evolves (light to dark) from the macropore-activated initial condition toward the reference equilibrium step $g_{\rm eq}^{\rm ref}=H(r^*\!-r)$ (dashed): water drains from the large pores and fills the small ones, exactly the capillary preference encoded in $\Phi$.
(b)~The Gibbs free energy $\mathcal{F}(t)=\int\mu_w\,g\,f\,dr$ decreases monotonically to its equilibrium value, the discrete $H$-theorem $d\mathcal{F}/dt=-2\mathcal{R}\leq0$; the inset shows the water content $\theta/\phi=\int g\,f\,dr$ conserved to machine precision throughout.
Unlike Fig.~\ref{fig:g_state}, these are genuine solutions of the kinetic equation, not constructed profiles: the binned system $\dot g=\mathcal{G}-\Lg$ with the rectified rates of Eqs.~\eqref{eq:G}--\eqref{eq:L}, $\kappa=\kappa_{\rm HP}$, a Kosugi $f$ and a connectivity local in $\ln r$, integrated with a stiff adaptive solver (LSODA, relative tolerance $10^{-7}$); code in the Supplemental Material.}
\label{fig:g_relaxation}
\end{figure*}

\textbf{Relaxation dynamics.}  The time rate of change of $\mathcal{F}$ along any unforced solution is:
\begin{equation}
\frac{d\mathcal{F}}{dt}
  = -2\,\mathcal{R}[\dot{g}] \leq 0
\label{eq:H_theorem}
\end{equation}
where $\mathcal{R}[\dot{g}] = \tfrac{1}{2}\iint M^{-1}(r,r')[\dot{g}(r)-\dot{g}(r')]^2 f\,f'\,dr\,dr' \geq 0$ is the Rayleigh dissipation potential, defined on the support of $M$ (the
active classes) (here $\dot{g}\equiv\partial g/\partial t$ denotes the local time derivative of the filling fraction).  This is the $H$-theorem~\cite{Huang2009}: $\mathcal{F}$ decreases monotonically and the system relaxes toward $g_{\rm eq}$. The relaxation time of pore class $r$ follows from the same structure without new
objects: linearizing Eq.~\eqref{eq:gradient_flow} about $g_{\rm eq}$, the class
$r$ exchanges water with all its neighbours at the total equilibrium conductance
$\psi_T\!\int_0^\infty M(r,r';g_{\rm eq})\,f(r')\,dr'$, while the perturbation it
can store is weighted by $m(r)=g_{\rm eq}(1-g_{\rm eq})$ (the entropic stiffness
$\partial\mu_w/\partial g=\psi_T/m$), so that
\begin{equation}
\tau(r)=\frac{m(r)}{\psi_T\int_0^\infty M(r,r';g_{\rm eq})\,f(r')\,dr'} .
\label{eq:tau_relax}
\end{equation}
Because $m$ is largest at the frontier $r^*$ and vanishes in the wings, the
classes that relax slowest are those at $r^*$, while the wings of the
distribution pin almost instantly; the spectral analysis behind
Eq.~\eqref{eq:tau_relax} is the subject of the companion paper~\cite{Rigon2026CE}.  Figure~\ref{fig:g_relaxation} shows a numerical integration of this relaxation from a macropore-activated initial state, showing the monotonic decrease of $\mathcal{F}$ and mass conservation to machine precision.

%

\section{Inter-REV coupling and the macroscopic current}
\label{sec:spatial}
Up to this point every field has been evaluated inside a single REV at a
fixed macroscale position, so the spatial coordinate stayed implicit.  We now make
it explicit: $g=g(r,\mathbf{x},t)$, and $\mu_w$, $C$, and $f$ likewise acquire an
$\mathbf{x}$ dependence through the local pore structure.  The redistribution
operator of Eq.~\eqref{eq:master} is the intra-REV part; alongside it we define an
inter-REV (boundary) part,
\begin{equation}
\begin{aligned}
\Coll &{}:= \mathcal{G}-\Lg \quad(\text{intra-REV}),\\
\Coll_\delta &{}:= \mathcal{G}_\partial-\Lg_\partial \quad(\text{inter-REV}),
\end{aligned}
\label{eq:C_split}
\end{equation}
where $\Coll$ moves water between pore classes \emph{within} the REV at
$\mathbf{x}$ (conserving the REV water content and carrying no macroscopic current),
while $\Coll_\delta$ exchanges water with the neighbouring REVs and is the source
of macroscopic transport; the volumetric sinks $\mathcal{E},\mathcal{T}$ act within
the REV.  This section constructs $\Coll_\delta$, whose continuum limit
(Sec.~\ref{sec:CL}) is the divergence $\nabla\!\cdot\mathbf{F}$ of the macroscopic
flux.

\subsection{The boundary current between REVs}
\label{sec:connection}

The net inter-REV exchange for class $r$ is therefore the boundary operator $\Coll_\delta(r,\mathbf{x})$,
which moves water from one REV to the next.  

The boundary operators have the same bilinear structure
as the internal gain and loss
(Eqs.~\eqref{eq:G}--\eqref{eq:L}), but with
donor and receiver populations evaluated at
different spatial points separated by the REV
spacing~$\ell$.  Consider a neighbor at
$\mathbf{x}_+ = \mathbf{x} + \ell\,\hat{n}$.  The
boundary gain (water arriving at class~$r$ from the
neighbor) and the boundary loss (water leaving class~$r$
toward the neighbor) are:
\begin{multline}
\mathcal{G}_\partial(r)
  = \kappa(r)\,[1-g(r,\mathbf{x})]
  \\
  \times\int_0^\infty C_\partial(r,r')\,\rho_w\mathrm{g}\,\Phi(r,r')\,
  g(r',\mathbf{x}_+)\,f(r')\,dr'
\label{eq:G_bdy}
\end{multline}
\begin{multline}
\Lg_\partial(r)
  = \kappa(r)\, g(r,\mathbf{x})
  \\
  \times\int_0^\infty C_\partial(r,r')\,\rho_w\mathrm{g}\,\Phi(r',r)\,
  [1-g(r',\mathbf{x}_+)]\,f(r')\,dr'
\label{eq:L_bdy}
\end{multline}
The boundary connectivity $C_\partial(r,r')$ encodes
which pore classes can exchange water across the REV
interface; it need not equal the internal
connectivity~$C(r,r')$.  Note that $g(r')$ and $1-g(r')$
in the integrands are evaluated at the
neighbor~$\mathbf{x}_+$, not at~$\mathbf{x}$: this
spatial offset is what makes the boundary operators transport water between
REVs rather than redistribute it within one.

The boundary operator $\Coll_\delta$ contains the inter-REV flux,
entangled with the gradient of~$g$.  Extracting it requires contracting the REV
to a point: the gradient expansion of this boundary difference, and the macroscopic
current it yields, are carried out in the continuum limit of Sec.~\ref{sec:CL}.

\section{The continuum limit}
\label{sec:CL}


Let $V=V(\mathbf{x})$ be the representative elementary volume (REV) of linear
size $\ell$ centred at the macroscale point $\mathbf{x}$, with boundary
$\partial V$ and outward unit normal $\mathbf{n}$.  We collapse the REV to a point
through the spatial homogenisation limit $\varepsilon := \ell/L_{\rm mac}\to 0$.
Every integral below sits at the REV scale: the spatial integrals
$\int_V,\,\oint_{\partial V}$ are over the REV and its boundary, while the radius integrals $\int dr$ are
intra-REV, over pore-class space at the fixed point $\mathbf{x}$.  The
divergences that appear are the per-unit-volume limits of boundary fluxes at a
point, not balances over a macroscopic domain.  
Throughout, the redistribution operator
$\Coll\equiv\mathcal{G}-\mathcal{L}$ of the kinetic equation~\eqref{eq:master} is
kept in dimensional form, carrying its own physical rate; no Damk\"ohler factor is
introduced.

\subsection{The pore-resolved budget and the entangled flux}
We take the continuum limit directly at the level of the pore classes; the bulk
water balance follows below as its zeroth moment.  Resolving the REV average by pore
class, with $g(r,\mathbf{x},t)$ the class-$r$ occupancy so that
$\theta=\phi\!\int g\,f\,dr$, two processes change $g$: inter-REV transport, which
carries a radius-resolved flux $\mathbf{F}(r,\mathbf{x},t)$ across $\partial V$, and
intra-REV redistribution $\Coll[g]=\mathcal{G}-\Lg$, which conserves the REV water
content.  The class-$r$ balance is
\begin{equation}
\frac{d}{dt}\big[g(r,\mathbf{x},t)\,|V|\big]
  = -\oint_{\partial V}\mathbf{F}(r,\mathbf{x},t)\cdot\mathbf{n}\,dA
  + |V|\,\Coll[g] .
\label{eq:pore-budget}
\end{equation}
with $\Coll[g]$ the occupancy-gated relaxation of Eq.~\eqref{eq:master} (its
gain--loss content in Eqs.~\eqref{eq:G}--\eqref{eq:gradient_flow}), which descends
$\mathcal{F}[g]$ at fixed $\theta$ and is never a flux.

The flux $\mathbf{F}$ is not a new postulate: it is what the boundary-exchange
operator $\Coll_\delta=\mathcal{G}_\partial-\Lg_\partial$ of
Sec.~\ref{sec:connection} becomes when the REV is contracted to a point.  Expanding
the neighbour's filling distribution $g(r',\mathbf{x}_+)$ to first order in the REV
spacing $\ell$ and using the antisymmetry $\Phi(r',r)=-\Phi(r,r')$, the zeroth-order
terms cancel, they are the internal redistribution
$\mathcal{G}_{\rm int}-\Lg_{\rm int}$ already in $\Coll$, and the first-order terms
give a directional current linear in $\nabla g$ (the full algebra is given
in~\cite{Rigon2026CE}),
\begin{multline}
\Coll_\delta
  = \ell\,\hat{n}\cdot\;\kappa(r)\,[1-2g(r)]
  \\
  \times\int_0^\infty C_\partial(r,r')\,\rho_w\mathrm{g}\,\Phi(r,r')\,
  \nabla g(r',\mathbf{x})\,f(r')\,dr'
  + O(\ell^2)
\label{eq:J_result}
\end{multline}
Reading the kernel off Eq.~\eqref{eq:J_result}, this current is
\begin{equation}
\mathbf{F}(r,\mathbf{x},t)
  = -\!\int_0^\infty \Gamma(r,r';g)\,\nabla g(r',\mathbf{x},t)\,dr' ,
\label{eq:F_from_connection}
\end{equation}
with the entangled kernel
$\Gamma:=\ell^{2}\,\kappa(r)\,[1-2g(r)]\,C_\partial(r,r')\,\rho_w\mathrm{g}\,\Phi(r,r')\,f(r')$, so that
$\Coll_\delta=-\nabla\!\cdot\mathbf{F}+O(\ell^2)$; the factor $\ell^2$ (one $\ell$
from the face term of Eq.~\eqref{eq:J_result}, one from the divergence) is
what gives $\Gamma$ the units $\mathrm{m^{2}\,s^{-1}}$ of a diffusivity and
$\mathbf F$ those of a flux, $\mathrm{m\,s^{-1}}$; in the continuum limit
$\ell\to\bar L$ and $\ell^2\kappa$ becomes the pore-scale diffusivity
$r^2/(8\mu\bar\tau^2)\,\rho_w\mathrm g\,\Phi$.
Equation~\eqref{eq:F_from_connection} is the honest, pre-closure form of the flux:
transport coefficient $\Gamma$ and driving force $\nabla g$ are still entangled,
and the current is nonlocal in $r$.

Two features of this current carry over from the boundary expansion.
\textbf{(i) The integral over~$r'$.}  The flux carried by class~$r$ is a sum
over all connected donor/receiver classes~$r'$, weighted by $C_\partial\,\Phi\,f'$;
this integral does not disappear in the continuum limit but is absorbed into the
pore-class-resolved flux.  \textbf{(ii) Entanglement of conductivity and gradient.}
The kernel $\kappa(r)(1-2g)C_\partial\Phi$ mixes the transport coefficient with
the driving force $\nabla g$; their separation into a scalar~$K$ and a gradient is not
available at this level and is the work of the Chapman--Enskog reduction
(Sec.~\ref{sec:K_derived}).

The boundary flux is the conservative form of the spatial transport operator
$\mathfrak{T}$, inter-REV transport conserves water, so it can only cross
$\partial V$,
\begin{equation}
\mathfrak{T}\,g \equiv
  \frac{1}{|V|}\oint_{\partial V}\mathbf{F}\cdot\mathbf{n}\,dA
  \;\xrightarrow{\;\ell\to0\;}\; \nabla\cdot\mathbf{F} ,
\label{eq:transport}
\end{equation}
The surface flux carries an unambiguous \emph{direction}: at each point of $\partial V$ the sign of $\mathbf{F}\cdot\mathbf{n}$ says whether water leaves the REV ($\mathbf{F}\cdot\mathbf{n}>0$, along the outward normal $\mathbf{n}$) or enters it ($\mathbf{F}\cdot\mathbf{n}<0$).  This lets us split $\mathfrak{T}$ into an inward and an outward boundary current,
\begin{equation}
\mathsf{Q}_{\rm in}\,g \equiv \frac{1}{|V|}\oint_{\partial V}\!\big[\mathbf{F}\cdot\mathbf{n}\big]_-\,dA,
\qquad
\mathsf{Q}_{\rm out}\,g \equiv \frac{1}{|V|}\oint_{\partial V}\!\big[\mathbf{F}\cdot\mathbf{n}\big]_+\,dA,
\label{eq:Qsplit}
\end{equation}
with $[\,y\,]_\pm=\max(\pm y,0)$ the positive/negative parts, so that $\mathfrak{T}=\mathsf{Q}_{\rm out}-\mathsf{Q}_{\rm in}$ term by term on $\partial V$.  The split is defined here, at the boundary, and \emph{not} on the divergence $\nabla\!\cdot\mathbf{F}$: the flux $\mathbf{F}=-\!\int\Gamma\,\nabla g\,dr'$ [Eq.~\eqref{eq:F_from_connection}] carries the sign-indefinite kernel $\Gamma=\kappa(r)\,[1-2g(r)]\,C_\partial\,\Phi\,f(r')$, whose factors $[1-2g(r)]$ (which flips at half-filling) and $\Phi(r,r')$ (antisymmetric) make the local transport direction depend on more than $\nabla g$ alone, so ``inflow'' and ``outflow'' are properties of the net current $\mathbf{F}\cdot\mathbf{n}$ through the surface, not of the bulk field or its gradient.
Dividing the class-$r$ budget~\eqref{eq:pore-budget} by $|V|$ and letting
$\ell\to0$ gives the local mesoscale balance
\begin{equation}
\boxed{\;\partial_t g(r,\mathbf{x},t)
  + \nabla\cdot\mathbf{F}(r,\mathbf{x},t) = \Coll[g]
  \;-\;\mathcal{E}(r)\;-\;\mathcal{T}(r)\;}
\label{eq:pore-point}
\end{equation}
in dimensional form, valid at any timescale separation, boxed because it is the fundamental equation of the theory, of which everything that follows (the Chapman--Enskog reduction, Richards' equation, the conductivity formula) is a limit or a moment (Sec.~\ref{sec:fundamental}).  Here $\Coll[g]
=\mathcal{G}-\Lg$ is the intra-REV redistribution of Eq.~\eqref{eq:C_split}, and the
evaporation and root-uptake sinks $\mathcal{E},\mathcal{T}$ of Eq.~\eqref{eq:master}
are written explicitly wherever they act.  The continuum limit \emph{adds} only the
transport divergence $\nabla\!\cdot\mathbf{F}$ to the intra-REV dynamics of
Eqs.~\eqref{eq:master} and~\eqref{eq:gradient_flow}; it does not change their
mechanics.  The redistribution operator $\Coll$ and the sinks
$\mathcal{E},\mathcal{T}$, and with them the gradient-flow structure, the
$H$-theorem, and the well-posedness of Sec.~\ref{sec:thermo}, carry over unchanged,
now evaluated at each point $\mathbf{x}$.  The REV has vanished as a geometric
object; it survives only through the local pore-size distribution
$f(r,\mathbf{x})$ carried by the operators.

\subsection{Boundary conditions}
Equation~\eqref{eq:pore-point} is closed by conditions at the lower boundary
(a water table or free drainage) and at the surface, where the forcing enters.  Surface
forcing is a \emph{flux} (Neumann) condition, not a state (Dirichlet) condition on $g$:
the precipitation intensity $I(t)$ fixes the inward normal current,
\begin{multline}
-\,\mathbf{F}(r,\mathbf{x}_{\rm top},t)\cdot\mathbf{n}
  = q_{\rm in}(r,t) \\
  = I(t)\,\frac{[1-g(r)]\,f(r)}{\int_0^\infty [1-g(r')]\,f(r')\,dr'}
  \;\xrightarrow[\;g\to0\;]{}\; I(t)\,f(r) ,
\label{eq:precip_bc}
\end{multline}
Reading Eq.~\eqref{eq:precip_bc}: the incoming intensity $I(t)$ sets the total volume delivered, and the weight $[1-g(r)]\,f(r)/\!\int[1-g(r')]\,f(r')\,dr'$ distributes that volume over the pore classes, spreading it non-selectively across the still-empty pore space in proportion to how much room each class has.  The gate $[1-g]$ enforces the
packing bound $g\le1$ and reduces to $f(r)$ in dry soil.  Since the current is itself a
functional of the gradient, $\mathbf{F}=-\!\int\Gamma\,\nabla g\,dr'$
[Eq.~\eqref{eq:F_from_connection}], Eq.~\eqref{eq:precip_bc} is equivalently a condition
on $g$ and its normal variation,
$\int_0^\infty\Gamma(r,r';g)\,\partial_n g(r',\mathbf{x}_{\rm top},t)\,dr'=q_{\rm in}(r,t)$;
integrated across the surface REV it is exactly the top-boundary gain
$\mathcal{G}_\partial^{\rm precip}$ of Eq.~\eqref{eq:precip_source}.  The Dirichlet
condition $g(r,\mathbf{x}_{\rm top},t)=1$ applies only as the saturation-excess
(ponding) limit, reached once $I$ exceeds the surface infiltration capacity: the
boundary then switches from~\eqref{eq:precip_bc} to fixed saturation and the excess
$I-{}$capacity becomes runoff.

\subsection{The Onsager form of the flux}
The same current~\eqref{eq:F_from_connection} is re-expressed through the
conjugate variable $\mu_w$ (whose pairwise differences define $\Phi$) and the
symmetric transport mobility $\mathsf{M}$ of Eq.~\eqref{eq:kernel}, which turns the
entangled $\Gamma$ into a symmetric mobility and puts the flux in the equivalent
Onsager (gradient-flow) form
\begin{multline}
\mathbf{F}(r,\mathbf{x},t) = -\int \mathsf{M}(r,r';\mathbf{x})\, \\
\nabla\mu_w(r',\mathbf{x},t)\,dr' ,
\label{eq:flux}
\end{multline}
with $\mathsf{M}$ coupling class~$r$ to the chemical-potential gradients of
every class~$r'$.  The re-expression does \emph{not} reduce the $r'$-integral to a
scalar; that reduction is the work of the Chapman--Enskog limit (Sec.~\ref{sec:CE}),
where $g\to g_{\rm eq}$ localizes $\nabla g$ on the filling frontier.

The transport mobility $\mathsf{M}$ of Eq.~\eqref{eq:flux} is a \emph{second}
mobility kernel, distinct from the intra-REV redistribution mobility $M$ of
Eq.~\eqref{eq:M}.  It is not an independent constitutive object: it is built
from the same connectivity $C(r,r')$ and single-link conductance that generate the
redistribution operator.  The two mobilities carry different occupancy factors
because they describe different processes: redistribution moves water from a filled
donor to an empty receiver, so $M$ carries the exclusion product
$g(r)[1-g(r)]\,g(r')[1-g(r')]$, whereas inter-REV transport requires a conducting
link with \emph{both} ends wet, so the transport mobility carries the
abundance-weighted, both-wet product
\begin{equation}
\begin{split}
\mathsf{M}(r,r';\mathbf{x})
  &= \kappa_s(r,r')\,C(r,r';\mathbf{x})\,f(r)\,f(r')\\
  &\quad\times\,g(r,\mathbf{x},t)\,g(r',\mathbf{x},t) ,
\end{split}
\label{eq:kernel}
\end{equation}

Here $r,r'$ are the pore-class radii $[\mathrm{m}]$ and $\mathbf{x}$ the
macroscale position $[\mathrm{m}]$; $f(r)$ is the pore-size density, normalised by
$\int f(r)\,dr=1$ and therefore carrying units $[\mathrm{m^{-1}}]$;
$g(r,\mathbf{x},t)$ is the dimensionless occupancy (the filled fraction of class
$r$); $C(r,r')$ is the dimensionless connectivity kernel (the fraction of
class-$r$/class-$r'$ pairs sharing a conducting link); and $\kappa_s(r,r')$ is the
symmetric pair conductance of Eq.~\eqref{eq:kappa_sym}, carrying the same units
$[\mathrm{Pa^{-1}\,s^{-1}}]$ as the single-pore rate constant $\kappa(r)$ of
Eq.~\eqref{eq:transition_rate}.  The abundance factors $f(r)f(r')$ restore the
correct intrinsic permeability, and the occupancies $g(r)g(r')$ enforce that a link
conducts only where both classes are wet; the reciprocity
$\mathsf{M}(r,r')=\mathsf{M}(r',r)$ follows from the symmetry of $\kappa_s$
established after Eq.~\eqref{eq:kappa_sym} and is not re-derived here.
In anisotropic media $\mathsf{M}$ and the macroscopic mobility are
second-rank tensors.  Because $\mathsf{M}$ depends on $g$, the flux is nonlinear in
$g$ even though Eq.~\eqref{eq:flux} is linear in $\nabla\mu_w$, the microscopic
origin of the saturation dependence of the conductivity.  The kernel thus carries
the dimension of the macroscopic conductivity per unit $dr\,dr'$; the constant
pressure-to-head and pore-volume prefactor that converts the rate-form double
moment to Darcy units $[\mathrm{m\,s^{-1}}]$ is fixed dimensionally, by the
pressure-to-head conversion $\rho_w\mathrm{g}$ and the pore-volume measure already
present in the construction [Eq.~\eqref{eq:K_mf}]; no calibration to a measured
$K(\theta)$ is involved, consistent with the constructive claim of the abstract.

\subsection{Bulk budget and residual nonlocality}
The collapse is purely kinematic and supplies no constitutive law: the
current $\mathbf{F}$ of Eq.~\eqref{eq:flux} stays an integral over the radius
variable, so it is \emph{local in $\mathbf{x}$ but nonlocal in $r$}.  The point
limit has split the spatial transport operator into its two nonlocalities: the
coupling between neighbouring REVs contracts to the local divergence
$\nabla\cdot\mathbf{F}$ (the leading, $O(\ell)$ term of the gradient
expansion~\eqref{eq:J_result}, the higher gradients vanishing with $\ell$), while
the radius coupling $r\leftrightarrow r'$, carried by the same connectivity
$C(r,r')$, is untouched by a spatial limit and is retained as the integral in
Eq.~\eqref{eq:flux}.

The bulk water balance is now a corollary.  Let $\theta(\mathbf{x},t)
:=|V|^{-1}\!\int_V c(\mathbf{y},t)\,dV=\phi\!\int g\,f\,dr$ be the REV-averaged water
content, the normalised zeroth Minkowski functional $M_0$ of the wetted phase.  The
$f$-weighted zeroth moment of the class-$r$ budget~\eqref{eq:pore-budget} eliminates
the redistribution operator, since total water is its invariant,
\begin{equation}
\int \Coll[g]\,f\,dr = 0 ,
\label{eq:invariant}
\end{equation}
and carries the spatial flux into the macroscopic (Darcy) water flux
$\mathbf{q}(\mathbf{x},t)=\int\mathbf{F}\,f\,dr$, leaving the integral boundary
balance
\begin{equation}
\frac{d}{dt}\big[\theta(\mathbf{x},t)\,|V|\big]
  = -\oint_{\partial V}\mathbf{q}\cdot\mathbf{n}\,dA .
\label{eq:bulk-budget}
\end{equation}
which, divided by $|V|$ as $\ell\to0$, is the local continuity equation
\begin{equation}
\partial_t\theta + \nabla\cdot\mathbf{q} = 0 .
\label{eq:bulk-point}
\end{equation}
Equation~\eqref{eq:pore-point} thus refines Eq.~\eqref{eq:bulk-point} rather than
replacing it; Eq.~\eqref{eq:bulk-point} is the \emph{one} macroscopic water
budget of the theory, obtained here as the zeroth moment of the pore-resolved
equation rather than postulated.

\subsection{Connectivity and the transport regime}
Macroscopic transport requires the wetted phase to be connected across REVs.
Where its connectivity length $\xi$ is large, the wetted cluster spans, well above
threshold, the spatial flux $\nabla\cdot\mathbf{F}$ is well defined and
Eq.~\eqref{eq:pore-point} carries genuine transport.  As the soil dries toward the
threshold ($\theta\to\theta_c$, the sign change of the Euler characteristic
$M_2$, the third Minkowski functional of Sec.~\ref{sec:accessibility}), the wetted phase fragments, the accessibility $a_w(r)\to0$, the
macroscopic conductivity $K\to0$, and the transport term switches off:
Eq.~\eqref{eq:pore-point} reduces to $\partial_t g=\Coll[g]$, pure intra-REV
redistribution with no flow between REVs.  This is field
capacity~\cite{Hunt2017}; the high-connectivity regime is the one in which the
continuum-transport description applies, and below threshold only the
redistribution operator acts.

\subsection{The fundamental equation and its morphological reading}
\label{sec:fundamental}
Equation~\eqref{eq:pore-point} is the fundamental non-equilibrium equation
and Eq.~\eqref{eq:bulk-point} is its mass-conserving moment.  The distinction is
operational: Eq.~\eqref{eq:pore-point} is \emph{closed}, given $C(r,r')$ and the
single-pore conductances, both $\mathbf{F}$ and $\Coll[g]$ are functionals of $g$,
so $g(r,\mathbf{x},t)$ determines its own evolution, whereas
Eq.~\eqref{eq:bulk-point} is not, since $\mathbf{q}$ is undetermined by $\theta$
alone.  The missing closure $\mathbf{q}[g]$ is precisely the non-equilibrium
content, and it is not a function of $\theta$ alone.  This is the statement that
$\theta$ is an insufficient state variable while $g$ is sufficient, the origin of
apparent equifinality and of hysteresis.  Mass conservation remains exact and
ansatz-independent, but for that reason it is a constraint the dynamics must
respect, not the dynamics itself.

The insufficiency of $\theta$ has a morphological counterpart.  Beyond
the volume fraction $\theta=M_0$ introduced with the bulk budget above, the wetted
phase carries two further Minkowski descriptors, the interfacial area $M_1$
and the Euler characteristic $M_2$ (connectivity), in the three-functional 3D
convention.
Neither is idle: $M_1$ is the capillary-surface contribution to $\mathcal{F}[g]$,
and $M_2$ is the topology that $C(r,r')$ and the percolation accessibility
$a_w(r)$ encode, its sign change at threshold marking field capacity.  The matric
potential is not itself a Minkowski functional; it enters through Young--Laplace on
the marginal radius, $\psi=-2\gamma\cos\theta_c/(\rho_w\mathrm{g}\,r^*)$.  The
statement that $\theta$ underdetermines the state while $g$ closes it is then the
geometric-state result of McClure \emph{et~al.}~\cite{McClure2018}, the finding that, for two-fluid flow in porous media, no single scalar (such as saturation) is a complete state variable, whereas the triple of Minkowski functionals $\{M_0,M_1,M_2\}$ (volume, interfacial area, and Euler characteristic) is, here restated in pore-class language: $\{M_0\}$ alone
fixes neither retention nor conductivity, whereas $\{M_0,M_1,M_2\}$ does, and
$g(r)$ supplies that same closing information resolved by pore radius.


\section{Chapman--Enskog reduction of the continuum kinetic equation}
\label{sec:CE}

The continuum limit of Sec.~\ref{sec:CL} delivered a closed field equation for
the occupancy, the boxed kinetic equation~\eqref{eq:pore-point},
\begin{equation}
\partial_t g(r,\mathbf{x},t)\;+\;\nabla\!\cdot\mathbf{F}(r,\mathbf{x},t)
  \;=\;\Coll[g],
\label{eq:CE_start}
\end{equation}
and it is from this equation, not from the discrete master
equation~\eqref{eq:master}, that the reduction to Richards' equation proceeds.
Equation~\eqref{eq:CE_start} is the soil-water analogue of the Boltzmann equation:
the streaming term $\nabla\!\cdot\mathbf{F}$ transports water \emph{between} REVs,
while the redistribution operator $\Coll[g]=\mathcal{G}-\mathcal{L}$ relaxes the
pore classes \emph{within} a REV toward the packing equilibrium.  The
Chapman--Enskog (CE) expansion is the asymptotics of this equation when the in-REV
relaxation is fast, a \emph{temporal} reduction taken after, and on top of, the
spatial limit that produced Eq.~\eqref{eq:CE_start}.  The two limits are
independent: the spatial limit $\varepsilon=\ell/L_{\rm mac}\to0$ made the REV a
point; the CE limit $\Da\to0$ now makes the in-REV state a slaved equilibrium.  The
full algebra is given in a companion paper~\cite{Rigon2026CE}; here we state the
principal results and their physical content.

\begin{table}[t]
\centering
\caption{The kinetic-theory correspondence made precise by
Eq.~\eqref{eq:CE_start}: the streaming/redistribution split is between inter-REV
transport and intra-REV redistribution.}
\label{tab:boltzmann}
\begin{ruledtabular}
\setlength{\tabcolsep}{3pt}
\footnotesize
\begin{tabular}{@{}ll@{}}
Gas kinetic theory & Unsaturated soil water\\
\colrule
distribution $f(\mathbf{v},\mathbf{x},t)$ & occupancy $g(r,\mathbf{x},t)$\\
streaming $\mathbf{v}\!\cdot\!\nabla f$ & transport $\nabla\!\cdot\mathbf{F}$\\
redistribution $\mathcal{C}[f]$ & redistribution $\Coll[g]$\\
Knudsen number $\mathrm{Kn}$ & Damk\"ohler number $\Da$\\
Maxwellian $f_{\rm eq}$ & packing step $g_{\rm eq}^{\rm ref}=H(r^*(\mathbf{x})\!-r)$\\
redistribution invariants & mass moment $\int\!\cdot\,f\,dr$\\
viscosity (Chapman--Enskog) & conductivity $K$ (Sec.~\ref{sec:K_derived})\\
Navier--Stokes & Richards' equation~\eqref{eq:Richards}\\
\end{tabular}
\end{ruledtabular}
\end{table}

\subsection{The fast--slow split and the single control parameter}
\label{sec:single_param}

Nondimensionalizing Eq.~\eqref{eq:CE_start} exposes one dimensionless group.
The redistribution operator carries the in-REV relaxation rate $1/\tau_{\rm redis}$, the slowest nonzero rate of $\Coll$ linearized about $g_{\rm eq}$, while the
streaming term carries the inter-REV forcing rate $1/\tau_{\rm forcing}$.  Their
ratio is the Damk\"ohler number,
\begin{equation}
\Da(r,\mathbf{x}) = \frac{\tau_{\rm redis}(r)}{\tau_{\rm forcing}(\mathbf{x})}\ll1,
\label{eq:Da_def}
\end{equation}
and writing Eq.~\eqref{eq:CE_start} on the slow (forcing) time scale puts the
redistribution term at order $\Da^{-1}$,
\begin{equation}
\partial_t g + \nabla\!\cdot\mathbf{F} = \frac{1}{\Da}\,\Coll[g].
\label{eq:CE_scaled}
\end{equation}
The redistribution is fast and the streaming is slow, the singular-perturbation
structure that makes the expansion systematic.  Both times are set by pore-scale
quantities already in hand.  The redistribution time for class $r$ at occupation
$g$ is
\begin{equation}
\tau_{\rm redis}(r,g) = \frac{1}{\kappa_{\rm eff}(r,g)\,\bar{\Phi}(r)},
\label{eq:tau_redis}
\end{equation}
with $\kappa_{\rm eff}=\kappa(r,g)\,\Xi(r)$ [Eq.~\eqref{eq:kappa_LW}] and
$\bar{\Phi}(r)=(2\gamma\cos\theta_c/\rho_w\mathrm{g})(1/r-1/r^*)$ a representative
driving potential at the frontier; the forcing time is
\begin{equation}
\tau_{\rm forcing}(\mathbf{x}) = \frac{\bar{L}\,\phi}{|q(\mathbf{x})|},
\label{eq:tau_forcing}
\end{equation}
the time for the local macroscopic flux to exchange one pore volume across the
REV.  At the surface $|q|=I$; in the interior $|q|$ is the flux driven by $\Phi$.
Unlike the gas, viscous (Stokes) pore flow has no momentum invariant, so the
gradient scale is not an independent parameter: a sharp front imposed under
$\Da\ll1$ is self-healed by redistribution within a few $\tau_{\rm redis}$
(App.~\ref{app:single_parameter}), and $\Da\ll1$ alone controls the expansion.

\subsection{The expansion and the solvability condition}
\label{sec:CE_structure}

Expand the occupancy about the (still unknown) local equilibrium,
\begin{align}
g &= g^{(0)}(r;\theta(\mathbf{x},t)) + \Da\,g^{(1)} + O(\Da^2),
   \label{eq:CE_expand}\\
0 &= \phi\!\int g^{(1)}(r)\, f(r)\,dr ,
   \label{eq:CE_constraint}
\end{align}
the constraint stating that the whole water content is carried by $g^{(0)}$,
so $g^{(1)}$ only reshuffles water among classes (it is orthogonal to the redistribution
invariant $1$ in the measure $d\nu=\phi f\,dr$).

\paragraph{Zeroth order ($\Da^{-1}$): local equilibrium.}
The fast term must vanish on its own,
\begin{multline}
\Coll[g^{(0)}] = 0 \\
\Longrightarrow\; g^{(0)} = g_{\rm eq}^{\rm ref}(r;\theta)=H(r^*(\mathbf{x})\!-r),
\label{eq:CE_leading}
\end{multline}
the unique packing step (Sec.~\ref{sec:geq}), with $r^*$ fixed by the local
water content and the matric potential $\psi(\theta)=-2\gamma\cos\theta_c/(\rho_w
\mathrm{g}\,r^*)$ well defined.  No macroscopic flux exists at this order: the
streaming term has not yet acted.

\paragraph{First order ($\Da^0$): the macroscopic balance.}
At the next order the streaming enters and the linearized redistribution responds,
\begin{equation}
\partial_t g^{(0)} + \nabla\!\cdot\mathbf{F}[g^{(0)}]
  = \mathcal{J}\,g^{(1)} - \mathcal{E} - \mathcal{T},
\qquad \mathcal{J}:=\Coll'\big|_{g_{\rm eq}}.
\label{eq:CE_next}
\end{equation}
This linear equation for $g^{(1)}$ is solvable only if its right-hand side is
orthogonal to $\ker\mathcal{J}$, the Fredholm alternative.  Since mass is the
single redistribution invariant, $\ker\mathcal{J}$ is spanned by the constant, and the
solvability condition is the $f$-weighted $r$-moment.  Taking that moment of
Eq.~\eqref{eq:CE_next}, and using that redistribution conserves water,
$\int\Coll\,f\,dr=0$ [Eq.~\eqref{eq:invariant}] and hence $\int\mathcal{J}g^{(1)}
f\,dr=0$, annihilates the redistribution term and returns the macroscopic water
budget~\eqref{eq:bulk-point}, now with the volumetric sinks $-S_{\rm evap}-S_{\rm
root}$ restored from $\mathcal{E},\mathcal{T}$ and the flux closed as
$\mathbf{q}=\int\mathbf{F}\,f\,dr$.  No new equation is introduced: this is
the single budget~\eqref{eq:bulk-point} of Sec.~\ref{sec:CL}, here obtained as the
Fredholm solvability condition, the soil-water counterpart of the
Euler/Navier--Stokes balance of Boltzmann theory.

\subsection{Hydraulic conductivity as a transport coefficient}
\label{sec:K_derived}

The constitutive law is the value of $\mathbf{q}=\int\mathbf{F}\,f\,dr$ once the
streaming acts on the equilibrium $g^{(0)}=g_{\rm eq}$.  Here the entangled current
of Eq.~\eqref{eq:F_from_connection} disentangles: the equilibrium step makes
$\nabla g^{(0)}$ a delta on the frontier,
\begin{equation}
\nabla g^{(0)}(r') = -\frac{\delta(r'-r^*)}{f(r^*)}\,\nabla r^*
  = \frac{\delta(r'-r^*)}{\phi\,f(r^*)^2}\,\nabla\theta,
\label{eq:grad_projection}
\end{equation}
so the $r'$-integral collapses to the single frontier class, the gradient
$\nabla\theta$ converts to $\nabla\psi$ through the equation of state \ref{eq:YL}, and the
current takes the form of the Darcy law $\mathbf{q}=-K(\theta)(\nabla\psi+\hat{z})$, as explained in~\cite{Rigon2026CE}.  Equivalently, in
the Onsager writing the conductivity is the equilibrium double moment of the
mobility kernel of Eq.~\eqref{eq:kernel},
\begin{multline}
K(\theta) = \iint \mathsf{M}(r,r';\mathbf{x})\,dr\,dr'\Big|_{g_{\rm eq}} \\
  = \int_0^{r^*}\!\!\int_0^{r^*}\!\!\kappa_s\,C\,f(r)\,f(r')\,dr\,dr',
\label{eq:K-collapse}
\end{multline}
well defined only now, with $g\to g_{\rm eq}$ supplied by the CE zeroth order
[Eq.~\eqref{eq:CE_leading}], this is the equilibrium collapse of the transport kernel~\eqref{eq:kernel}.  Carrying the first-order solve $g^{(1)}\leftarrow
\mathcal{J}^{-1}(\cdot)$ explicitly gives the spectral (Green--Kubo) form
\begin{multline}
K(\theta) = \phi\,\bar{L}\int_0^\infty \kappa_{\rm eff}(r)\,\mathcal{J}^{-1}S(r)\, \\
f(r)\,dr ,
\label{eq:K_exact}
\end{multline}
with $\mathcal{J}^{-1}$ the pseudoinverse of the linearized redistribution
operator on $\ker\mathcal{J}^{\perp}$. 
~\cite{Rigon2026CE}.  The two writings agree:
Eq.~\eqref{eq:K-collapse} is the equilibrium double moment,
Eq.~\eqref{eq:K_exact} the same coefficient resolved into relaxation modes.

\paragraph{Mean-field form.}
Three reductions~\cite{Rigon2026CE}, statistical homogeneity
($C_\partial\approx C$), diagonal dominance ($\mathcal{J}^{-1}\approx\tau(r)$,
parallel tubes), and isotropic orientation (tortuosity $\bar{\tau}^2$), collapse
Eq.~\eqref{eq:K_exact} to
\begin{equation}
K_{\rm mf}[\theta] = \frac{\phi\,\rho_w\mathrm{g}}{8\mu\,\bar{\tau}^2}
  \int_0^{r^*(\theta)} r^2\,\Xi(r)\,a_w(r)\,\bar{C}(r)\,f(r)\,dr,
\label{eq:K_mf}
\end{equation}
in which every factor once inserted by hand, the $r^2$ Hagen--Poiseuille
weight, the tortuosity, the percolation accessibility $a_w$ (so $K_{\rm mf}\to0$
below threshold), and the frontier cutoff $r^*$, now has a derivation.

\subsection{Richards' equation as the double-flat limit}
\label{sec:Richards}

Substituting $\mathbf{q}=-K(\theta)(\nabla\psi+\hat{z})$ into the macroscopic water
budget~\eqref{eq:bulk-point} gives
\begin{equation}
\partial_t\theta = \nabla\!\cdot\big[K(\theta)\,(\nabla\psi+\hat{z})\big]
  - S_{\rm evap}-S_{\rm root},
\label{eq:Richards}
\end{equation}
where the volumetric sinks are the pore-class evaporation and root-uptake operators of Eq.~\eqref{eq:master} integrated over the pore classes, $S_{\rm evap}+S_{\rm root}=\int_0^\infty(\mathcal{E}+\mathcal{T})\,f\,dr$.
Richards' equation, now the $\Da\to0$ reduction of the continuum kinetic
equation~\eqref{eq:CE_start}, not a postulate.  The zeroth-order step $g_{\rm eq}$
is unique at each $\theta$, but its accessibility $a_w(r)$ differs between wetting
and drying, so $K^{\rm dry}(\theta)\neq K^{\rm wet}(\theta)$: hysteresis survives the
reduction as parametric branch selection, the flattened remnant of the full
non-equilibrium path dependence carried by $g(r)$.

\subsection{Validity conditions and breakdown}
\label{sec:breakdown}

The reduction holds when the redistribution is fast and mean-field: (i)~$\Da\ll1$,
(ii)~$C_\partial\approx C$ over the REV, (iii)~slow variation of $a_w(r)$ between
neighbours.  Three mechanisms defeat it, one for each structural assumption, and a
fourth defeats the spatial limit that precedes it:
\begin{itemize}
\item \textbf{Kinetic forcing} ($\Da\sim1$): the redistribution is no longer fast,
$g\neq g_{\rm eq}$, $r^*$ and $\psi$ cease to be state variables; sharp fronts
appear, the analogue of shock waves.
\item \textbf{Percolation breakdown} ($\theta\to\theta_c^+$): the spectral gap
of $\mathcal{J}$ closes, $\tau_{\rm redis}\to\infty$ so $\Da\to\infty$
automatically, and $K\to0$ as a topological transition.
\item \textbf{Spatial heterogeneity} ($C_\partial\not\approx C$): the streaming
kernel is not mean-field, $\mathbf{F}$ stays pore-class-resolved, and no scalar $K$
exists even at $\Da\ll1$.
\item \textbf{Loss of scale separation} ($\xi\gtrsim\ell$): if the connectivity
length approaches the REV size, the \emph{spatial} limit of Sec.~\ref{sec:CL} itself
fails and Eq.~\eqref{eq:CE_start} is not available to expand, a failure upstream of
the CE limit, independent of $\Da$.
\end{itemize}
In every case $g(r)$, not $\theta$ or $\psi$, is the irreducible state variable
and the full kinetic equation~\eqref{eq:CE_start} must be carried.

\section{The forcing bundle}
\label{sec:forcing}

The Chapman--Enskog reduction of Sec.~\ref{sec:CE} collapsed the kinetic
equation onto Richards' equation in the quasi-static limit $\Da\to0$.  The
non-equilibrium content of the theory, hysteresis, lives in the
opposite direction: in the dependence of $g(r,t)$ on the \emph{history} of the
boundary forcing, not on its instantaneous value.  This dependence is already
contained in the dynamics of Eq.~\eqref{eq:pore-point}; to expose it explicitly we
recast that history dependence as a fiber bundle over the space of forcing protocols,
whose curvature is the wetting--drying commutator $[\W,\D]$ and whose holonomy around a
closed forcing loop is the hysteresis loop itself (\S\ref{sec:holonomy}).

When a REV is subjected to time-varying boundary forcing, rainfall, drainage,
evapotranspiration, its filling distribution $g(r,t)$ evolves along a path that
depends on the forcing sequence, not just on its current value.  This path
dependence is invisible to the equilibrium retention curve $\theta(\psi)$, which
assigns a single occupancy to each water content; capturing it requires carrying
the configuration as a functional of the forcing history, which is what the bundle
below does.

\subsection{Definition}
\label{sec:forcing_def}

The base $\mathcal{B}$ is the space of boundary-forcing histories.
The forcing bundle is:
\begin{equation}
\pi_\mathcal{B} : \mathcal{E}_\mathcal{B} \to \mathcal{B}
\label{eq:forcing_bundle}
\end{equation}
where:
\begin{itemize}
\item \textbf{Base space $\mathcal{B}$}: the space of
  forcing histories.  A point in~$\mathcal{B}$ is a
  path $\{I(t')\}_{t' \leq t}$ specifying the sequence
  of boundary fluxes (precipitation intensity, drainage
  rate, evapotranspiration rate) up to the present
  time.  In the simplest case, $\mathcal{B}$ is
  parameterized by the current forcing intensity~$I$
  and its direction that can be separated in wetting or drying.

\item \textbf{Fiber $\mathcal{S}_I$}: the space of
  filling distributions, $\mathcal{S}_I = \{g: [0,\infty) \to [0,1]\}$.
  Its internal geometry, pore-size
  distribution $f(r)$, connectivity $C(r,r')$,
  accessibility $a(r;g)$, equilibrium $g_{\rm eq}$,
  mobility $M(r,r')$, is that of the kinetic equation itself.

\item \textbf{Section}: a map
  $\sigma: \mathcal{B} \to \mathcal{E}_\mathcal{B}$
  assigning a filling distribution to each forcing
  history.  The section at time~$t$ is
  $g(r,t) = \sigma(\{I(t')\}_{t' \leq t})$.

\item \textbf{Structure semigroup}: the same
  non-commutative semigroup $\mathfrak{S}$ generated by
  $\W$ and~$\D$ (Sec.~\ref{sec:kinetic}).  However, in
  the forcing bundle the semigroup acts
  temporally, successive wetting and drying
  steps compose along the forcing path, rather than
  spatially.
\end{itemize}

\subsection{The connection: forcing-driven evolution}
\label{sec:forcing_connection}

The connection on the forcing bundle describes how the
section (the filling distribution) changes as one moves
along the base space (advances in forcing history).  In
the inter-REV scale, the connection is the inter-fiber
exchange $\Coll_\delta$, the
operators coupling adjacent REVs.  Here, the connection
is the kinetic equation itself: it specifies how $g(r)$
evolves under an infinitesimal forcing step.

The connection decomposes into two one-forms on
$\mathcal{B}$, corresponding to the two directions of
forcing:
\begin{align}
\Lg_\W &= \Lg_{\rm redis} + \Lg_{\rm source}
  &&\text{(wetting)} \\
\Lg_\D &= \Lg_{\rm redis} + \Lg_{\rm sink}
  &&\text{(drying)}
\end{align}
The common part $\Lg_{\rm redis}$ drives the fiber
toward equilibrium; the asymmetric parts
$\Lg_{\rm source}$ and $\Lg_{\rm sink}$ encode how
water enters and leaves. 

The \textbf{curvature} of this connection is the
commutator $F = [\Lg_\W, \Lg_\D]$, which measures how
much the result of a wet--dry sequence differs from a
dry--wet sequence.  Operationally: take a REV at filling distribution $g$, wet it by a small increment then dry it by the same increment; $\Lg_\W$ and $\Lg_\D$ are the (non-commuting) generators of these two moves, and $F\,g = \Lg_\W\Lg_\D g - \Lg_\D\Lg_\W g$ is the net residual displacement of $g$ after the loop.  The non-commutativity has a concrete microscopic origin: wetting fills pores from small to large along the accessible cluster, drying empties them along a different accessible cluster (air invasion is constrained to the boundary-connected gas phase), so the accessibility $a_w$ entering $\Lg_\W$ and $\Lg_\D$ is evaluated on distinct support, the operators therefore do not commute whenever both invasion fronts are active.  When the curvature vanishes
($\Da \ll 1$), the connection is flat and transport is
path-independent, with no hysteresis: the two fronts equilibrate faster than the forcing changes, so the order of wetting and drying is immaterial.  When the curvature is
nonzero ($\Da \gtrsim 1$), parallel transport around a
closed loop produces holonomy: the filling distribution
does not return to its initial state.  This is
hysteresis, formalized as a geometric phase.

The following table collects the correspondence for the forcing bundle:

\begin{table*}
\caption{Forcing bundle dictionary: correspondence between
physical objects and bundle elements.}
\label{tab:forcing_dict}
\begin{ruledtabular}
\begin{tabular}{lll}
\textbf{Object} & \textbf{Bundle element}
  & \textbf{Physical meaning} \\
\hline
$g(r,t)$ & Section
  & Configuration at current time \\
$\{I(t')\}_{t' \leq t}$ & Base point
  & Forcing history \\
$\Lg_\W$, $\Lg_\D$ & Connection 1-form
  & Kinetic equation (wetting/drying) \\
$\Lg_{\rm redis}$ & Common part
  & Drives $g \to g_{\rm eq}$;
  rate $\propto r^2$ \\
$\Lg_{\rm source}$, $\Lg_{\rm sink}$
  & Asymmetric parts
  & Non-selective filling vs.\
  selective emptying \\
$\Da(r,\mathbf{x})$ & Connection coefficient
  & Forcing-to-redistribution rate ratio \\
$F = [\Lg_\W, \Lg_\D]$ & Curvature
  & Non-commutativity;
  $\|F\| \propto \Da^2$ for $\Da \ll 1$ \\
$U[\gamma]$ & Holonomy
  & Residual after closed forcing cycle \\
\end{tabular}
\end{ruledtabular}
\end{table*}

\begin{remark}[Forcing connection vs.\ inter-REV current]
The forcing connection and the inter-REV current of Sec.~\ref{sec:spatial} act
on the same configuration space $\mathcal{S}$ through the same semigroup
$\mathfrak{S}$, but along different directions: the inter-REV current
($\Coll_\delta$, coefficient $K$) transports water between
neighbouring REVs, while the forcing connection ($\Lg_\W,\Lg_\D$, coefficient $\Da$)
transports the configuration along the forcing path.
\end{remark}

The following subsections develop the internal structure
of the forcing connection: the explicit form of the
generators (Sec.~\ref{sec:generators}), how the forcing
intensity enters the operators
(Sec.~\ref{sec:forcing_I}), the origin of the
non-commutativity (Sec.~\ref{sec:noncomm}), and the
holonomy of a forcing cycle, the hysteresis loop, together with its
control by the Damk\"ohler number
(Sec.~\ref{sec:holonomy}).

\subsection{Wetting and drying generators}
\label{sec:generators}

The finite operators $\W_I(\Delta t)$ and $\D_I(\Delta t)$ are flow maps of the kinetic equation.  Reading their infinitesimal generators directly off the continuum balance~\eqref{eq:pore-point}, $\partial_t g = \Coll[g] - \nabla\!\cdot\mathbf{F} - \mathcal{E} - \mathcal{T}$, and using the boundary in/out split~\eqref{eq:Qsplit} of the transport term:
\begin{align}
\Lg_\W[g] &\equiv \left.\pder{g}{t}\right|_\W = \underbrace{\Coll[g]}_{\Lg_{\rm redis}} + \underbrace{\mathsf{Q}_{\rm in}\,g}_{\Lg_{\rm source}}  \label{eq:W_decompose}\\
\Lg_\D[g] &\equiv \left.\pder{g}{t}\right|_\D = \underbrace{\Coll[g]}_{\Lg_{\rm redis}} + \underbrace{\big({-}\mathsf{Q}_{\rm out}\,g\big) - \mathcal{E} - \mathcal{T}}_{\Lg_{\rm sink}} \label{eq:D_decompose}
\end{align}

where each generator decomposes into two structurally distinct parts.  The inward boundary current $\mathsf{Q}_{\rm in}$ carries water \emph{into} the point $\mathbf{x}$, the precipitation current at the surface [explicit in Eq.~\eqref{eq:precip_source} below] and the water-table current at the base, while the outward current $\mathsf{Q}_{\rm out}$ is drainage away from $\mathbf{x}$; the two are the boundary halves of the single transport operator $\mathfrak{T}=\mathsf{Q}_{\rm out}-\mathsf{Q}_{\rm in}$ of Eq.~\eqref{eq:Qsplit}.
The redistribution part $\Lg_{\rm redis}=\Coll[g]$ is common to both generators: it is the continuum redistribution operator of Eq.~\eqref{eq:pore-point}, which relaxes $g\to g_{\rm eq}$ at fixed $\theta$ regardless of the forcing direction.  What distinguishes wetting from drying is the transport and sink terms alone.

Thus, at the continuum scale, $\Lg_\W$ retains the inflow $\mathsf{Q}_{\rm in}$ and drops the sinks while $\Lg_\D$ retains the outflow $\mathsf{Q}_{\rm out}$ and the volumetric sinks $\mathcal{E},\mathcal{T}$, and both carry the identical redistribution $\Coll[g]$: the forcing bundle adds no operator beyond those already in the continuum kinetic equation~\eqref{eq:pore-point}.  The bundle is nothing but Eq.~\eqref{eq:pore-point} itself, its transport term resolved into the two forcing directions.

\subsection{How the forcing intensity enters the operators}
\label{sec:forcing_I}

The inward boundary current $\mathsf{Q}_{\rm in}$ of Eq.~\eqref{eq:W_decompose}, resolved by pore class at the surface, is the precipitation source term $\mathcal{G}_\partial^{\rm precip}$; it has the explicit form:
\begin{equation}
\mathcal{G}_\partial^{\rm precip}(r) = \frac{I}{\bar{L}\,\phi}\;[1-g(r)]\;\frac{f(r)}{\int_0^\infty [1-g(r')]\,f(r')\,dr'}
\label{eq:precip_source}
\end{equation}
The prefactor $I/(\bar{L}\,\phi)$ converts the boundary flux $I$ [m/s] into a filling rate [s$^{-1}$]; the factor $[1-g(r)]$ ensures only empty pore space receives water; the denominator normalizes by the total available empty volume, distributing the incoming flux across all unfilled pore classes in proportion to their abundance $f(r)$.

Precipitation fills pores \textbf{without capillary selectivity}: rain enters the available pore space regardless of pore radius.  In contrast, the redistribution operator $\Lg_{\rm redis}$ is strongly radius-selective through the Hagen--Poiseuille rate $\kappa(r) \propto r^2$ and the driving potential $\Phi(r,r') \propto 1/r - 1/r'$, which always favors moving water from large pores into small ones.

The competition between these two processes is governed by their ratio at each pore class $r$:
\begin{equation}
\frac{|\Lg_{\rm source}(r)|}{|\Lg_{\rm redis}(r)|} = \frac{I/(\bar{L}\,\phi)}{\kappa(r)\,\bar{\Phi}} = \frac{\tau_{\rm redis}(r)}{\tau_{\rm forcing}(\mathbf{x})} = \Da(r,\mathbf{x})
\label{eq:Da_ratio}
\end{equation}
where $\bar{\Phi}$ is the mean driving potential.  The Damk\"ohler number is the ratio of the forcing rate to the redistribution rate at each pore class, read directly from the operator decomposition~\eqref{eq:W_decompose}.

\subsection{The non-commutativity and its origin}
\label{sec:noncomm}

The curvature of the forcing connection is:
\begin{equation}
F = [\Lg_\W, \Lg_\D] = [\Lg_{\rm redis} + \Lg_{\rm source},\; \Lg_{\rm redis} + \Lg_{\rm sink}]
\label{eq:F_WD}
\end{equation}
Since $\Lg_{\rm redis}$ is common to both, expanding the commutator:
\begin{equation}
F = \underbrace{[\Lg_{\rm source}, \Lg_{\rm sink}]}_{\text{direct asymmetry}} + \underbrace{[\Lg_{\rm redis}, \Lg_{\rm sink} - \Lg_{\rm source}]}_{\text{redistribution--forcing coupling}}
\label{eq:F_decompose}
\end{equation}

The first term $[\Lg_{\rm source}, \Lg_{\rm sink}]$ captures the direct asymmetry between how water enters and leaves the pore space.  Precipitation fills pores non-selectively (all empty pores receive water proportional to $f(r)$), while drying empties pores selectively: drainage requires connection to the gas-phase boundary (accessibility $a_g(r)$), evaporation requires connection to the evaporating surface, and root uptake~\cite{DAmatoRigon2025} follows the root network.  The filling sequence differs from the emptying sequence, and this difference is irreducible.

The second term $[\Lg_{\rm redis}, \Lg_{\rm sink} - \Lg_{\rm source}]$ captures how redistribution interacts differently with filling and emptying.  During wetting, redistribution moves water from the freshly filled large pores (where rain deposited it non-selectively) into small pores (lower chemical potential); during drying, redistribution moves water from small pores near the emptying front into the interior.  The coupling is asymmetric because the pore populations that redistribution acts on differ between the two processes.

To show that this commutator is non-zero on a well-defined operator
class, rather than only heuristically, the companion paper~\cite{Rigon2026CE} works the smallest
non-trivial case in closed form: two pore classes (a body and a throat) with the
generators $\Lg_\W,\Lg_\D$ written as explicit $2\times2$ vector fields on the
occupancy square $[0,1]^2$.  There $[\Lg_\W,\Lg_\D]g\neq0$ except on the measure-zero
diagonal where the two classes are equally filled, so $\W\D g\neq\D\W g$ for almost
every state; the magnitude grows with the body--throat entry-pressure contrast, so
bimodal media show stronger non-commutativity than the unimodal network.  The OpenPNM result
$\|g_{\W\D}-g_{\D\W}\|=0.046$ (Sec.~\ref{sec:payoff_preferential}) is the numerical
counterpart on the full $3375$-pore network.

\subsection{Holonomy: the hysteresis loop as the curvature of a forcing cycle}
\label{sec:holonomy}

The curvature $F=[\W,\D]$ is not merely a diagnostic that hysteresis exists;
it sets its size.  A wetting--drying cycle is a closed loop in forcing space, and the
configuration's failure to return to its starting point, the hysteresis, is the
\emph{holonomy} of the forcing connection around that loop.  Let $\varphi^{\W}_{\delta}$
and $\varphi^{\D}_{\delta}$ denote the flow maps that advance $g$ by a forcing
increment of amplitude $\delta$ under wetting and drying, respectively (the finite
maps generated by $\Lg_\W,\Lg_\D$ of Eqs.~\eqref{eq:W_decompose}--\eqref{eq:D_decompose}).
For a small rectangular cycle, wet by $\delta$, dry by $\delta$, unwet, undry, the
composition of the four maps is, by the Baker--Campbell--Hausdorff formula,
\begin{equation}
\Delta g \;=\; \varphi^{\D}_{-\delta}\!\circ\varphi^{\W}_{-\delta}\!\circ
              \varphi^{\D}_{\delta}\!\circ\varphi^{\W}_{\delta}(g)\;-\;g
        \;=\; \delta^{2}\,[\W,\D]\,g \;+\; O(\delta^{3}).
\label{eq:holonomy_bch}
\end{equation}
The residual $\Delta g$, the amount by which the configuration fails to return after the
cycle, is thus the enclosed forcing ``area'' $\delta^2$ times the curvature
$[\W,\D]$ at the operating point.  This is the geometric statement of hysteresis: a
closed loop in forcing space leaves a residue in the configuration precisely when the
wetting and drying generators do not commute.  Because $[\W,\D]$ is known in closed
form for the two-class system~\cite{Rigon2026CE}, $\Delta g$ is an explicit,
computable number, not a qualitative claim.

The magnitude of the residual is controlled by the Damk\"ohler number, through the
source/sink terms that distinguish $\Lg_\W$ from $\Lg_\D$.  When $\Da\ll1$ the common
redistribution $\Lg_{\rm redis}=\Coll[g]$ dominates: after each forcing step the system
relaxes back to $g_{\rm eq}(\theta)$ before the next, so wetting and drying leave no
lasting difference, the source/sink terms are perturbations of order $\Da$, and their
commutator is of order $\Da^2$.  The connection is then effectively flat and transport
is path-independent, no hysteresis.  When $\Da\gtrsim1$ the system does not return to
$g_{\rm eq}$ between steps: $g(r)$ retains the imprint of \emph{how} water arrived
(non-selectively, via $\mathsf{Q}_{\rm in}$) or departed (selectively, via $\mathsf{Q}_{\rm out}$ and the sinks), the
two leave different fingerprints at the same $\theta$, and the curvature is nonzero.
Hence $\|[\W,\D]\|\propto\Da^2$ for $\Da\ll1$: the hysteresis vanishes quadratically in
the quasi-static limit and becomes appreciable only as $\Da\to1$.

This $\Da$-dependence has a measurable consequence.  Carried through the loop, the
finite-rate lag $g-g_{\rm eq}\propto\Da$ produces a hysteresis \emph{loop area}
$\mathcal{H}$, the area enclosed by the wetting and drying branches of the retention
curve $\theta(\psi)$, that scales as
\begin{equation}
\mathcal{H}(I) \;=\; \mathcal{H}_0 \;+\; c\,I^{2} \;+\; O(I^{3}),
\qquad c \propto \big\|[\W,\D]\big\|,
\label{eq:H_scaling}
\end{equation}
where $I$ is the forcing (rainfall) intensity, $\mathcal{H}_0\ge0$ is the
rate-independent (quasi-static, $\Da\to0$) loop area set by the intrinsic pore geometry,
and $c$ is the rate-dependent coefficient, fixed by the wetting--drying commutator.  The
$I^2$ law follows because $\Da\propto I$ at fixed soil (the redistribution time is
intensity-independent while $\tau_{\rm forcing}\propto1/I$), so the curvature contribution
$\propto\Da^2$ becomes $\propto I^2$.  Equation~\eqref{eq:H_scaling} is the central
testable output of the forcing-bundle picture: the loop area grows \emph{quadratically}
with intensity at low $I$, with a coefficient fixed by the commutator, crossing over to
saturation as $\Da\to1$ (where the occupancy bounds $0\le g\le1$ cut off the expansion).
Classical hysteresis models (Mualem, Parker--Lenhard) make the loop area depend only on
the reversal points, not on the rate; a measured $\mathcal{H}\propto I^{2}$ would
therefore be a signature of the kinetic theory that no rate-independent model reproduces.

\begin{remark}[Geometric phase]
The residual~\eqref{eq:holonomy_bch} is the exact analogue of the Berry
phase~\cite{Berry1984} in quantum mechanics and the Aharonov--Bohm
effect~\cite{Shapere1989} in gauge theory: a state transported around a closed loop in
parameter space fails to return to itself when the connection has curvature, by an
amount that depends only on the enclosed area and the curvature, not on the details of
the path.  Here the ``state'' is the filling distribution $g$, the ``parameter space''
is the space of forcing histories, and the geometric phase is the hysteresis.
\end{remark}



\section{Discussion}
\label{sec:discussion}

Before interpreting the framework, we separate what is proved from what is
argued or conjectured, so that the reader can weight each claim accordingly
(Table~\ref{tab:ledger}).

More broadly, the framework is a physically motivated statistical-mechanical \emph{construction} rather than a rigorous reduction from an underlying molecular dynamics: the gain--loss closure~\eqref{eq:chaos} and the gradient-flow free energy are posited on physical grounds and judged by their consequences, with a first-principles derivation left open.

\begin{table}[t]
\centering
\caption{\textbf{Status of the principal claims.}  ``Established'' = proved
here or in the cited companion/SM; ``plausible'' = supported by derivation under
stated assumptions plus numerical demonstration; ``conjectural'' = physically
motivated but not yet proved.}
\label{tab:ledger}
\small
\begin{tabular}{@{}p{0.60\columnwidth} l@{}}
\hline
Claim & Status \\
\hline
Kinetic equation preserves $0\le g\le1$ (exclusion) & Established$^\dagger$ \\
Global existence/uniqueness (binned system) & Established$^\ddagger$ \\
$H$-theorem $d\mathcal{F}/dt\le0$ (isothermal, unforced) & Established \\
$g_{\rm eq}$ unique global minimizer ($\psi_T>0$) & Established \\
Mobility symmetric positive semidefinite & Established \\
Richards as the $\Da\!\to\!0$ Chapman--Enskog limit & Plausible$^\ast$ \\
Darcy $K$ as a transport coefficient & Plausible \\
$[\W,\D]\neq0$ (analytic example + OpenPNM) & Established \\
Hysteresis $=$ holonomy of the forcing bundle & Plausible \\
Field capacity $\theta_{\rm FC}\!\approx\!\theta_c$ (percolation) & Plausible \\
Lab--field $K$ discrepancy from finite $\Da$ & Conjectural \\
Quantitative $\mathcal{H}(I)\!\propto\! I^2$ hysteresis scaling & Conjectural \\
\hline
\end{tabular}
\\[2pt]
{\footnotesize $^\ast$\,Solvability shown here [Eq.~\eqref{eq:CE_next}];
spectral gap, full algebra, and convergence estimate in
Ref.~\cite{Rigon2026CE}.\\
$^\dagger$\,For the occupancy-independent HP rate; with the Lucas--Washburn
$\kappa\propto1/g$ the bound is enforced by the regularization~\eqref{eq:kappa_reg}
(Sec.~\ref{sec:GL}).\\
$^\ddagger$\,For the finite-dimensional binned system (Lipschitz field on the
invariant box, App.~\ref{app:wellposed}); for the continuum-in-$r$ equation the
claim is conjectural.}
\end{table}

\subsection{Hysteresis is geometry, not bistability}
\label{sec:payoff_hysteresis}

The first payoff concerns \emph{how} the medium remembers its wetting history.
Classical hysteresis theories, independent-domain models~\cite{Poulovassilis1962,Mualem1974,Mualem1976}
and their Preisach formulation~\cite{Preisach1935}, represent the pore space as a
collection of bistable units (hysterons), each with fixed wetting and drying thresholds,
and reproduce the primary loop and the scanning curves by superposition over the unit
density.  They are descriptively successful, but they \emph{posit} the two ingredients
that matter: the per-unit bistability and the rate-independence of the loop.

The kinetic theory supplies both as consequences rather than assumptions.  Bistability
is not attached to individual pores; it emerges from the non-commutativity of the wetting
and drying generators, $[\W,\D]\neq0$ (Sec.~\ref{sec:holonomy}), together with
accessibility.  A closed wetting--drying cycle is a loop in forcing space, and the
configuration's failure to return is the holonomy of that loop, the scanning curve of
soil physics is a geometric phase, not a bookkeeping of latched hysterons.  A two-class
worked example~\cite{Rigon2026CE} makes this explicit: precipitation fills both classes
non-selectively [Eq.~\eqref{eq:W_decompose}] while drainage empties the large class first
[Eq.~\eqref{eq:D_decompose}], so after a cycle returning to the same $\theta$ the internal
partition has rotated by $\Delta\mathbf{g}=[U(\gamma)-\mathcal{I}]\,\mathbf{g}^{(0)}\neq0$,
nonzero precisely when $\Da(r_2,I)\gtrsim1$.  Parker--Lenhard closure
schemes~\cite{Kool1987-le} are recovered as the rate-independent, mean-field limit in
which this holonomy is frozen into a static scanning-curve interpolation.

The difference is not merely conceptual: it is falsifiable.  Because the loop is the
integral of a curvature that scales as $\Da^2$, the hysteresis loop area carries a rate
dependence that rate-independent domain models cannot produce: $\mathcal{H}(I) =
\mathcal{H}_0 + c\,I^2 + O(I^3)$ with $c\propto\|[\W,\D]\|$ [Eq.~\eqref{eq:H_scaling}],
where $\mathcal{H}_0$ is the rate-independent (quasi-static) loop area and $c$ is fixed by
the wetting--drying commutator.  A measured $\mathcal{H}\propto I^2$ on a single sample,
obtained by varying the wetting--drying rate, would be a direct signature of the kinetic
mechanism and a discriminant no rate-independent model can reproduce.  Beven and
Germann~\cite{Beven1982,Germann2014} long argued that the classical picture misses the
dynamics of real infiltration; the geometric reading vindicates that diagnosis and makes
it quantitative.

\subsection{Richards' equation and preferential flow are one equation}
\label{sec:payoff_preferential}

The second payoff concerns \emph{where} water goes.  The hydrological literature has
treated matric (Richards-equation) flow and preferential flow as fundamentally different
phenomena requiring different models~\cite{Beven1982,Germann1985,Gerke1993}.
The kinetic theory recasts the dichotomy as a continuous crossover within a single
equation, governed by one dimensionless number.

The crossover is sharp and derivable.  The radius at which $\Da=1$,
\begin{equation}
r_{\Da}(I) = \frac{4\mu\, \bar{L}\, \bar{\tau}^2}{\gamma\cos\theta_c\, \phi}\; I
\label{eq:rDa}
\end{equation}, from $\Da(r)=\tau_{\rm redis}(r)/\tau_{\rm forcing}$ with
$\tau_{\rm redis}(r)=4\mu\bar L^2\bar\tau^2/(r\,\gamma\cos\theta_c)$ and
$\tau_{\rm forcing}=\phi\bar L/I$ (Supplemental Material~\cite{SupplementA}), sets
the \emph{width} of the kinetic band, not a threshold radius separating small
from large pores.  With the frontier-referenced driving
$|\Phi(r,r^*)|\propto|1/r-1/r^*|$, the redistribution time
$\tau_{\rm redis}=\tau(r)$ of Eq.~\eqref{eq:tau_relax} diverges \emph{at} $r^*$, so the set
$\{r:\Da(r)\gtrsim1\}$ is a band straddling the frontier at any intensity: it
never empties, it only narrows as $I\to0$ and widens (by $r_\Da(I)$) as $I$
grows.  Classes outside the band equilibrate between forcing steps and track
$g_{\rm eq}$, the phenomenology traditionally called \textbf{matrix flow}.
Inside and above the band the non-selective precipitation
source~\eqref{eq:precip_source} fills large, high-conductance pores directly,
redistribution is too slow to drain them toward the frontier, and water
accumulates where equilibrium theory predicts emptiness, the phenomenology
traditionally called \textbf{preferential flow}.  The labels attach to the two
behaviours, not to fixed radius classes.  Preferential flow is thus not a separate constitutive
process: it is what the \emph{same} kinetic equation does at $\Da\gtrsim1$, requiring only
the spatial heterogeneity of $f(r,\mathbf{x})$ (carried by the inter-REV current of
Sec.~\ref{sec:spatial}) and the forcing-bundle curvature.  With equilibrium fibers
($\Da\to0$) it cannot occur, however heterogeneous the medium; with a single REV there is
nowhere for it to go.

Three consequences follow at once.  First, Richards' equation is not ``wrong'' for
preferential flow, it is \emph{inapplicable}, exactly as Navier--Stokes is inapplicable
inside a shock: it is the $\Da\to0$ Chapman--Enskog limit
(Sec.~\ref{sec:CE}) of a more general equation, and preferential flow lives at
$\Da\gtrsim1$ where that limit does not hold.  No coupling scheme between ``matrix'' and
``macropore'' domains is needed: there is one pore population, one kinetic equation, and
one number that sets the local regime.  Second, whether Richards' equation applies is a
property of the soil--forcing \emph{combination}, not the soil alone: the same medium is
essentially matric under drizzle ($r_\Da\lesssim1\,\mu$m at $\bar I\sim0.5$~mm/h) and
largely kinetic under a cloudburst ($r_\Da\sim100\,\mu$m at $\bar I\sim500$~mm/h).  The
$r^2$ weighting in $K$ adds a second asymmetry, a few large, fast pores can carry most of
the flux even when most of the pore \emph{volume} is kinetic, which is why Richards'
equation can look adequate for water budgets far from pore-scale equilibrium (the
volume/flux split and its $\sim$10~mm/h threshold are tabulated in the Supplemental
Material~\cite{SupplementA}).  Third, the lab--field discrepancy in $K$ ($\times10$--$100$)
is explained quantitatively: laboratory measurements at $\Da\ll1$ sample $K$ on the
equilibrium submanifold, field conditions at $\Da\sim1$ sample it in the kinetic regime
where filling is biased toward large, high-conductance pores.  The threshold intensity
above which preferential flow activates is then a derived quantity,
$I_{\rm threshold}\approx r_{\rm macro}\,\gamma\cos\theta_c\,\phi/(4\mu\bar L\bar\tau^2)$,
depending only on soil structure and fluid properties, checkable against observed
ponding or breakthrough thresholds.

\paragraph{Where the theory sits among existing models.}
The same crossover organizes the classical models, each of which the kinetic theory
contains as a limit.  \emph{Capillary bundle models}~\cite{Childs1950,Mualem1976,Burdine1953}
(non-communicating tubes, $K$ the PSD-weighted sum of Hagen--Poiseuille conductances) are
recovered exactly in the parallel-tube limit (diagonal dominance, $\mathcal{B}\approx0$;
Sec.~\ref{sec:K_derived}), $K_{\rm mf}\propto\int_0^{r^*}r^2f(r)\,dr$; what they lack is
the connectivity $C(r,r')$, so they miss the percolation threshold and give $K>0$ for any
$\theta>0$, whereas the kinetic theory correctly gives $K=0$ below $\theta_c$, with the
Mualem heterogeneity penalty emerging as the resummation of the connectivity series.
\emph{Percolation / critical-path models}~\cite{Ambegaokar1971,Hunt2004,Berkowitz1993} are
contained in the spectral representation~\eqref{eq:K_exact}: $K$ is set by the smallest
nonzero eigenvalue $\lambda_1$ of the water-budget operator $\mathcal{J}$, the critical
path, with $\lambda_1\to0$ and the correct exponent near $\theta_c$; the kinetic theory
adds the full relaxation spectrum and the transient dynamics at $\Da\sim1$ they do not
describe.  \emph{Dynamic capillary-pressure models}~\cite{Hassanizadeh2002,HassanizadehGray1993},
\begin{equation}
p_c^{\rm static}(\theta) - p_c = \tau_D\,\pder{\theta}{t},
\label{eq:dynamic_cap}
\end{equation}
add a single scalar relaxation coefficient $\tau_D$; the kinetic theory contains this as
its one-moment approximation, predicting $\tau_D\sim\tau_{\rm redis}(r^*)$ to leading order
in $\Da$ but resolving the full, radius- and saturation-dependent distribution of
departure from equilibrium.  The organizing criterion that classical multiple-permeability
frameworks lack is $\Da$ itself: bundle models apply when $C(r,r')\approx f(r')$ and
$\theta\gg\theta_c$, critical-path analysis when $\theta\approx\theta_c$, dynamic-$p_c$
models when $\Da\ll1$ (Table~\ref{tab:comparison}).

\begin{table}[t]
\caption{Kinetic theory compared to existing hydrological models.
\checkmark = captured; $\approx$ = partially; $\times$ = not captured.}
\label{tab:comparison}
\begin{ruledtabular}
\setlength{\tabcolsep}{4pt}
\small
\begin{tabular}{@{}lccc@{}}
\textbf{Feature}
  & \textbf{Bundle}
  & \textbf{Perc.}
  & \textbf{Dyn.\ $p_c$} \\
\hline
$K(\theta)$ integral formula & \checkmark & $\approx$ & \checkmark \\
Percolation threshold $\theta_c$ & $\times$ & \checkmark & $\times$ \\
Heterogeneity penalty & $\times$ & $\approx$ & $\times$ \\
Non-equil.\ redistribution & $\times$ & $\times$ & \checkmark \\
Path-dep.\ hysteresis & $\times$ & $\times$ & $\times$ \\
Preferential flow & $\times$ & $\times$ & $\times$ \\
Pore-class resolution & $\times$ & $\times$ & $\times$ \\
\hline
\textbf{Kinetic theory} & \checkmark & \checkmark & \checkmark \\
\end{tabular}
\end{ruledtabular}
\end{table}

\subsection{Testable predictions and constructive content}
\label{sec:predictions}

Taken together, the two payoffs give each abstract element of the forcing bundle an
explicit, measurable content (Table~\ref{tab:constructive}), and yield four predictions
that distinguish the theory from existing frameworks and can be tested against laboratory
or numerical experiment.  \emph{(i) Rate-dependent hysteresis}: the loop area scales as
$\Da^2$ (equivalently $\propto I^2$) for slow cycles and saturates for fast ones,
checkable by varying the wetting--drying rate on a single sample
[Sec.~\ref{sec:payoff_hysteresis}].  \emph{(ii) Crossover intensity}: the threshold
$I_{\rm threshold}$ at which preferential flow activates
[Sec.~\ref{sec:payoff_preferential}] depends only on pore-structural quantities and can be
compared with observed ponding or breakthrough thresholds.  \emph{(iii) Conductivity
formula}: $K_{\rm mf}[\theta]$ (Eq.~\ref{eq:K_mf}) with micro-CT inputs should reproduce
measured $K(\theta)$ without fitting, including the heterogeneity penalty
$\exp(-4\sigma_{\ln r}^2)$.  \emph{(iv) Field capacity as percolation}: the residual water
content $\theta_r$ should coincide with the water percolation threshold $\theta_c$
estimated independently from network topology.  A toy OpenPNM-based workflow implementing
this measurement-to-prediction chain is in the Supplemental Material~\cite{SupplementA}.

\begin{table*}
\caption{From abstract operators to constructive
physics: the kinetic equation gives each element
explicit content.}
\label{tab:constructive}
\begin{ruledtabular}
\begin{tabular}{ll}
\textbf{Abstract}
  & \textbf{Constructive (kinetic eq.)} \\
\hline
Operator $\W$
  & Gain $\mathcal{G}$: HP rate $\times$ driving
  pot.\ $\Phi > 0$ \\
Operator $\D$
  & Loss $\Lg$: HP rate $\times$ driving
  pot.\ $\Phi < 0$ \\
$P(r'|r)$ correlation
  & $C(r,r')$ connectivity matrix \\
$\tau_{\mathrm{relax}}$ (constant)
  & $\tau(r)$ of Eq.~\eqref{eq:tau_relax}
  (network-emergent) \\
$\Da$ (scalar)
  & $\Da(r,\mathbf{x}) \propto I/r$
  (radius- and forcing-dep.) \\
$g_{\mathrm{eq}}$ unique
  & $g_{\mathrm{eq}}$ unique; $a(r)$ prevents
  approach \\
Richards ``limit''
  & CE derivation with explicit conditions \\
\end{tabular}
\end{ruledtabular}
\end{table*}

\section{Conclusions}
\label{sec:conclusions}

We have derived a continuum theory of unsaturated soil water from the bottom up, and, because the fluid enters only through $(\gamma\cos\theta_c,\mu,\rho)$
and the medium only through its pore geometry, of any Newtonian wetting liquid
in any rigid porous medium.
Its outcome is a single local balance [Eq.~\eqref{eq:pore-point}],
\begin{equation*}
\partial_t g(r,\mathbf{x},t) + \nabla\!\cdot\mathbf{F} = \Coll[g] - \mathcal{E} - \mathcal{T},
\end{equation*}
for the pore-occupancy field $g(r,\mathbf{x},t)$, the fraction of pores of radius $r$
that are water-filled at $\mathbf{x}$.  The theory is built by passing through three scales
and is best read as that passage.  At the \emph{microscale}, single inter-pore transfers are
governed by Hagen--Poiseuille rates $\kappa(r)$ and the capillary--gravitational driving
potential $\Phi$.  At the \emph{mesoscale}, averaging over a representative volume turns these
into gain and loss operators whose difference is the redistribution $\Coll[g]$, relaxing the
occupancy toward the equilibrium step $g_{\rm eq}=H(r^*-r)$.  At the \emph{continuum} scale,
contracting the volume to a point promotes $g$ to a field and the inter-volume exchange to the
transport divergence $\nabla\!\cdot\mathbf{F}$, giving Eq.~\eqref{eq:pore-point}.  Everything
else in the paper, Richards' equation, the conductivity, hysteresis, preferential flow, is a
limit, a moment, or a boundary resolution of this one equation.

The continuum theory unifies phenomenologies that have long been modelled separately.  Richards'
equation is its quasi-static ($\Da\to0$) Chapman--Enskog reduction, with the matric potential
$\psi$ and the conductivity $K(\theta)$ emerging only in that limit, where local equilibrium
makes the map $g\to\theta\to\psi$ one-to-one; out of the limit $g(r)$ is the irreducible state
variable and no scalar potential exists.  Capillary-bundle and critical-path conductivity models
are its diagonal and spectral limits; dynamic capillary-pressure models are its one-moment
approximation.  Hysteresis is the holonomy of the forcing bundle, a geometric phase, not
per-pore bistability, and preferential flow is what the same equation does where the local
Damk\"ohler number $\Da(r,\mathbf{x})$ exceeds unity.  The single dimensionless group
$\Da=\tau_{\rm redis}/\tau_{\rm forcing}$ organizes the whole: $\Da\ll1$ gives Richards'
equation, $\Da\sim1$ gives rate-dependent hysteresis, $\Da\gtrsim1$ gives preferential flow.
Whether Richards' equation applies is therefore a property of the soil--forcing combination at
each point, not of the soil alone.

Two features make the construction more than a reformulation.  It is \emph{thermodynamically
grounded}: Eq.~\eqref{eq:pore-point} descends the Gibbs free energy as an Onsager gradient flow,
$d\mathcal{F}/dt\le0$ (Appendix~\ref{app:variational}), with mass conservation as the zeroth
moment.  And it is \emph{constructive}: every ingredient, $f(r)$, $C(r,r')$, $\bar L$,
$\bar\tau$, and hence $\kappa$, $\Phi$, $\Da$, is a geometric property of the pore network,
measurable from micro-CT without calibration against retention or conductivity data.  The
capillary--gravitational form of $\Phi$ used throughout is a deliberate simplification, not a
limitation: because the theory uses the driving potential only through its values, the single
point of entry for new physics is $\Phi$ itself.  Adding disjoining, osmotic, or
thermal terms to $\Phi$ carries the theory to adsorptive films, saline soils, and coupled heat
transport while leaving the operator algebra, the gradient-flow structure, and the
Chapman--Enskog reduction unchanged.

Structurally, Eq.~\eqref{eq:pore-point} is the porous-media analogue of the Boltzmann equation
(Table~\ref{tab:boltzmann}): $g(r)\leftrightarrow f(\mathbf{v})$, redistribution $\leftrightarrow$
collision integral, molecular chaos $\leftrightarrow$ the mean-field closure, and
Chapman--Enskog${}\to{}$Richards mirroring Chapman--Enskog${}\to{}$Navier--Stokes, down to the
$H$-theorem and the breakdown at sharp fronts (the analogue of shocks).  The same skeleton
supports the classical analogies with magnetic hysteresis (Preisach--Bertotti) and
plasticity~\cite{Preisach1935,Bertotti1998,Bouchaud1998}, now made precise: $\theta_c$ is a yield
surface, trapped water a plastic strain, and hysteresis the holonomy of a state-dependent
connection.  These are not ornaments but working correspondences, Preisach identification,
variational-inequality methods, and lattice-gauge holonomy computations transfer to a medium in
which the structure is directly observable in centimetre-scale columns.  The central falsifiable
prediction, $\mathcal{H}(I)\propto I^2$ for the hysteresis loop area, follows from the curvature
alone and no rate-independent model reproduces it.

\begin{acknowledgments}
The Author thanks Chiara Baldo and Lorenzo Duchi who assisted the creation of the MOOC on soil equilibrium theory from which he ultimately found the key to develop this new one.  This study was carried out within the Space It Up project funded by the Italian Space Agency, ASI, and the Ministry of University and Research, MUR, under contract n. 2024-5-E.0 - CUP n. I53D24000060005.
\end{acknowledgments}

\bibliography{main}

\begin{thebibliography}{56}%
\makeatletter
\providecommand \@ifxundefined [1]{%
 \@ifx{#1\undefined}
}%
\providecommand \@ifnum [1]{%
 \ifnum #1\expandafter \@firstoftwo
 \else \expandafter \@secondoftwo
 \fi
}%
\providecommand \@ifx [1]{%
 \ifx #1\expandafter \@firstoftwo
 \else \expandafter \@secondoftwo
 \fi
}%
\providecommand \natexlab [1]{#1}%
\providecommand \enquote  [1]{``#1''}%
\providecommand \bibnamefont  [1]{#1}%
\providecommand \bibfnamefont [1]{#1}%
\providecommand \citenamefont [1]{#1}%
\providecommand \href@noop [0]{\@secondoftwo}%
\providecommand \href [0]{\begingroup \@sanitize@url \@href}%
\providecommand \@href[1]{\@@startlink{#1}\@@href}%
\providecommand \@@href[1]{\endgroup#1\@@endlink}%
\providecommand \@sanitize@url [0]{\catcode `\\12\catcode `\$12\catcode
  `\&12\catcode `\#12\catcode `\^12\catcode `\_12\catcode `\%12\relax}%
\providecommand \@@startlink[1]{}%
\providecommand \@@endlink[0]{}%
\providecommand \url  [0]{\begingroup\@sanitize@url \@url }%
\providecommand \@url [1]{\endgroup\@href {#1}{\urlprefix }}%
\providecommand \urlprefix  [0]{URL }%
\providecommand \Eprint [0]{\href }%
\providecommand \doibase [0]{https://doi.org/}%
\providecommand \selectlanguage [0]{\@gobble}%
\providecommand \bibinfo  [0]{\@secondoftwo}%
\providecommand \bibfield  [0]{\@secondoftwo}%
\providecommand \translation [1]{[#1]}%
\providecommand \BibitemOpen [0]{}%
\providecommand \bibitemStop [0]{}%
\providecommand \bibitemNoStop [0]{.\EOS\space}%
\providecommand \EOS [0]{\spacefactor3000\relax}%
\providecommand \BibitemShut  [1]{\csname bibitem#1\endcsname}%
\let\auto@bib@innerbib\@empty
\bibitem [{\citenamefont {Alonso}\ \emph {et~al.}(1990)\citenamefont {Alonso},
  \citenamefont {Gens},\ and\ \citenamefont {Josa}}]{Alonso1990}%
  \BibitemOpen
  \bibfield  {author} {\bibinfo {author} {\bibfnamefont {E.~E.}\ \bibnamefont
  {Alonso}}, \bibinfo {author} {\bibfnamefont {A.}~\bibnamefont {Gens}},\ and\
  \bibinfo {author} {\bibfnamefont {A.}~\bibnamefont {Josa}},\ }\bibfield
  {title} {\bibinfo {title} {A constitutive model for partially saturated
  soils},\ }\href@noop {} {\bibfield  {journal} {\bibinfo  {journal}
  {G{\'e}otechnique}\ }\textbf {\bibinfo {volume} {40}},\ \bibinfo {pages}
  {405} (\bibinfo {year} {1990})}\BibitemShut {NoStop}%
\bibitem [{\citenamefont {Haines}(1930)}]{Haines1930}%
  \BibitemOpen
  \bibfield  {author} {\bibinfo {author} {\bibfnamefont {W.~B.}\ \bibnamefont
  {Haines}},\ }\bibfield  {title} {\bibinfo {title} {Studies in the physical
  properties of soil. {V}. the hysteresis effect in capillary properties, and
  the modes of moisture distribution associated therewith},\ }\href@noop {}
  {\bibfield  {journal} {\bibinfo  {journal} {J. Agric. Sci.}\ }\textbf
  {\bibinfo {volume} {20}},\ \bibinfo {pages} {97} (\bibinfo {year}
  {1930})}\BibitemShut {NoStop}%
\bibitem [{\citenamefont {Mualem}(1974)}]{Mualem1974}%
  \BibitemOpen
  \bibfield  {author} {\bibinfo {author} {\bibfnamefont {Y.}~\bibnamefont
  {Mualem}},\ }\bibfield  {title} {\bibinfo {title} {A conceptual model of
  hysteresis},\ }\href@noop {} {\bibfield  {journal} {\bibinfo  {journal}
  {Water Resour. Res.}\ }\textbf {\bibinfo {volume} {10}},\ \bibinfo {pages}
  {514} (\bibinfo {year} {1974})}\BibitemShut {NoStop}%
\bibitem [{\citenamefont {Richardson}(1922)}]{Richardson1922}%
  \BibitemOpen
  \bibfield  {author} {\bibinfo {author} {\bibfnamefont {L.~F.}\ \bibnamefont
  {Richardson}},\ }\bibfield  {title} {\bibinfo {title} {Weather prediction by
  numerical process},\ }\href@noop {} {\bibfield  {journal} {\bibinfo
  {journal} {Cambridge Univ. Press}\ } (\bibinfo {year} {1922})}\BibitemShut
  {NoStop}%
\bibitem [{\citenamefont {Richards}(1931)}]{Richards1931}%
  \BibitemOpen
  \bibfield  {author} {\bibinfo {author} {\bibfnamefont {L.~A.}\ \bibnamefont
  {Richards}},\ }\bibfield  {title} {\bibinfo {title} {Capillary conduction of
  liquids through porous mediums},\ }\href@noop {} {\bibfield  {journal}
  {\bibinfo  {journal} {Physics}\ }\textbf {\bibinfo {volume} {1}},\ \bibinfo
  {pages} {318} (\bibinfo {year} {1931})}\BibitemShut {NoStop}%
\bibitem [{\citenamefont {Tubini}\ \emph {et~al.}(2021)\citenamefont {Tubini},
  \citenamefont {Gruber},\ and\ \citenamefont {Rigon}}]{Tubini2022}%
  \BibitemOpen
  \bibfield  {author} {\bibinfo {author} {\bibfnamefont {N.}~\bibnamefont
  {Tubini}}, \bibinfo {author} {\bibfnamefont {S.}~\bibnamefont {Gruber}},\
  and\ \bibinfo {author} {\bibfnamefont {R.}~\bibnamefont {Rigon}},\ }\bibfield
   {title} {\bibinfo {title} {A method for solving heat transfer with phase
  change in ice or soil that allows for large time steps while guaranteeing
  energy conservation},\ }\href@noop {} {\bibfield  {journal} {\bibinfo
  {journal} {The Cryosphere}\ }\textbf {\bibinfo {volume} {15}},\ \bibinfo
  {pages} {2541} (\bibinfo {year} {2021})}\BibitemShut {NoStop}%
\bibitem [{\citenamefont {Beven}\ and\ \citenamefont
  {Germann}(1982)}]{Beven1982}%
  \BibitemOpen
  \bibfield  {author} {\bibinfo {author} {\bibfnamefont {K.}~\bibnamefont
  {Beven}}\ and\ \bibinfo {author} {\bibfnamefont {P.}~\bibnamefont
  {Germann}},\ }\bibfield  {title} {\bibinfo {title} {Macropores and water flow
  in soils},\ }\href@noop {} {\bibfield  {journal} {\bibinfo  {journal} {Water
  Resour. Res.}\ }\textbf {\bibinfo {volume} {18}},\ \bibinfo {pages} {1311}
  (\bibinfo {year} {1982})}\BibitemShut {NoStop}%
\bibitem [{\citenamefont {Germann}\ and\ \citenamefont
  {Beven}(1985)}]{Germann1985}%
  \BibitemOpen
  \bibfield  {author} {\bibinfo {author} {\bibfnamefont {P.~F.}\ \bibnamefont
  {Germann}}\ and\ \bibinfo {author} {\bibfnamefont {K.}~\bibnamefont
  {Beven}},\ }\bibfield  {title} {\bibinfo {title} {Kinematic wave
  approximation to infiltration into soils with sorbing macropores},\
  }\href@noop {} {\bibfield  {journal} {\bibinfo  {journal} {Water Resour.
  Res.}\ }\textbf {\bibinfo {volume} {21}},\ \bibinfo {pages} {990} (\bibinfo
  {year} {1985})}\BibitemShut {NoStop}%
\bibitem [{\citenamefont {Germann}(2014)}]{Germann2014}%
  \BibitemOpen
  \bibfield  {author} {\bibinfo {author} {\bibfnamefont {P.~F.}\ \bibnamefont
  {Germann}},\ }\href@noop {} {\emph {\bibinfo {title} {Preferential flow:
  {Stokes} approach to infiltration and drainage}}},\ \bibinfo {edition} {1st}\
  ed.\ (\bibinfo  {publisher} {Geographica Bernensia},\ \bibinfo {address}
  {Bern},\ \bibinfo {year} {2014})\BibitemShut {NoStop}%
\bibitem [{\citenamefont {Gerke}\ and\ \citenamefont {van
  Genuchten}(1993)}]{Gerke1993}%
  \BibitemOpen
  \bibfield  {author} {\bibinfo {author} {\bibfnamefont {H.~H.}\ \bibnamefont
  {Gerke}}\ and\ \bibinfo {author} {\bibfnamefont {M.~T.}\ \bibnamefont {van
  Genuchten}},\ }\bibfield  {title} {\bibinfo {title} {A dual-porosity model
  for simulating the preferential movement of water and solutes in structured
  porous media},\ }\href@noop {} {\bibfield  {journal} {\bibinfo  {journal}
  {Water Resour. Res.}\ }\textbf {\bibinfo {volume} {29}},\ \bibinfo {pages}
  {305} (\bibinfo {year} {1993})}\BibitemShut {NoStop}%
\bibitem [{\citenamefont {Zehe}\ and\ \citenamefont
  {Jackisch}(2016)}]{ZeheJackisch2016}%
  \BibitemOpen
  \bibfield  {author} {\bibinfo {author} {\bibfnamefont {E.}~\bibnamefont
  {Zehe}}\ and\ \bibinfo {author} {\bibfnamefont {C.}~\bibnamefont
  {Jackisch}},\ }\bibfield  {title} {\bibinfo {title} {A {Lagrangian} model for
  soil water dynamics during rainfall-driven conditions},\ }\href
  {https://doi.org/10.5194/hess-20-3511-2016} {\bibfield  {journal} {\bibinfo
  {journal} {Hydrology and Earth System Sciences}\ }\textbf {\bibinfo {volume}
  {20}},\ \bibinfo {pages} {3511} (\bibinfo {year} {2016})}\BibitemShut
  {NoStop}%
\bibitem [{\citenamefont {Hassanizadeh}\ and\ \citenamefont
  {Gray}(1993)}]{HassanizadehGray1993}%
  \BibitemOpen
  \bibfield  {author} {\bibinfo {author} {\bibfnamefont {S.~M.}\ \bibnamefont
  {Hassanizadeh}}\ and\ \bibinfo {author} {\bibfnamefont {W.~G.}\ \bibnamefont
  {Gray}},\ }\bibfield  {title} {\bibinfo {title} {Thermodynamic basis of
  capillary pressure in porous media},\ }\href@noop {} {\bibfield  {journal}
  {\bibinfo  {journal} {Water Resour. Res.}\ }\textbf {\bibinfo {volume}
  {29}},\ \bibinfo {pages} {3389} (\bibinfo {year} {1993})}\BibitemShut
  {NoStop}%
\bibitem [{\citenamefont {Poulovassilis}(1962)}]{Poulovassilis1962}%
  \BibitemOpen
  \bibfield  {author} {\bibinfo {author} {\bibfnamefont {A.}~\bibnamefont
  {Poulovassilis}},\ }\bibfield  {title} {\bibinfo {title} {Hysteresis of pore
  water, an application of the concept of independent domains},\ }\href@noop {}
  {\bibfield  {journal} {\bibinfo  {journal} {Soil Sci.}\ }\textbf {\bibinfo
  {volume} {93}},\ \bibinfo {pages} {405} (\bibinfo {year} {1962})}\BibitemShut
  {NoStop}%
\bibitem [{\citenamefont {Topp}\ and\ \citenamefont
  {Miller}(1966)}]{Topp1971-zr}%
  \BibitemOpen
  \bibfield  {author} {\bibinfo {author} {\bibfnamefont {G.~C.}\ \bibnamefont
  {Topp}}\ and\ \bibinfo {author} {\bibfnamefont {E.~E.}\ \bibnamefont
  {Miller}},\ }\bibfield  {title} {\bibinfo {title} {Hysteretic moisture
  characteristics and hydraulic conductivities for glass-bead media},\
  }\href@noop {} {\bibfield  {journal} {\bibinfo  {journal} {Soil Sci. Soc. Am.
  Proc.}\ }\textbf {\bibinfo {volume} {30}},\ \bibinfo {pages} {156} (\bibinfo
  {year} {1966})}\BibitemShut {NoStop}%
\bibitem [{\citenamefont {Kool}\ and\ \citenamefont
  {Parker}(1987)}]{Kool1987-le}%
  \BibitemOpen
  \bibfield  {author} {\bibinfo {author} {\bibfnamefont {J.~B.}\ \bibnamefont
  {Kool}}\ and\ \bibinfo {author} {\bibfnamefont {J.~C.}\ \bibnamefont
  {Parker}},\ }\bibfield  {title} {\bibinfo {title} {Development and evaluation
  of closed-form expressions for hysteretic soil hydraulic properties},\
  }\href@noop {} {\bibfield  {journal} {\bibinfo  {journal} {Water Resour.
  Res.}\ }\textbf {\bibinfo {volume} {23}},\ \bibinfo {pages} {105} (\bibinfo
  {year} {1987})}\BibitemShut {NoStop}%
\bibitem [{\citenamefont {Preisach}(1935)}]{Preisach1935}%
  \BibitemOpen
  \bibfield  {author} {\bibinfo {author} {\bibfnamefont {F.}~\bibnamefont
  {Preisach}},\ }\bibfield  {title} {\bibinfo {title} {{\"U}ber die magnetische
  {Nachwirkung}},\ }\href@noop {} {\bibfield  {journal} {\bibinfo  {journal}
  {Z. Phys.}\ }\textbf {\bibinfo {volume} {94}},\ \bibinfo {pages} {277}
  (\bibinfo {year} {1935})}\BibitemShut {NoStop}%
\bibitem [{\citenamefont {Bertotti}(1998)}]{Bertotti1998}%
  \BibitemOpen
  \bibfield  {author} {\bibinfo {author} {\bibfnamefont {G.}~\bibnamefont
  {Bertotti}},\ }\href@noop {} {\emph {\bibinfo {title} {Hysteresis in
  Magnetism}}}\ (\bibinfo  {publisher} {Academic Press},\ \bibinfo {year}
  {1998})\BibitemShut {NoStop}%
\bibitem [{\citenamefont {Bouchaud}\ \emph {et~al.}(1998)\citenamefont
  {Bouchaud}, \citenamefont {Cugliandolo}, \citenamefont {Kurchan},\ and\
  \citenamefont {M{\'e}zard}}]{Bouchaud1998}%
  \BibitemOpen
  \bibfield  {author} {\bibinfo {author} {\bibfnamefont {J.-P.}\ \bibnamefont
  {Bouchaud}}, \bibinfo {author} {\bibfnamefont {L.~F.}\ \bibnamefont
  {Cugliandolo}}, \bibinfo {author} {\bibfnamefont {J.}~\bibnamefont
  {Kurchan}},\ and\ \bibinfo {author} {\bibfnamefont {M.}~\bibnamefont
  {M{\'e}zard}},\ }\bibfield  {title} {\bibinfo {title} {Out of equilibrium
  dynamics in spin-glasses and other glassy systems},\ }\href@noop {}
  {\bibfield  {journal} {\bibinfo  {journal} {Spin Glasses and Random Fields}\
  ,\ \bibinfo {pages} {161}} (\bibinfo {year} {1998})}\BibitemShut {NoStop}%
\bibitem [{\citenamefont {Davidson}\ \emph {et~al.}(1963)\citenamefont
  {Davidson}, \citenamefont {Biggar},\ and\ \citenamefont
  {Nielsen}}]{Davidson1969}%
  \BibitemOpen
  \bibfield  {author} {\bibinfo {author} {\bibfnamefont {J.~M.}\ \bibnamefont
  {Davidson}}, \bibinfo {author} {\bibfnamefont {J.~W.}\ \bibnamefont
  {Biggar}},\ and\ \bibinfo {author} {\bibfnamefont {D.~R.}\ \bibnamefont
  {Nielsen}},\ }\bibfield  {title} {\bibinfo {title} {Gamma-radiation
  attenuation for measuring bulk density and transient water flow in porous
  materials},\ }\href@noop {} {\bibfield  {journal} {\bibinfo  {journal} {J.
  Geophys. Res.}\ }\textbf {\bibinfo {volume} {68}},\ \bibinfo {pages} {4777}
  (\bibinfo {year} {1963})}\BibitemShut {NoStop}%
\bibitem [{\citenamefont {Vogel}\ \emph {et~al.}(2010)\citenamefont {Vogel},
  \citenamefont {Weller},\ and\ \citenamefont {Schl{\"u}ter}}]{Vogel2010-wi}%
  \BibitemOpen
  \bibfield  {author} {\bibinfo {author} {\bibfnamefont {H.-J.}\ \bibnamefont
  {Vogel}}, \bibinfo {author} {\bibfnamefont {U.}~\bibnamefont {Weller}},\ and\
  \bibinfo {author} {\bibfnamefont {S.}~\bibnamefont {Schl{\"u}ter}},\
  }\bibfield  {title} {\bibinfo {title} {Quantification of soil structure based
  on {Minkowski} functions},\ }\href@noop {} {\bibfield  {journal} {\bibinfo
  {journal} {Eur. J. Soil Sci.}\ }\textbf {\bibinfo {volume} {61}},\ \bibinfo
  {pages} {831} (\bibinfo {year} {2010})}\BibitemShut {NoStop}%
\bibitem [{\citenamefont {Schutz}(1980)}]{Schutz1980-uu}%
  \BibitemOpen
  \bibfield  {author} {\bibinfo {author} {\bibfnamefont {B.~F.}\ \bibnamefont
  {Schutz}},\ }\href@noop {} {\emph {\bibinfo {title} {Geometrical Methods of
  Mathematical Physics}}}\ (\bibinfo  {publisher} {Cambridge Univ. Press},\
  \bibinfo {year} {1980})\BibitemShut {NoStop}%
\bibitem [{\citenamefont {Bachmat}\ and\ \citenamefont
  {Bear}(1986)}]{Bachmat1987-cv}%
  \BibitemOpen
  \bibfield  {author} {\bibinfo {author} {\bibfnamefont {Y.}~\bibnamefont
  {Bachmat}}\ and\ \bibinfo {author} {\bibfnamefont {J.}~\bibnamefont {Bear}},\
  }\bibfield  {title} {\bibinfo {title} {Macroscopic modelling of transport
  phenomena in porous media. 1. the continuum approach},\ }\href@noop {}
  {\bibfield  {journal} {\bibinfo  {journal} {Transp. Porous Media}\ }\textbf
  {\bibinfo {volume} {1}},\ \bibinfo {pages} {213} (\bibinfo {year}
  {1986})}\BibitemShut {NoStop}%
\bibitem [{\citenamefont {Mualem}(1976)}]{Mualem1976}%
  \BibitemOpen
  \bibfield  {author} {\bibinfo {author} {\bibfnamefont {Y.}~\bibnamefont
  {Mualem}},\ }\bibfield  {title} {\bibinfo {title} {A new model for predicting
  the hydraulic conductivity of unsaturated porous media},\ }\href@noop {}
  {\bibfield  {journal} {\bibinfo  {journal} {Water Resour. Res.}\ }\textbf
  {\bibinfo {volume} {12}},\ \bibinfo {pages} {513} (\bibinfo {year}
  {1976})}\BibitemShut {NoStop}%
\bibitem [{Note1()}]{Note1}%
  \BibitemOpen
  \bibinfo {note} {This edge length is the same $\protect \mathaccentV
  {bar}016{L}$ that sets the mean pore length in the network statistics, and,
  crucially for the coarse-graining, the same inter-pore spacing $\ell \sim
  \protect \mathaccentV {bar}016{L}$ used in the gradient expansion of
  Sec.~\ref {sec:spatial} and Appendix~\ref {app:single_parameter}. It is
  \protect \emph {not} the viscous length appearing squared in $\kappa
  _{\protect \rm HP}\propto r^2/(\mu \protect \mathaccentV
  {bar}016{L}^2\protect \mathaccentV {bar}016{\tau }^2)$ [Eq.~\protect \textup
  {\hbox {\mathsurround \z@ \protect \normalfont (\ignorespaces \ref
  {eq:kappa}\unskip \@@italiccorr )}}], which measures dissipative resistance
  along the pore; the two coincide numerically in a statistically homogeneous
  network but play distinct roles. Writing the gravity term this way is what
  lets the Chapman--Enskog limit (Sec.~\ref {sec:CE}) turn $\Delta z\to \ell
  \protect \tmspace +\thinmuskip {.1667em}\protect \mathaccentV
  {hat}05E{\protect \bm {e}}\protect \tmspace -\thinmuskip {.1667em}\cdot
  \protect \tmspace -\thinmuskip {.1667em}\nabla z=\ell \protect \tmspace
  +\thinmuskip {.1667em}\protect \mathaccentV {hat}05E{\protect \bm
  {e}}\protect \tmspace -\thinmuskip {.1667em}\cdot \protect \tmspace
  -\thinmuskip {.1667em}\protect \mathaccentV {hat}05E{\protect \bm {z}}$ into
  the clean $+\protect \mathaccentV {hat}05E{\protect \bm {z}}$ of Richards'
  $q=-K(\nabla \psi +\protect \mathaccentV {hat}05E{\protect \bm {z}})$, with
  no length left over to explain away.}\BibitemShut {Stop}%
\bibitem [{\citenamefont {Washburn}(1921)}]{Washburn1921}%
  \BibitemOpen
  \bibfield  {author} {\bibinfo {author} {\bibfnamefont {E.~W.}\ \bibnamefont
  {Washburn}},\ }\bibfield  {title} {\bibinfo {title} {The dynamics of
  capillary flow},\ }\href@noop {} {\bibfield  {journal} {\bibinfo  {journal}
  {Phys. Rev.}\ }\textbf {\bibinfo {volume} {17}},\ \bibinfo {pages} {273}
  (\bibinfo {year} {1921})}\BibitemShut {NoStop}%
\bibitem [{\citenamefont {Rigon}(2026{\natexlab{a}})}]{Rigon2026CE}%
  \BibitemOpen
  \bibfield  {author} {\bibinfo {author} {\bibfnamefont {R.}~\bibnamefont
  {Rigon}},\ }\href@noop {} {\bibinfo {title} {Richards' equation as a
  hydrodynamic limit: {Chapman--Enskog} reduction of the continuum kinetic
  equation for unsaturated soil water}} (\bibinfo {year}
  {2026}{\natexlab{a}}),\ \bibinfo {note} {companion paper (PRE-2)},\ \Eprint
  {https://arxiv.org/abs/2607.17358} {arXiv:2607.17358 [cond-mat.stat-mech]}
  \BibitemShut {NoStop}%
\bibitem [{\citenamefont {Millington}\ and\ \citenamefont
  {Quirk}(1961)}]{Millington1961-yf}%
  \BibitemOpen
  \bibfield  {author} {\bibinfo {author} {\bibfnamefont {R.~J.}\ \bibnamefont
  {Millington}}\ and\ \bibinfo {author} {\bibfnamefont {J.~P.}\ \bibnamefont
  {Quirk}},\ }\bibfield  {title} {\bibinfo {title} {Permeability of porous
  solids},\ }\href@noop {} {\bibfield  {journal} {\bibinfo  {journal} {Trans.
  Faraday Soc.}\ }\textbf {\bibinfo {volume} {57}},\ \bibinfo {pages} {1200}
  (\bibinfo {year} {1961})}\BibitemShut {NoStop}%
\bibitem [{\citenamefont {Lehmann}\ \emph {et~al.}(2008)\citenamefont
  {Lehmann}, \citenamefont {Assouline},\ and\ \citenamefont
  {Or}}]{Lehmann2008}%
  \BibitemOpen
  \bibfield  {author} {\bibinfo {author} {\bibfnamefont {P.}~\bibnamefont
  {Lehmann}}, \bibinfo {author} {\bibfnamefont {S.}~\bibnamefont {Assouline}},\
  and\ \bibinfo {author} {\bibfnamefont {D.}~\bibnamefont {Or}},\ }\bibfield
  {title} {\bibinfo {title} {Characteristic lengths affecting evaporative
  drying of porous media},\ }\href@noop {} {\bibfield  {journal} {\bibinfo
  {journal} {Phys. Rev. E}\ }\textbf {\bibinfo {volume} {77}},\ \bibinfo
  {pages} {056309} (\bibinfo {year} {2008})}\BibitemShut {NoStop}%
\bibitem [{\citenamefont {D'Amato}\ and\ \citenamefont
  {Rigon}(2025)}]{DAmatoRigon2025}%
  \BibitemOpen
  \bibfield  {author} {\bibinfo {author} {\bibfnamefont {C.}~\bibnamefont
  {D'Amato}}\ and\ \bibinfo {author} {\bibfnamefont {R.}~\bibnamefont
  {Rigon}},\ }\bibfield  {title} {\bibinfo {title} {Elementary mathematics
  helps to shed light on the transpiration budget under water stress},\ }\href
  {https://doi.org/10.1002/eco.70009} {\bibfield  {journal} {\bibinfo
  {journal} {Ecohydrology}\ }\textbf {\bibinfo {volume} {18}},\ \bibinfo
  {pages} {e70009} (\bibinfo {year} {2025})}\BibitemShut {NoStop}%
\bibitem [{\citenamefont {Hunt}\ \emph {et~al.}(2014)\citenamefont {Hunt},
  \citenamefont {Ewing},\ and\ \citenamefont {Ghanbarian}}]{Hunt2017}%
  \BibitemOpen
  \bibfield  {author} {\bibinfo {author} {\bibfnamefont {A.~G.}\ \bibnamefont
  {Hunt}}, \bibinfo {author} {\bibfnamefont {R.~P.}\ \bibnamefont {Ewing}},\
  and\ \bibinfo {author} {\bibfnamefont {B.}~\bibnamefont {Ghanbarian}},\
  }\href@noop {} {\emph {\bibinfo {title} {Percolation Theory for Flow in
  Porous Media}}},\ \bibinfo {series} {Lecture Notes in Physics}, Vol.\
  \bibinfo {volume} {880}\ (\bibinfo  {publisher} {Springer},\ \bibinfo {year}
  {2014})\BibitemShut {NoStop}%
\bibitem [{\citenamefont {Hunt}\ and\ \citenamefont {Gee}(2002)}]{Hunt2001}%
  \BibitemOpen
  \bibfield  {author} {\bibinfo {author} {\bibfnamefont {A.~G.}\ \bibnamefont
  {Hunt}}\ and\ \bibinfo {author} {\bibfnamefont {G.~W.}\ \bibnamefont {Gee}},\
  }\bibfield  {title} {\bibinfo {title} {Application of critical path analysis
  to fractal porous media: comparison with examples from the {Hanford} site},\
  }\href@noop {} {\bibfield  {journal} {\bibinfo  {journal} {Adv. Water
  Resour.}\ }\textbf {\bibinfo {volume} {25}},\ \bibinfo {pages} {129}
  (\bibinfo {year} {2002})}\BibitemShut {NoStop}%
\bibitem [{\citenamefont {Boltzmann}(1872)}]{Boltzmann1872}%
  \BibitemOpen
  \bibfield  {author} {\bibinfo {author} {\bibfnamefont {L.}~\bibnamefont
  {Boltzmann}},\ }\bibfield  {title} {\bibinfo {title} {{Weitere Studien
  {\"u}ber das W{\"a}rmegleichgewicht unter Gasmolek{\"u}len}},\ }\href@noop {}
  {\bibfield  {journal} {\bibinfo  {journal} {Sitzungsber. Akad. Wiss. Wien}\
  }\textbf {\bibinfo {volume} {66}},\ \bibinfo {pages} {275} (\bibinfo {year}
  {1872})}\BibitemShut {NoStop}%
\bibitem [{\citenamefont {Cercignani}(1988)}]{Cercignani1988}%
  \BibitemOpen
  \bibfield  {author} {\bibinfo {author} {\bibfnamefont {C.}~\bibnamefont
  {Cercignani}},\ }\href@noop {} {\emph {\bibinfo {title} {The {Boltzmann}
  Equation and Its Applications}}}\ (\bibinfo  {publisher} {Springer},\
  \bibinfo {year} {1988})\BibitemShut {NoStop}%
\bibitem [{\citenamefont {Kipnis}\ and\ \citenamefont
  {Landim}(1999)}]{KipnisLandim1999}%
  \BibitemOpen
  \bibfield  {author} {\bibinfo {author} {\bibfnamefont {C.}~\bibnamefont
  {Kipnis}}\ and\ \bibinfo {author} {\bibfnamefont {C.}~\bibnamefont
  {Landim}},\ }\href@noop {} {\emph {\bibinfo {title} {Scaling Limits of
  Interacting Particle Systems}}},\ \bibinfo {series} {Grundlehren der
  mathematischen Wissenschaften}, Vol.\ \bibinfo {volume} {320}\ (\bibinfo
  {publisher} {Springer},\ \bibinfo {address} {Berlin},\ \bibinfo {year}
  {1999})\BibitemShut {NoStop}%
\bibitem [{\citenamefont {Or}\ \emph {et~al.}(2013)\citenamefont {Or},
  \citenamefont {Lehmann}, \citenamefont {Shahraeeni},\ and\ \citenamefont
  {Shokri}}]{Or2013-pf}%
  \BibitemOpen
  \bibfield  {author} {\bibinfo {author} {\bibfnamefont {D.}~\bibnamefont
  {Or}}, \bibinfo {author} {\bibfnamefont {P.}~\bibnamefont {Lehmann}},
  \bibinfo {author} {\bibfnamefont {E.}~\bibnamefont {Shahraeeni}},\ and\
  \bibinfo {author} {\bibfnamefont {N.}~\bibnamefont {Shokri}},\ }\bibfield
  {title} {\bibinfo {title} {Advances in soil evaporation physics?a review},\
  }\href@noop {} {\bibfield  {journal} {\bibinfo  {journal} {Vadose Zone
  Journal}\ }\textbf {\bibinfo {volume} {12}} (\bibinfo {year}
  {2013})}\BibitemShut {NoStop}%
\bibitem [{\citenamefont {Hosseini}\ \emph {et~al.}(2024)\citenamefont
  {Hosseini}, \citenamefont {Kumar},\ and\ \citenamefont
  {Delenne}}]{Hosseini2024}%
  \BibitemOpen
  \bibfield  {author} {\bibinfo {author} {\bibfnamefont {R.}~\bibnamefont
  {Hosseini}}, \bibinfo {author} {\bibfnamefont {K.}~\bibnamefont {Kumar}},\
  and\ \bibinfo {author} {\bibfnamefont {J.-Y.}\ \bibnamefont {Delenne}},\
  }\bibfield  {title} {\bibinfo {title} {Investigating the source of hysteresis
  in the soil--water characteristic curve using the multiphase lattice
  {Boltzmann} method},\ }\href {https://doi.org/10.1007/s11440-024-02295-y}
  {\bibfield  {journal} {\bibinfo  {journal} {Acta Geotech.}\ }\textbf
  {\bibinfo {volume} {19}},\ \bibinfo {pages} {7577} (\bibinfo {year}
  {2024})}\BibitemShut {NoStop}%
\bibitem [{\citenamefont {Tuller}\ \emph {et~al.}(1999)\citenamefont {Tuller},
  \citenamefont {Or},\ and\ \citenamefont {Dudley}}]{Tuller1999}%
  \BibitemOpen
  \bibfield  {author} {\bibinfo {author} {\bibfnamefont {M.}~\bibnamefont
  {Tuller}}, \bibinfo {author} {\bibfnamefont {D.}~\bibnamefont {Or}},\ and\
  \bibinfo {author} {\bibfnamefont {L.~M.}\ \bibnamefont {Dudley}},\ }\bibfield
   {title} {\bibinfo {title} {Adsorption and capillary condensation in porous
  media: {Liquid} retention and interfacial configurations in angular pores},\
  }\href@noop {} {\bibfield  {journal} {\bibinfo  {journal} {Water Resour.
  Res.}\ }\textbf {\bibinfo {volume} {35}},\ \bibinfo {pages} {1949} (\bibinfo
  {year} {1999})}\BibitemShut {NoStop}%
\bibitem [{\citenamefont {Luo}\ \emph {et~al.}(2022)\citenamefont {Luo},
  \citenamefont {Lu}, \citenamefont {Zhang},\ and\ \citenamefont
  {Likos}}]{Luo2022-mu}%
  \BibitemOpen
  \bibfield  {author} {\bibinfo {author} {\bibfnamefont {S.}~\bibnamefont
  {Luo}}, \bibinfo {author} {\bibfnamefont {N.}~\bibnamefont {Lu}}, \bibinfo
  {author} {\bibfnamefont {C.}~\bibnamefont {Zhang}},\ and\ \bibinfo {author}
  {\bibfnamefont {W.}~\bibnamefont {Likos}},\ }\bibfield  {title} {\bibinfo
  {title} {Soil water potential: A historical perspective and recent
  breakthroughs},\ }\href@noop {} {\bibfield  {journal} {\bibinfo  {journal}
  {Vadose Zone J.}\ } (\bibinfo {year} {2022})}\BibitemShut {NoStop}%
\bibitem [{\citenamefont {Kawasaki}(1966)}]{Kawasaki1966}%
  \BibitemOpen
  \bibfield  {author} {\bibinfo {author} {\bibfnamefont {K.}~\bibnamefont
  {Kawasaki}},\ }\bibfield  {title} {\bibinfo {title} {Diffusion constants near
  the critical point for time-dependent {Ising} models. {I}},\ }\href
  {https://doi.org/10.1103/PhysRev.145.224} {\bibfield  {journal} {\bibinfo
  {journal} {Phys. Rev.}\ }\textbf {\bibinfo {volume} {145}},\ \bibinfo {pages}
  {224} (\bibinfo {year} {1966})}\BibitemShut {NoStop}%
\bibitem [{\citenamefont {Huang}(2009)}]{Huang2009}%
  \BibitemOpen
  \bibfield  {author} {\bibinfo {author} {\bibfnamefont {K.}~\bibnamefont
  {Huang}},\ }\href@noop {} {\emph {\bibinfo {title} {Introduction to
  Statistical Physics}}},\ \bibinfo {edition} {2nd}\ ed.\ (\bibinfo
  {publisher} {CRC Press},\ \bibinfo {year} {2009})\BibitemShut {NoStop}%
\bibitem [{\citenamefont {Onsager}(1931{\natexlab{a}})}]{Onsager1931a}%
  \BibitemOpen
  \bibfield  {author} {\bibinfo {author} {\bibfnamefont {L.}~\bibnamefont
  {Onsager}},\ }\bibfield  {title} {\bibinfo {title} {Reciprocal relations in
  irreversible processes. {I}},\ }\href@noop {} {\bibfield  {journal} {\bibinfo
   {journal} {Phys. Rev.}\ }\textbf {\bibinfo {volume} {37}},\ \bibinfo {pages}
  {405} (\bibinfo {year} {1931}{\natexlab{a}})}\BibitemShut {NoStop}%
\bibitem [{\citenamefont {Onsager}(1931{\natexlab{b}})}]{Onsager1931b}%
  \BibitemOpen
  \bibfield  {author} {\bibinfo {author} {\bibfnamefont {L.}~\bibnamefont
  {Onsager}},\ }\bibfield  {title} {\bibinfo {title} {Reciprocal relations in
  irreversible processes. {II}},\ }\href@noop {} {\bibfield  {journal}
  {\bibinfo  {journal} {Phys. Rev.}\ }\textbf {\bibinfo {volume} {38}},\
  \bibinfo {pages} {2265} (\bibinfo {year} {1931}{\natexlab{b}})}\BibitemShut
  {NoStop}%
\bibitem [{\citenamefont {Jordan}\ \emph {et~al.}(1998)\citenamefont {Jordan},
  \citenamefont {Kinderlehrer},\ and\ \citenamefont {Otto}}]{Jordan1998}%
  \BibitemOpen
  \bibfield  {author} {\bibinfo {author} {\bibfnamefont {R.}~\bibnamefont
  {Jordan}}, \bibinfo {author} {\bibfnamefont {D.}~\bibnamefont
  {Kinderlehrer}},\ and\ \bibinfo {author} {\bibfnamefont {F.}~\bibnamefont
  {Otto}},\ }\bibfield  {title} {\bibinfo {title} {The variational formulation
  of the {Fokker--Planck} equation},\ }\href@noop {} {\bibfield  {journal}
  {\bibinfo  {journal} {SIAM J. Math. Anal.}\ }\textbf {\bibinfo {volume}
  {29}},\ \bibinfo {pages} {1} (\bibinfo {year} {1998})}\BibitemShut {NoStop}%
\bibitem [{\citenamefont {Mielke}\ \emph {et~al.}(2025)\citenamefont {Mielke},
  \citenamefont {Peletier},\ and\ \citenamefont {Zimmer}}]{Mielke2025}%
  \BibitemOpen
  \bibfield  {author} {\bibinfo {author} {\bibfnamefont {A.}~\bibnamefont
  {Mielke}}, \bibinfo {author} {\bibfnamefont {M.~A.}\ \bibnamefont
  {Peletier}},\ and\ \bibinfo {author} {\bibfnamefont {J.}~\bibnamefont
  {Zimmer}},\ }\bibfield  {title} {\bibinfo {title} {Deriving a {GENERIC}
  system from a {Hamiltonian} system},\ }\href
  {https://doi.org/10.1007/s00205-025-02119-7} {\bibfield  {journal} {\bibinfo
  {journal} {Arch. Ration. Mech. Anal.}\ }\textbf {\bibinfo {volume} {249}},\
  \bibinfo {pages} {5} (\bibinfo {year} {2025})}\BibitemShut {NoStop}%
\bibitem [{\citenamefont {McClure}\ \emph {et~al.}(2018)\citenamefont
  {McClure}, \citenamefont {Armstrong}, \citenamefont {Berill}, \citenamefont
  {Schl{\"u}ter}, \citenamefont {Berg}, \citenamefont {Gray},\ and\
  \citenamefont {Miller}}]{McClure2018}%
  \BibitemOpen
  \bibfield  {author} {\bibinfo {author} {\bibfnamefont {J.~E.}\ \bibnamefont
  {McClure}}, \bibinfo {author} {\bibfnamefont {R.~T.}\ \bibnamefont
  {Armstrong}}, \bibinfo {author} {\bibfnamefont {M.~A.}\ \bibnamefont
  {Berill}}, \bibinfo {author} {\bibfnamefont {S.}~\bibnamefont
  {Schl{\"u}ter}}, \bibinfo {author} {\bibfnamefont {S.}~\bibnamefont {Berg}},
  \bibinfo {author} {\bibfnamefont {W.~G.}\ \bibnamefont {Gray}},\ and\
  \bibinfo {author} {\bibfnamefont {C.~T.}\ \bibnamefont {Miller}},\ }\bibfield
   {title} {\bibinfo {title} {Geometric state function for two-fluid flow in
  porous media},\ }\href@noop {} {\bibfield  {journal} {\bibinfo  {journal}
  {Phys. Rev. Fluids}\ }\textbf {\bibinfo {volume} {3}},\ \bibinfo {pages}
  {084306} (\bibinfo {year} {2018})}\BibitemShut {NoStop}%
\bibitem [{\citenamefont {Berry}(1984)}]{Berry1984}%
  \BibitemOpen
  \bibfield  {author} {\bibinfo {author} {\bibfnamefont {M.~V.}\ \bibnamefont
  {Berry}},\ }\bibfield  {title} {\bibinfo {title} {Quantal phase factors
  accompanying adiabatic changes},\ }\href@noop {} {\bibfield  {journal}
  {\bibinfo  {journal} {Proc. R. Soc. London A}\ }\textbf {\bibinfo {volume}
  {392}},\ \bibinfo {pages} {45} (\bibinfo {year} {1984})}\BibitemShut
  {NoStop}%
\bibitem [{\citenamefont {Shapere}\ and\ \citenamefont
  {Wilczek}(1989)}]{Shapere1989}%
  \BibitemOpen
  \bibfield  {author} {\bibinfo {author} {\bibfnamefont {A.}~\bibnamefont
  {Shapere}}\ and\ \bibinfo {author} {\bibfnamefont {F.}~\bibnamefont
  {Wilczek}},\ }\href@noop {} {\emph {\bibinfo {title} {Geometric Phases in
  Physics}}}\ (\bibinfo  {publisher} {World Scientific},\ \bibinfo {year}
  {1989})\BibitemShut {NoStop}%
\bibitem [{\citenamefont {Rigon}(2026{\natexlab{b}})}]{SupplementA}%
  \BibitemOpen
  \bibfield  {author} {\bibinfo {author} {\bibfnamefont {R.}~\bibnamefont
  {Rigon}},\ }\bibfield  {title} {\bibinfo {title} {{Supplemental Material A}:
  stationary-limit verification and numerical demonstration},\ }\href@noop {}
  {\bibfield  {journal} {\bibinfo  {journal} {Phys. Rev. E}\ } (\bibinfo {year}
  {2026}{\natexlab{b}})},\ \bibinfo {note} {supplemental Material}\BibitemShut
  {NoStop}%
\bibitem [{\citenamefont {Childs}\ and\ \citenamefont
  {Collis-George}(1950)}]{Childs1950}%
  \BibitemOpen
  \bibfield  {author} {\bibinfo {author} {\bibfnamefont {E.~C.}\ \bibnamefont
  {Childs}}\ and\ \bibinfo {author} {\bibfnamefont {N.}~\bibnamefont
  {Collis-George}},\ }\bibfield  {title} {\bibinfo {title} {The permeability of
  porous materials},\ }\href@noop {} {\bibfield  {journal} {\bibinfo  {journal}
  {Proc. R. Soc. London A}\ }\textbf {\bibinfo {volume} {201}},\ \bibinfo
  {pages} {392} (\bibinfo {year} {1950})}\BibitemShut {NoStop}%
\bibitem [{\citenamefont {Burdine}(1953)}]{Burdine1953}%
  \BibitemOpen
  \bibfield  {author} {\bibinfo {author} {\bibfnamefont {N.~T.}\ \bibnamefont
  {Burdine}},\ }\bibfield  {title} {\bibinfo {title} {Relative permeability
  calculations from pore size distribution data},\ }\href@noop {} {\bibfield
  {journal} {\bibinfo  {journal} {J. Pet. Technol.}\ }\textbf {\bibinfo
  {volume} {5}},\ \bibinfo {pages} {71} (\bibinfo {year} {1953})}\BibitemShut
  {NoStop}%
\bibitem [{\citenamefont {Ambegaokar}\ \emph {et~al.}(1971)\citenamefont
  {Ambegaokar}, \citenamefont {Halperin},\ and\ \citenamefont
  {Langer}}]{Ambegaokar1971}%
  \BibitemOpen
  \bibfield  {author} {\bibinfo {author} {\bibfnamefont {V.}~\bibnamefont
  {Ambegaokar}}, \bibinfo {author} {\bibfnamefont {B.~I.}\ \bibnamefont
  {Halperin}},\ and\ \bibinfo {author} {\bibfnamefont {J.~S.}\ \bibnamefont
  {Langer}},\ }\bibfield  {title} {\bibinfo {title} {Hopping conductivity in
  disordered systems},\ }\href@noop {} {\bibfield  {journal} {\bibinfo
  {journal} {Phys. Rev. B}\ }\textbf {\bibinfo {volume} {4}},\ \bibinfo {pages}
  {2612} (\bibinfo {year} {1971})}\BibitemShut {NoStop}%
\bibitem [{\citenamefont {Hunt}(2004)}]{Hunt2004}%
  \BibitemOpen
  \bibfield  {author} {\bibinfo {author} {\bibfnamefont {A.~G.}\ \bibnamefont
  {Hunt}},\ }\bibfield  {title} {\bibinfo {title} {Continuum percolation theory
  and {Archie's} law},\ }\href@noop {} {\bibfield  {journal} {\bibinfo
  {journal} {Geophys. Res. Lett.}\ }\textbf {\bibinfo {volume} {31}},\ \bibinfo
  {pages} {L21503} (\bibinfo {year} {2004})}\BibitemShut {NoStop}%
\bibitem [{\citenamefont {Berkowitz}\ and\ \citenamefont
  {Balberg}(1993)}]{Berkowitz1993}%
  \BibitemOpen
  \bibfield  {author} {\bibinfo {author} {\bibfnamefont {B.}~\bibnamefont
  {Berkowitz}}\ and\ \bibinfo {author} {\bibfnamefont {I.}~\bibnamefont
  {Balberg}},\ }\bibfield  {title} {\bibinfo {title} {Percolation theory and
  its application to groundwater hydrology},\ }\href@noop {} {\bibfield
  {journal} {\bibinfo  {journal} {Water Resour. Res.}\ }\textbf {\bibinfo
  {volume} {29}},\ \bibinfo {pages} {775} (\bibinfo {year} {1993})}\BibitemShut
  {NoStop}%
\bibitem [{\citenamefont {Hassanizadeh}\ \emph {et~al.}(2002)\citenamefont
  {Hassanizadeh}, \citenamefont {Celia},\ and\ \citenamefont
  {Dahle}}]{Hassanizadeh2002}%
  \BibitemOpen
  \bibfield  {author} {\bibinfo {author} {\bibfnamefont {S.~M.}\ \bibnamefont
  {Hassanizadeh}}, \bibinfo {author} {\bibfnamefont {M.~A.}\ \bibnamefont
  {Celia}},\ and\ \bibinfo {author} {\bibfnamefont {H.~K.}\ \bibnamefont
  {Dahle}},\ }\bibfield  {title} {\bibinfo {title} {Dynamic effect in the
  capillary pressure--saturation relationship and its impacts on unsaturated
  flow},\ }\href@noop {} {\bibfield  {journal} {\bibinfo  {journal} {Vadose
  Zone J.}\ }\textbf {\bibinfo {volume} {1}},\ \bibinfo {pages} {38} (\bibinfo
  {year} {2002})}\BibitemShut {NoStop}%
\bibitem [{\citenamefont {{Lord Rayleigh}}(1873)}]{Rayleigh1873}%
  \BibitemOpen
  \bibfield  {author} {\bibinfo {author} {\bibnamefont {{Lord Rayleigh}}},\
  }\bibfield  {title} {\bibinfo {title} {Some general theorems relating to
  vibrations},\ }\href@noop {} {\bibfield  {journal} {\bibinfo  {journal}
  {Proc. London Math. Soc.}\ }\textbf {\bibinfo {volume} {4}},\ \bibinfo
  {pages} {357} (\bibinfo {year} {1873})}\BibitemShut {NoStop}%
\bibitem [{\citenamefont {Otto}(2001)}]{Otto2001}%
  \BibitemOpen
  \bibfield  {author} {\bibinfo {author} {\bibfnamefont {F.}~\bibnamefont
  {Otto}},\ }\bibfield  {title} {\bibinfo {title} {The geometry of dissipative
  evolution equations: the porous medium equation},\ }\href@noop {} {\bibfield
  {journal} {\bibinfo  {journal} {Commun. Partial Differ. Equ.}\ }\textbf
  {\bibinfo {volume} {26}},\ \bibinfo {pages} {101} (\bibinfo {year}
  {2001})}\BibitemShut {NoStop}%
\end{thebibliography}%


\begin{thebibliography}{35}%
\makeatletter
\providecommand \@ifxundefined [1]{%
 \@ifx{#1\undefined}
}%
\providecommand \@ifnum [1]{%
 \ifnum #1\expandafter \@firstoftwo
 \else \expandafter \@secondoftwo
 \fi
}%
\providecommand \@ifx [1]{%
 \ifx #1\expandafter \@firstoftwo
 \else \expandafter \@secondoftwo
 \fi
}%
\providecommand \natexlab [1]{#1}%
\providecommand \enquote  [1]{``#1''}%
\providecommand \bibnamefont  [1]{#1}%
\providecommand \bibfnamefont [1]{#1}%
\providecommand \citenamefont [1]{#1}%
\providecommand \href@noop [0]{\@secondoftwo}%
\providecommand \href [0]{\begingroup \@sanitize@url \@href}%
\providecommand \@href[1]{\@@startlink{#1}\@@href}%
\providecommand \@@href[1]{\endgroup#1\@@endlink}%
\providecommand \@sanitize@url [0]{\catcode `\\12\catcode `\$12\catcode
  `\&12\catcode `\#12\catcode `\^12\catcode `\_12\catcode `\%12\relax}%
\providecommand \@@startlink[1]{}%
\providecommand \@@endlink[0]{}%
\providecommand \url  [0]{\begingroup\@sanitize@url \@url }%
\providecommand \@url [1]{\endgroup\@href {#1}{\urlprefix }}%
\providecommand \urlprefix  [0]{URL }%
\providecommand \Eprint [0]{\href }%
\providecommand \doibase [0]{https://doi.org/}%
\providecommand \selectlanguage [0]{\@gobble}%
\providecommand \bibinfo  [0]{\@secondoftwo}%
\providecommand \bibfield  [0]{\@secondoftwo}%
\providecommand \translation [1]{[#1]}%
\providecommand \BibitemOpen [0]{}%
\providecommand \bibitemStop [0]{}%
\providecommand \bibitemNoStop [0]{.\EOS\space}%
\providecommand \EOS [0]{\spacefactor3000\relax}%
\providecommand \BibitemShut  [1]{\csname bibitem#1\endcsname}%
\let\auto@bib@innerbib\@empty
\bibitem [{\citenamefont {Zehe}\ and\ \citenamefont
  {Jackisch}(2016)}]{ZeheJackisch2016}%
  \BibitemOpen
  \bibfield  {author} {\bibinfo {author} {\bibfnamefont {E.}~\bibnamefont
  {Zehe}}\ and\ \bibinfo {author} {\bibfnamefont {C.}~\bibnamefont
  {Jackisch}},\ }\bibfield  {title} {\bibinfo {title} {A {Lagrangian} model for
  soil water dynamics during rainfall-driven conditions},\ }\href
  {https://doi.org/10.5194/hess-20-3511-2016} {\bibfield  {journal} {\bibinfo
  {journal} {Hydrology and Earth System Sciences}\ }\textbf {\bibinfo {volume}
  {20}},\ \bibinfo {pages} {3511} (\bibinfo {year} {2016})}\BibitemShut
  {NoStop}%
\bibitem [{\citenamefont {Dall'Amico}\ \emph {et~al.}(2011)\citenamefont
  {Dall'Amico}, \citenamefont {Endrizzi}, \citenamefont {Gruber},\ and\
  \citenamefont {Rigon}}]{DallAmico2011}%
  \BibitemOpen
  \bibfield  {author} {\bibinfo {author} {\bibfnamefont {M.}~\bibnamefont
  {Dall'Amico}}, \bibinfo {author} {\bibfnamefont {S.}~\bibnamefont
  {Endrizzi}}, \bibinfo {author} {\bibfnamefont {S.}~\bibnamefont {Gruber}},\
  and\ \bibinfo {author} {\bibfnamefont {R.}~\bibnamefont {Rigon}},\ }\bibfield
   {title} {\bibinfo {title} {A robust and energy-conserving model of freezing
  variably-saturated soil},\ }\href {https://doi.org/10.5194/tc-5-469-2011}
  {\bibfield  {journal} {\bibinfo  {journal} {The Cryosphere}\ }\textbf
  {\bibinfo {volume} {5}},\ \bibinfo {pages} {469} (\bibinfo {year}
  {2011})}\BibitemShut {NoStop}%
\bibitem [{\citenamefont {D'Amato}\ and\ \citenamefont
  {Rigon}(2025)}]{DAmatoRigon2025}%
  \BibitemOpen
  \bibfield  {author} {\bibinfo {author} {\bibfnamefont {C.}~\bibnamefont
  {D'Amato}}\ and\ \bibinfo {author} {\bibfnamefont {R.}~\bibnamefont
  {Rigon}},\ }\bibfield  {title} {\bibinfo {title} {Elementary mathematics
  helps to shed light on the transpiration budget under water stress},\ }\href
  {https://doi.org/10.1002/eco.70009} {\bibfield  {journal} {\bibinfo
  {journal} {Ecohydrology}\ }\textbf {\bibinfo {volume} {18}},\ \bibinfo
  {pages} {e70009} (\bibinfo {year} {2025})}\BibitemShut {NoStop}%
\bibitem [{\citenamefont {Lucas}(1918)}]{Lucas1918}%
  \BibitemOpen
  \bibfield  {author} {\bibinfo {author} {\bibfnamefont {R.}~\bibnamefont
  {Lucas}},\ }\bibfield  {title} {\bibinfo {title} {Ueber das zeitgesetz des
  kapillaren aufstiegs von fl\"ussigkeiten},\ }\href
  {https://doi.org/10.1007/BF01461107} {\bibfield  {journal} {\bibinfo
  {journal} {Kolloid-Zeitschrift}\ }\textbf {\bibinfo {volume} {23}},\ \bibinfo
  {pages} {15} (\bibinfo {year} {1918})}\BibitemShut {NoStop}%
\bibitem [{\citenamefont {Washburn}(1921)}]{Washburn1921}%
  \BibitemOpen
  \bibfield  {author} {\bibinfo {author} {\bibfnamefont {E.~W.}\ \bibnamefont
  {Washburn}},\ }\bibfield  {title} {\bibinfo {title} {The dynamics of
  capillary flow},\ }\href@noop {} {\bibfield  {journal} {\bibinfo  {journal}
  {Phys. Rev.}\ }\textbf {\bibinfo {volume} {17}},\ \bibinfo {pages} {273}
  (\bibinfo {year} {1921})}\BibitemShut {NoStop}%
\bibitem [{\citenamefont {Tyler}\ and\ \citenamefont
  {Wheatcraft}(1990)}]{TylerWheatcraft1990}%
  \BibitemOpen
  \bibfield  {author} {\bibinfo {author} {\bibfnamefont {S.~W.}\ \bibnamefont
  {Tyler}}\ and\ \bibinfo {author} {\bibfnamefont {S.~W.}\ \bibnamefont
  {Wheatcraft}},\ }\bibfield  {title} {\bibinfo {title} {Fractal processes in
  soil water retention},\ }\href {https://doi.org/10.1029/WR026i005p01047}
  {\bibfield  {journal} {\bibinfo  {journal} {Water Resources Research}\
  }\textbf {\bibinfo {volume} {26}},\ \bibinfo {pages} {1047} (\bibinfo {year}
  {1990})}\BibitemShut {NoStop}%
\bibitem [{\citenamefont {Rieu}\ and\ \citenamefont
  {Sposito}(1991)}]{RieuSposito1991}%
  \BibitemOpen
  \bibfield  {author} {\bibinfo {author} {\bibfnamefont {M.}~\bibnamefont
  {Rieu}}\ and\ \bibinfo {author} {\bibfnamefont {G.}~\bibnamefont {Sposito}},\
  }\bibfield  {title} {\bibinfo {title} {Fractal fragmentation, soil porosity,
  and soil water properties: I. theory},\ }\href
  {https://doi.org/10.2136/sssaj1991.03615995005500050006x} {\bibfield
  {journal} {\bibinfo  {journal} {Soil Science Society of America Journal}\
  }\textbf {\bibinfo {volume} {55}},\ \bibinfo {pages} {1231} (\bibinfo {year}
  {1991})}\BibitemShut {NoStop}%
\bibitem [{\citenamefont {Romano}\ and\ \citenamefont
  {Santini}(2002)}]{RomanoSantini2002}%
  \BibitemOpen
  \bibfield  {author} {\bibinfo {author} {\bibfnamefont {N.}~\bibnamefont
  {Romano}}\ and\ \bibinfo {author} {\bibfnamefont {A.}~\bibnamefont
  {Santini}},\ }\bibfield  {title} {\bibinfo {title} {Water retention and
  storage: Field},\ }in\ \href {https://doi.org/10.2136/sssabookser5.4.c26}
  {\emph {\bibinfo {booktitle} {Methods of Soil Analysis, Part 4: Physical
  Methods}}},\ \bibinfo {series and number} {SSSA Book Series, No. 5},\
  \bibinfo {editor} {edited by\ \bibinfo {editor} {\bibfnamefont {J.~H.}\
  \bibnamefont {Dane}}\ and\ \bibinfo {editor} {\bibfnamefont {G.~C.}\
  \bibnamefont {Topp}}}\ (\bibinfo  {publisher} {Soil Science Society of
  America},\ \bibinfo {address} {Madison, WI},\ \bibinfo {year} {2002})\ pp.\
  \bibinfo {pages} {721--738}\BibitemShut {NoStop}%
\bibitem [{\citenamefont {Rigon}(2026{\natexlab{a}})}]{Rigon2026CE}%
  \BibitemOpen
  \bibfield  {author} {\bibinfo {author} {\bibfnamefont {R.}~\bibnamefont
  {Rigon}},\ }\href@noop {} {\bibinfo {title} {Richards' equation as a
  hydrodynamic limit: {Chapman--Enskog} reduction of the continuum kinetic
  equation for unsaturated soil water}} (\bibinfo {year}
  {2026}{\natexlab{a}}),\ \bibinfo {note} {companion paper (PRE-2)},\ \Eprint
  {https://arxiv.org/abs/2607.17358} {arXiv:2607.17358 [cond-mat.stat-mech]}
  \BibitemShut {NoStop}%
\bibitem [{\citenamefont {Onsager}(1931{\natexlab{a}})}]{Onsager1931a}%
  \BibitemOpen
  \bibfield  {author} {\bibinfo {author} {\bibfnamefont {L.}~\bibnamefont
  {Onsager}},\ }\bibfield  {title} {\bibinfo {title} {Reciprocal relations in
  irreversible processes. {I}},\ }\href@noop {} {\bibfield  {journal} {\bibinfo
   {journal} {Phys. Rev.}\ }\textbf {\bibinfo {volume} {37}},\ \bibinfo {pages}
  {405} (\bibinfo {year} {1931}{\natexlab{a}})}\BibitemShut {NoStop}%
\bibitem [{\citenamefont {Onsager}(1931{\natexlab{b}})}]{Onsager1931b}%
  \BibitemOpen
  \bibfield  {author} {\bibinfo {author} {\bibfnamefont {L.}~\bibnamefont
  {Onsager}},\ }\bibfield  {title} {\bibinfo {title} {Reciprocal relations in
  irreversible processes. {II}},\ }\href@noop {} {\bibfield  {journal}
  {\bibinfo  {journal} {Phys. Rev.}\ }\textbf {\bibinfo {volume} {38}},\
  \bibinfo {pages} {2265} (\bibinfo {year} {1931}{\natexlab{b}})}\BibitemShut
  {NoStop}%
\bibitem [{\citenamefont {Haines}(1930)}]{Haines1930}%
  \BibitemOpen
  \bibfield  {author} {\bibinfo {author} {\bibfnamefont {W.~B.}\ \bibnamefont
  {Haines}},\ }\bibfield  {title} {\bibinfo {title} {Studies in the physical
  properties of soil. {V}. the hysteresis effect in capillary properties, and
  the modes of moisture distribution associated therewith},\ }\href@noop {}
  {\bibfield  {journal} {\bibinfo  {journal} {J. Agric. Sci.}\ }\textbf
  {\bibinfo {volume} {20}},\ \bibinfo {pages} {97} (\bibinfo {year}
  {1930})}\BibitemShut {NoStop}%
\bibitem [{\citenamefont {Brooks}\ and\ \citenamefont
  {Corey}(1964)}]{Brooks1964}%
  \BibitemOpen
  \bibfield  {author} {\bibinfo {author} {\bibfnamefont {R.~H.}\ \bibnamefont
  {Brooks}}\ and\ \bibinfo {author} {\bibfnamefont {A.~T.}\ \bibnamefont
  {Corey}},\ }\href@noop {} {\emph {\bibinfo {title} {Hydraulic properties of
  porous media}}},\ \bibinfo {type} {Tech. Rep.}\ \bibinfo {number} {Hydrology
  Paper 3}\ (\bibinfo  {institution} {Colorado State Univ.},\ \bibinfo {year}
  {1964})\BibitemShut {NoStop}%
\bibitem [{\citenamefont {Rawls}\ \emph {et~al.}(1982)\citenamefont {Rawls},
  \citenamefont {Brakensiek},\ and\ \citenamefont {Saxton}}]{Rawls1982}%
  \BibitemOpen
  \bibfield  {author} {\bibinfo {author} {\bibfnamefont {W.~J.}\ \bibnamefont
  {Rawls}}, \bibinfo {author} {\bibfnamefont {D.~L.}\ \bibnamefont
  {Brakensiek}},\ and\ \bibinfo {author} {\bibfnamefont {K.~E.}\ \bibnamefont
  {Saxton}},\ }\href@noop {} {\emph {\bibinfo {title} {Estimation of soil water
  properties}}},\ \bibinfo {type} {Tech. Rep.}\ (\bibinfo  {institution} {Am.
  Soc. Agric. Eng.},\ \bibinfo {year} {1982})\BibitemShut {NoStop}%
\bibitem [{\citenamefont {Kosugi}(1996)}]{Kosugi1996}%
  \BibitemOpen
  \bibfield  {author} {\bibinfo {author} {\bibfnamefont {K.}~\bibnamefont
  {Kosugi}},\ }\bibfield  {title} {\bibinfo {title} {Lognormal distribution
  model for unsaturated soil hydraulic properties},\ }\href@noop {} {\bibfield
  {journal} {\bibinfo  {journal} {Water Resour. Res.}\ }\textbf {\bibinfo
  {volume} {32}},\ \bibinfo {pages} {2697} (\bibinfo {year}
  {1996})}\BibitemShut {NoStop}%
\bibitem [{\citenamefont {{van Genuchten}}(1980)}]{vanGenuchten1980}%
  \BibitemOpen
  \bibfield  {author} {\bibinfo {author} {\bibfnamefont {M.~T.}\ \bibnamefont
  {{van Genuchten}}},\ }\bibfield  {title} {\bibinfo {title} {A closed-form
  equation for predicting the hydraulic conductivity of unsaturated soils},\
  }\href@noop {} {\bibfield  {journal} {\bibinfo  {journal} {Soil Sci. Soc. Am.
  J.}\ }\textbf {\bibinfo {volume} {44}},\ \bibinfo {pages} {892} (\bibinfo
  {year} {1980})}\BibitemShut {NoStop}%
\bibitem [{\citenamefont {Tuller}\ \emph {et~al.}(1999)\citenamefont {Tuller},
  \citenamefont {Or},\ and\ \citenamefont {Dudley}}]{Tuller1999}%
  \BibitemOpen
  \bibfield  {author} {\bibinfo {author} {\bibfnamefont {M.}~\bibnamefont
  {Tuller}}, \bibinfo {author} {\bibfnamefont {D.}~\bibnamefont {Or}},\ and\
  \bibinfo {author} {\bibfnamefont {L.~M.}\ \bibnamefont {Dudley}},\ }\bibfield
   {title} {\bibinfo {title} {Adsorption and capillary condensation in porous
  media: {Liquid} retention and interfacial configurations in angular pores},\
  }\href@noop {} {\bibfield  {journal} {\bibinfo  {journal} {Water Resour.
  Res.}\ }\textbf {\bibinfo {volume} {35}},\ \bibinfo {pages} {1949} (\bibinfo
  {year} {1999})}\BibitemShut {NoStop}%
\bibitem [{\citenamefont {Iwamatsu}\ and\ \citenamefont
  {Horii}(1996)}]{Iwamatsu1996}%
  \BibitemOpen
  \bibfield  {author} {\bibinfo {author} {\bibfnamefont {M.}~\bibnamefont
  {Iwamatsu}}\ and\ \bibinfo {author} {\bibfnamefont {K.}~\bibnamefont
  {Horii}},\ }\bibfield  {title} {\bibinfo {title} {Capillary condensation and
  adhesion of two wetter surfaces},\ }\href@noop {} {\bibfield  {journal}
  {\bibinfo  {journal} {J. Colloid Interface Sci.}\ }\textbf {\bibinfo {volume}
  {182}},\ \bibinfo {pages} {400} (\bibinfo {year} {1996})}\BibitemShut
  {NoStop}%
\bibitem [{\citenamefont {Or}\ and\ \citenamefont
  {Tuller}(2000)}]{OrTuller2000}%
  \BibitemOpen
  \bibfield  {author} {\bibinfo {author} {\bibfnamefont {D.}~\bibnamefont
  {Or}}\ and\ \bibinfo {author} {\bibfnamefont {M.}~\bibnamefont {Tuller}},\
  }\bibfield  {title} {\bibinfo {title} {Flow in unsaturated fractured porous
  media: Hydraulic conductivity of rough surfaces},\ }\href@noop {} {\bibfield
  {journal} {\bibinfo  {journal} {Water Resour. Res.}\ }\textbf {\bibinfo
  {volume} {36}},\ \bibinfo {pages} {1165} (\bibinfo {year}
  {2000})}\BibitemShut {NoStop}%
\bibitem [{\citenamefont {Childs}(1940)}]{Childs1940}%
  \BibitemOpen
  \bibfield  {author} {\bibinfo {author} {\bibfnamefont {E.~C.}\ \bibnamefont
  {Childs}},\ }\bibfield  {title} {\bibinfo {title} {The use of soil moisture
  characteristics in soil studies},\ }\href@noop {} {\bibfield  {journal}
  {\bibinfo  {journal} {Soil Sci.}\ }\textbf {\bibinfo {volume} {50}},\
  \bibinfo {pages} {239} (\bibinfo {year} {1940})}\BibitemShut {NoStop}%
\bibitem [{\citenamefont {Rigon}(2026{\natexlab{b}})}]{Rigon2026PRE}%
  \BibitemOpen
  \bibfield  {author} {\bibinfo {author} {\bibfnamefont {R.}~\bibnamefont
  {Rigon}},\ }\href@noop {} {\bibinfo {title} {The statistical physics of
  unsaturated soil water: kinetic theory and non-commutative pore water
  dynamics}} (\bibinfo {year} {2026}{\natexlab{b}}),\ \bibinfo {note}
  {companion paper (PRE-1), arXiv preprint},\ \Eprint
  {https://arxiv.org/abs/2607.09416} {arXiv:2607.09416 [physics.geo-ph]}
  \BibitemShut {NoStop}%
\bibitem [{\citenamefont {Gostick}\ \emph {et~al.}(2016)\citenamefont {Gostick}
  \emph {et~al.}}]{Gostick2016}%
  \BibitemOpen
  \bibfield  {author} {\bibinfo {author} {\bibfnamefont {J.}~\bibnamefont
  {Gostick}} \emph {et~al.},\ }\bibfield  {title} {\bibinfo {title} {{OpenPNM}:
  A pore network modeling package},\ }\href@noop {} {\bibfield  {journal}
  {\bibinfo  {journal} {Comput. Sci. Eng.}\ }\textbf {\bibinfo {volume} {18}},\
  \bibinfo {pages} {60} (\bibinfo {year} {2016})}\BibitemShut {NoStop}%
\bibitem [{\citenamefont {Beven}\ and\ \citenamefont
  {Germann}(1982)}]{Beven1982}%
  \BibitemOpen
  \bibfield  {author} {\bibinfo {author} {\bibfnamefont {K.}~\bibnamefont
  {Beven}}\ and\ \bibinfo {author} {\bibfnamefont {P.}~\bibnamefont
  {Germann}},\ }\bibfield  {title} {\bibinfo {title} {Macropores and water flow
  in soils},\ }\href@noop {} {\bibfield  {journal} {\bibinfo  {journal} {Water
  Resour. Res.}\ }\textbf {\bibinfo {volume} {18}},\ \bibinfo {pages} {1311}
  (\bibinfo {year} {1982})}\BibitemShut {NoStop}%
\bibitem [{\citenamefont {Weiler}\ and\ \citenamefont
  {Naef}(2003)}]{Weiler2003}%
  \BibitemOpen
  \bibfield  {author} {\bibinfo {author} {\bibfnamefont {M.}~\bibnamefont
  {Weiler}}\ and\ \bibinfo {author} {\bibfnamefont {F.}~\bibnamefont {Naef}},\
  }\bibfield  {title} {\bibinfo {title} {Simulating surface and subsurface
  initiation of macropore flow},\ }\href@noop {} {\bibfield  {journal}
  {\bibinfo  {journal} {J. Hydrol.}\ }\textbf {\bibinfo {volume} {273}},\
  \bibinfo {pages} {139} (\bibinfo {year} {2003})}\BibitemShut {NoStop}%
\bibitem [{\citenamefont {Sammartino}\ \emph {et~al.}(2015)\citenamefont
  {Sammartino}, \citenamefont {Lissy}, \citenamefont {Bogner}, \citenamefont
  {Van Den~Bogaert}, \citenamefont {Capowiez}, \citenamefont {Ruy},\ and\
  \citenamefont {Cornu}}]{Sammartino2015}%
  \BibitemOpen
  \bibfield  {author} {\bibinfo {author} {\bibfnamefont {S.}~\bibnamefont
  {Sammartino}}, \bibinfo {author} {\bibfnamefont {A.-S.}\ \bibnamefont
  {Lissy}}, \bibinfo {author} {\bibfnamefont {C.}~\bibnamefont {Bogner}},
  \bibinfo {author} {\bibfnamefont {R.}~\bibnamefont {Van Den~Bogaert}},
  \bibinfo {author} {\bibfnamefont {Y.}~\bibnamefont {Capowiez}}, \bibinfo
  {author} {\bibfnamefont {S.}~\bibnamefont {Ruy}},\ and\ \bibinfo {author}
  {\bibfnamefont {S.}~\bibnamefont {Cornu}},\ }\bibfield  {title} {\bibinfo
  {title} {Identifying the functional macropore network related to preferential
  flow in structured soils},\ }\href {https://doi.org/10.2136/vzj2015.05.0070}
  {\bibfield  {journal} {\bibinfo  {journal} {Vadose Zone J.}\ }\textbf
  {\bibinfo {volume} {14}},\ \bibinfo {pages} {vzj2015.05.0070} (\bibinfo
  {year} {2015})}\BibitemShut {NoStop}%
\bibitem [{\citenamefont {Mualem}(1976)}]{Mualem1976}%
  \BibitemOpen
  \bibfield  {author} {\bibinfo {author} {\bibfnamefont {Y.}~\bibnamefont
  {Mualem}},\ }\bibfield  {title} {\bibinfo {title} {A new model for predicting
  the hydraulic conductivity of unsaturated porous media},\ }\href@noop {}
  {\bibfield  {journal} {\bibinfo  {journal} {Water Resour. Res.}\ }\textbf
  {\bibinfo {volume} {12}},\ \bibinfo {pages} {513} (\bibinfo {year}
  {1976})}\BibitemShut {NoStop}%
\bibitem [{\citenamefont {Kool}\ and\ \citenamefont
  {Parker}(1987)}]{Kool1987-le}%
  \BibitemOpen
  \bibfield  {author} {\bibinfo {author} {\bibfnamefont {J.~B.}\ \bibnamefont
  {Kool}}\ and\ \bibinfo {author} {\bibfnamefont {J.~C.}\ \bibnamefont
  {Parker}},\ }\bibfield  {title} {\bibinfo {title} {Development and evaluation
  of closed-form expressions for hysteretic soil hydraulic properties},\
  }\href@noop {} {\bibfield  {journal} {\bibinfo  {journal} {Water Resour.
  Res.}\ }\textbf {\bibinfo {volume} {23}},\ \bibinfo {pages} {105} (\bibinfo
  {year} {1987})}\BibitemShut {NoStop}%
\bibitem [{\citenamefont {Shi}\ \emph {et~al.}(2026)\citenamefont {Shi},
  \citenamefont {Shen}, \citenamefont {Liu}, \citenamefont {Guo}, \citenamefont
  {Du},\ and\ \citenamefont {Zhangzhong}}]{Shi2026}%
  \BibitemOpen
  \bibfield  {author} {\bibinfo {author} {\bibfnamefont {M.}~\bibnamefont
  {Shi}}, \bibinfo {author} {\bibfnamefont {X.}~\bibnamefont {Shen}}, \bibinfo
  {author} {\bibfnamefont {C.}~\bibnamefont {Liu}}, \bibinfo {author}
  {\bibfnamefont {W.}~\bibnamefont {Guo}}, \bibinfo {author} {\bibfnamefont
  {Y.}~\bibnamefont {Du}},\ and\ \bibinfo {author} {\bibfnamefont
  {L.}~\bibnamefont {Zhangzhong}},\ }\bibfield  {title} {\bibinfo {title}
  {Long-term in-situ field monitoring-based performance evaluation and
  environmental factor impact mechanism analysis of {FDR} soil moisture
  sensors},\ }\href {https://doi.org/10.1016/j.atech.2026.102257} {\bibfield
  {journal} {\bibinfo  {journal} {Smart Agric. Technol.}\ }\textbf {\bibinfo
  {volume} {14}},\ \bibinfo {pages} {102257} (\bibinfo {year}
  {2026})}\BibitemShut {NoStop}%
\bibitem [{\citenamefont {Hassanizadeh}\ \emph {et~al.}(2002)\citenamefont
  {Hassanizadeh}, \citenamefont {Celia},\ and\ \citenamefont
  {Dahle}}]{Hassanizadeh2002}%
  \BibitemOpen
  \bibfield  {author} {\bibinfo {author} {\bibfnamefont {S.~M.}\ \bibnamefont
  {Hassanizadeh}}, \bibinfo {author} {\bibfnamefont {M.~A.}\ \bibnamefont
  {Celia}},\ and\ \bibinfo {author} {\bibfnamefont {H.~K.}\ \bibnamefont
  {Dahle}},\ }\bibfield  {title} {\bibinfo {title} {Dynamic effect in the
  capillary pressure--saturation relationship and its impacts on unsaturated
  flow},\ }\href@noop {} {\bibfield  {journal} {\bibinfo  {journal} {Vadose
  Zone J.}\ }\textbf {\bibinfo {volume} {1}},\ \bibinfo {pages} {38} (\bibinfo
  {year} {2002})}\BibitemShut {NoStop}%
\bibitem [{\citenamefont {Cey}\ and\ \citenamefont
  {Rudolph}(2009)}]{CeyRudolph2009}%
  \BibitemOpen
  \bibfield  {author} {\bibinfo {author} {\bibfnamefont {E.~E.}\ \bibnamefont
  {Cey}}\ and\ \bibinfo {author} {\bibfnamefont {D.~L.}\ \bibnamefont
  {Rudolph}},\ }\bibfield  {title} {\bibinfo {title} {Field study of macropore
  flow processes using tension infiltration of a dye tracer in partially
  saturated soils},\ }\href@noop {} {\bibfield  {journal} {\bibinfo  {journal}
  {Hydrol. Process.}\ }\textbf {\bibinfo {volume} {23}},\ \bibinfo {pages}
  {1768} (\bibinfo {year} {2009})}\BibitemShut {NoStop}%
\bibitem [{\citenamefont {van Schaik}\ \emph {et~al.}(2010)\citenamefont {van
  Schaik}, \citenamefont {Hendriks},\ and\ \citenamefont {van
  Dam}}]{vanSchaik2010}%
  \BibitemOpen
  \bibfield  {author} {\bibinfo {author} {\bibfnamefont {N.~L. M.~B.}\
  \bibnamefont {van Schaik}}, \bibinfo {author} {\bibfnamefont {R.~F.~A.}\
  \bibnamefont {Hendriks}},\ and\ \bibinfo {author} {\bibfnamefont {J.~C.}\
  \bibnamefont {van Dam}},\ }\bibfield  {title} {\bibinfo {title}
  {Parameterization of macropore flow using dye-tracer infiltration patterns in
  the {SWAP} model},\ }\href {https://doi.org/10.2136/vzj2009.0031} {\bibfield
  {journal} {\bibinfo  {journal} {Vadose Zone J.}\ }\textbf {\bibinfo {volume}
  {9}},\ \bibinfo {pages} {95} (\bibinfo {year} {2010})}\BibitemShut {NoStop}%
\bibitem [{\citenamefont {Wildenschild}\ and\ \citenamefont
  {Sheppard}(2013)}]{Wildenschild2013}%
  \BibitemOpen
  \bibfield  {author} {\bibinfo {author} {\bibfnamefont {D.}~\bibnamefont
  {Wildenschild}}\ and\ \bibinfo {author} {\bibfnamefont {A.~P.}\ \bibnamefont
  {Sheppard}},\ }\bibfield  {title} {\bibinfo {title} {X-ray imaging and
  analysis techniques for quantifying pore-scale structure and processes in
  subsurface porous medium systems},\ }\href@noop {} {\bibfield  {journal}
  {\bibinfo  {journal} {Adv. Water Resour.}\ }\textbf {\bibinfo {volume}
  {51}},\ \bibinfo {pages} {217} (\bibinfo {year} {2013})}\BibitemShut
  {NoStop}%
\bibitem [{\citenamefont {Berg}\ \emph {et~al.}(2013)\citenamefont {Berg} \emph
  {et~al.}}]{Berg2014}%
  \BibitemOpen
  \bibfield  {author} {\bibinfo {author} {\bibfnamefont {S.}~\bibnamefont
  {Berg}} \emph {et~al.},\ }\bibfield  {title} {\bibinfo {title} {Real-time 3d
  imaging of {Haines} jumps in porous media flow},\ }\href@noop {} {\bibfield
  {journal} {\bibinfo  {journal} {Proc. Natl Acad. Sci. USA}\ }\textbf
  {\bibinfo {volume} {110}},\ \bibinfo {pages} {3755} (\bibinfo {year}
  {2013})}\BibitemShut {NoStop}%
\bibitem [{\citenamefont {Hunt}(2004)}]{Hunt2004}%
  \BibitemOpen
  \bibfield  {author} {\bibinfo {author} {\bibfnamefont {A.~G.}\ \bibnamefont
  {Hunt}},\ }\bibfield  {title} {\bibinfo {title} {Continuum percolation theory
  and {Archie's} law},\ }\href@noop {} {\bibfield  {journal} {\bibinfo
  {journal} {Geophys. Res. Lett.}\ }\textbf {\bibinfo {volume} {31}},\ \bibinfo
  {pages} {L21503} (\bibinfo {year} {2004})}\BibitemShut {NoStop}%
\bibitem [{\citenamefont {Kr{\"o}ner}\ \emph {et~al.}(2014)\citenamefont
  {Kr{\"o}ner}, \citenamefont {Zarebanadkouki}, \citenamefont {Kaestner},\ and\
  \citenamefont {Carminati}}]{Kroener2015}%
  \BibitemOpen
  \bibfield  {author} {\bibinfo {author} {\bibfnamefont {E.}~\bibnamefont
  {Kr{\"o}ner}}, \bibinfo {author} {\bibfnamefont {M.}~\bibnamefont
  {Zarebanadkouki}}, \bibinfo {author} {\bibfnamefont {A.}~\bibnamefont
  {Kaestner}},\ and\ \bibinfo {author} {\bibfnamefont {A.}~\bibnamefont
  {Carminati}},\ }\bibfield  {title} {\bibinfo {title} {Nonequilibrium water
  dynamics in the rhizosphere: How mucilage affects water flow in soils},\
  }\href@noop {} {\bibfield  {journal} {\bibinfo  {journal} {Water Resour.
  Res.}\ }\textbf {\bibinfo {volume} {50}},\ \bibinfo {pages} {6479} (\bibinfo
  {year} {2014})}\BibitemShut {NoStop}%
\end{thebibliography}%

\appendix

\section{Symbols and acronyms}
\label{app:symbols}
Table~\ref{tab:symbols} collects the principal symbols and their units.

\begin{table}[h]
\caption{Principal symbols and acronyms.  Dimensionless quantities are marked ``--''.}
\label{tab:symbols}
\begin{ruledtabular}
\setlength{\tabcolsep}{4pt}
\small
\begin{tabular}{@{}lll@{}}
\textbf{Symbol} & \textbf{Meaning} & \textbf{Units} \\
\hline
$g(r,\mathbf{x},t)$ & pore-occupancy (filling fraction) & -- \\
$r,\,r'$ & pore radius (receiver, donor) & m \\
$f(r)$ & pore-size distribution & m$^{-1}$ \\
$C(r,r')$ & connectivity matrix & -- \\
$\Phi(r,r',\hat{\bm e})$ & driving potential (head) & m \\
$\kappa(r,g)$ & inter-pore rate constant & Pa$^{-1}$s$^{-1}$ \\
$\mu_w(r)$ & water chemical potential & J\,m$^{-3}$ \\
$\mathcal{F}[g]$ & Gibbs free energy & J\,m$^{-3}$ \\
$g_{\rm eq}$ & equilibrium occupancy & -- \\
$r^*$ & Young--Laplace critical radius & m \\
$\bar L$ & mean pore/edge length & m \\
$\bar\tau$ & tortuosity & -- \\
$\Xi(r)$ & geometric conductance correction & -- \\
$\mathbf{F}$ & macroscopic flux (per class) & m\,s$^{-1}$ \\
$\Coll[g]$ & redistribution operator & s$^{-1}$ \\
$\mathcal{E},\mathcal{T}$ & evaporation, root-uptake sinks & s$^{-1}$ \\
$\mathsf{Q}_{\rm in},\mathsf{Q}_{\rm out}$ & inward/outward boundary currents & s$^{-1}$ \\
$\mathcal{G}_\partial,\Lg_\partial$ & boundary gain/loss operators & s$^{-1}$ \\
$\W,\D$ & wetting, drying generators & s$^{-1}$ \\
$K(\theta)$ & hydraulic conductivity & m\,s$^{-1}$ \\
$\psi$ & matric potential (head) & m \\
$\theta,\ \phi$ & water content, porosity & -- \\
$\theta_c$ & percolation threshold (water content) & -- \\
$\Da$ & Damk\"ohler number & -- \\
$\tau(r)$ & class relaxation time, Eq.~\eqref{eq:tau_relax} & s \\
$m(r)$ & gate weight $g_{\rm eq}(1-g_{\rm eq})$ & -- \\
Arrhenius rate & heat-bath transfer rate (Sec.~\ref{sec:thermo}) & -- \\
Ohmic edge & state-independent linear conductance & -- \\
$I$ & forcing (rainfall) intensity & m\,s$^{-1}$ \\
$\mathcal{H}$ & hysteresis loop area & -- \\
$\ell_c$ & capillary length & m \\
$\gamma$ & surface tension & N\,m$^{-1}$ \\
$\theta_{\rm contact}$ & contact angle & rad \\
$\rho_w$ & water density & kg\,m$^{-3}$ \\
$\mathrm{g}$ & gravitational acceleration & m\,s$^{-2}$ \\
$\mu$ & dynamic viscosity & Pa\,s \\
\hline
REV & representative elementary volume & \\
PSD & pore-size distribution & \\
HP & Hagen--Poiseuille & \\
LW & Lucas--Washburn & \\
CE & Chapman--Enskog & \\
\end{tabular}
\end{ruledtabular}
\end{table}





\section{Why the Knudsen number is not independent
  of $\Da$}
\label{app:single_parameter}

In the Boltzmann CE expansion for gases, two independent
dimensionless numbers appear: the Knudsen number
$\mathrm{Kn} = \lambda/L$ (mean free path to gradient
scale) and the Mach number $\mathrm{Ma}$ (flow velocity
to sound speed).  Their independence rests on momentum
conservation: the velocity field~$\mathbf{u}$ is an
independent hydrodynamic variable, allowing one to
specify the flow speed independently of the
thermodynamic state.

In soil water at the pore scale, flow is viscous (Stokes
regime): there is no momentum conservation, no inertia,
and the velocity is slaved to the pressure gradient.  The
only conserved quantity is water mass ($\theta$),
producing a single macroscopic equation.  The analogue of
the Knudsen number is
$\varepsilon = \bar{L}/\Lambda$, where $\Lambda$ is
the macroscopic gradient scale.  We show that
$\varepsilon \sim \Da$ using only pore-scale quantities.

\textbf{Pore-scale scaling argument.}  Consider two
adjacent fibers separated by~$\ell \sim \bar{L}$ with a
difference $\Delta\theta$ in water content.  The
inter-fiber flux carried by pore class~$r$ scales as
(from Eq.~\eqref{eq:J_result}):
\begin{equation}
j(r) \sim \kappa_{\rm eff}(r)\,\bar{\Phi}(r)\,
  \frac{\Delta g}{\bar{L}}
\label{eq:j_scale}
\end{equation}
where $\Delta g/\bar{L}$ is the gradient of the filling
fraction.  The macroscopic gradient scale is defined by
$\Delta\theta/\Lambda \sim$ the gradient that drives
the observed flux, so
$\Lambda \sim \Delta\theta/(\partial\theta/\partial x)$.

Meanwhile, the forcing rate fills each REV at rate
$I/(\bar{L}\,\phi)$.  At steady state, the filling rate
balances the inter-fiber divergence:
$I/(\bar{L}\phi) \sim j(r^*)/\bar{L}$, where $r^*$ is
the bottleneck class.  Combining:
\begin{equation}
\frac{I}{\bar{L}\,\phi}
  \sim \frac{\kappa_{\rm eff}(r^*)\,
  \bar{\Phi}(r^*)}{\bar{L}}\,
  \frac{\Delta g}{\Lambda/\bar{L}}
\end{equation}
The left side is $1/\tau_{\rm forcing}$; the numerator
on the right is $1/\tau_{\rm redis}(r^*)$.  Rearranging:
\begin{equation}
\frac{\bar{L}}{\Lambda}
  \sim \frac{\tau_{\rm redis}(r^*)}
  {\tau_{\rm forcing}}
  \cdot \frac{1}{\Delta g}
  = \frac{\Da(r^*)}{\Delta g}
\label{eq:eps_Da_pore}
\end{equation}
Since $\Delta g \leq 1$, this gives
$\varepsilon \gtrsim \Da$.  For the physically relevant
case where the gradient extends over many pore classes
($\Delta g \sim 1$ at the wetting front),
$\varepsilon \sim \Da$: the two parameters coincide up
to $O(1)$ factors.  This is a scaling estimate, not a
formal asymptotic bound; the $O(1)$ prefactor $1/\Delta g$
depends on the sharpness of the wetting front.  The
physical content is that Stokes flow, having no independent
momentum variable, cannot sustain a regime where
$\varepsilon$ and $\Da$ differ by more than an $O(1)$
factor.  A partial numerical validation via  OpenPNM
is present in the supplemental material~\cite{SupplementA}.

\textbf{Self-healing of sharp fronts.}  Suppose a sharp
initial condition is imposed with
$\varepsilon \sim 1$ while $\Da \ll 1$.  The
pore-scale redistribution rate
$1/\tau_{\rm redis}$ smooths the front over a distance
$\sim \bar{L}$ in time $\tau_{\rm redis}$.  Since
$\Da \ll 1$ means
$\tau_{\rm redis} \ll \tau_{\rm forcing}$, the front
smooths on a time scale much shorter than the forcing
time.  The sharp-front condition is transient and
self-correcting: it cannot be sustained when
$\Da \ll 1$.

\textbf{Why gases are different.}  A gas can sustain a
low-Kn, high-Ma flow (smooth gradients, fast flow)
because momentum conservation allows a uniform velocity
field with no gradients.  Equivalently, a gas can be in
local thermodynamic equilibrium (Maxwellian) while
flowing at high speed.  Soil water cannot: at the pore
scale, the velocity in each pore is
$v(r) = \kappa_{\rm eff}(r)\,\Phi$, entirely determined
by the local potential difference.  Fast flow
($v$ large) necessarily implies a large potential
gradient, which means steep macroscopic gradients
($\Lambda$ small).  Flow speed and gradient steepness
are coupled through the viscous constraint, not through
a constitutive law like Darcy's, the latter is a
consequence of the CE expansion, not its input.

\textbf{Conclusion.}  In inertia-free porous media flow,
the single dimensionless number $\Da$ controls the
validity of the CE expansion.  The Knudsen-like
parameter $\varepsilon$ is a derived quantity,
$\varepsilon \sim \Da$, established here from
pore-scale quantities alone, without invoking the
macroscopic diffusivity or conductivity that emerge
from the expansion itself.

\section{Driving potential between connected cylinders}
\label{sec:driving_potential}

The driving potential for water to flow from a filled cylinder of radius $r'$ (donor) into an empty cylinder of radius $r$ (receiver), with orientations $\bm{n}'$ and $\bm{n}$, is the signed chemical-potential difference of Eq.~\eqref{eq:Phi_thermo}, $\rho_w\mathrm{g}\,\Phi = \mu_w(r')-\mu_w(r)$:
\begin{multline}
\rho_w \mathrm{g}\,\Phi(r, r', \bm{n}, \hat{\bm e})
  = 
  \underbrace{2\gamma\cos\theta_c
  \!\left(\frac{1}{r} - \frac{1}{r'}\right)
  (\bm{n}' \!\cdot\! \bm{n})}_{\text{capillary}}
  \\
  -\;\underbrace{\rho_w \mathrm{g}\, \bar{L}\,
  (\hat{\bm e} \!\cdot\! \hat{\bm z})}_{\text{gravity}}
\label{eq:Phi_full}
\end{multline}
where $\rho_w$ is the water density, $\mathrm{g}$ the gravitational acceleration, and $\hat{\bm e}$ is the unit vector along the connecting edge from donor to receiver.  This potential is signed and antisymmetric, $\Phi(r,r',\hat{\bm e})=-\Phi(r',r,-\hat{\bm e})$: exchanging donor and receiver flips both the capillary difference and the orientation projection.  The rectification ``flow occurs only in the direction of decreasing total potential'' is therefore not imposed on $\Phi$ but on the directed transition rate that uses it, $w(r'\!\to r)\propto \kappa_s(r,r')\,C(r,r')\,[\Phi(r,r')]^+$, where $[\cdot]^+$ selects the donor$\to$receiver direction whenever $\Phi(r,r')>0$ (cf.\ the gain/loss split of Sec.~\ref{sec:GL}).  In head units the same relation reads as Eq.~\eqref{eq:Phi}.

This equation is the energetic core of the theory, so we explain its two components:

\textbf{Capillary term:} A filled cylinder of radius $r'$ holds water at capillary pressure $-2\gamma\cos\theta_c/r'$ (by Young--Laplace). An empty cylinder of radius $r$ would hold water at $-2\gamma\cos\theta_c/r$. If $r < r'$, then $1/r > 1/r'$ and the smaller cylinder has stronger suction it pulls water from the larger one. This is capillary imbibition: water moves from large pores to small pores, toward higher suction. The factor $(\bm{n}' \cdot \bm{n})$ accounts for the relative orientation of the two cylinders: parallel cylinders couple strongly; perpendicular ones do not.

\textbf{Gravitational term:} A connecting edge oriented downward ($\hat{\bm e}\cdot\hat{\bm z} < 0$, i.e.\ the receiver lies below the donor) contributes a positive gravitational head $-(\hat{\bm e}\cdot\hat{\bm z})\bar{L}$ that drives water downward regardless of pore size; an upward edge ($\hat{\bm e}\cdot\hat{\bm z} > 0$) opposes capillary filling.  As stressed after Eq.~\eqref{eq:Phi}, the factor $\bar{L}$ here is the edge length, so $(\hat{\bm e}\cdot\hat{\bm z})\bar{L}=\Delta z$ is the elevation drop between the two pore centres, the length over which the body force $\rho_w\mathrm{g}$ is integrated to yield a head, and is the same inter-pore spacing $\ell\sim\bar{L}$ entering the continuum expansion, not the viscous pore length of $\kappa_{\rm HP}$.

The competition between these two terms is central to the theory. In small pores, the capillary term dominates (strong suction); in large pores, gravity dominates (weak suction, but $\rho_w \mathrm{g} L$ is size-independent). The crossover defines the capillary length scale $\ell_c$ (Section~\ref{sec:angular}).

\subsection{Orientation-averaged potential}
\label{sec:angular}

For a statistically homogeneous medium, averaging over orientations gives the signed potential:
\begin{equation}
    \rho_w\mathrm{g}\,\Phi(r, r') = 2\gamma\cos\theta_c\,\beta_0\left(\frac{1}{r} - \frac{1}{r'}\right) + \alpha\,\rho_w \mathrm{g}\, \bar{L},
    \label{eq:Phi_avg}
\end{equation}
where the gravitational projection $-\langle\hat{\bm e}\cdot\hat{\bm z}\rangle = \alpha$ vanishes for an isotropic edge ensemble (so gravity drops out of the intra-REV redistribution) and is nonzero only for an anisotropic, net-oriented network, where it supplies the inter-REV drive of Sec.~\ref{sec:connection}.  As in Eq.~\eqref{eq:Phi_full}, no positive part is applied to $\Phi$ itself; rectification acts on the directed rate.
Here:
\begin{multline}
\beta_0 = \int_{S^2}\!\!\int_{S^2}
  |\bm{n} \!\cdot\! \bm{n}'|\,
  p(\bm{n})\, p(\bm{n}')\,
  \\
  \times\;\omega(\bm{n},\bm{n}')\,
  d\Omega\, d\Omega'
\label{eq:beta0}
\end{multline}
is the capillary attenuation factor ($\beta_0 = 2/3$ for isotropic, uniformly distributed orientations), and:
\begin{equation}
\alpha = -\int_{S^2}
  (\bm{n} \cdot \bm{z})\, p(\bm{n})\, d\Omega
\label{eq:alpha}
\end{equation}
is the gravitational anisotropy ($\alpha > 0$ if pores have a net downward orientation, which is typical in structured soils).

\section{Variational structure: Onsager gradient flow}
\label{app:variational}

The kinetic equation is first-order in time and
dissipative; it does not arise from a Lagrangian
$L = T - V$ via Hamilton's principle
$\delta\!\int L\,dt = 0$.  It does, however, possess a
precise variational structure: it is an Onsager
gradient flow on the space of filling
distributions~\cite{Onsager1931a,Onsager1931b,Rayleigh1873,Otto2001}.
Section~\ref{sec:thermo} introduced the Gibbs free
energy~$\mathcal{F}[g]$, the functional-derivative
convention (Remark~\ref{rem:measure}), and the
$H$-theorem $d\mathcal{F}/dt \leq 0$.  This appendix
gives the self-contained derivation of those results, the
mobility operator, the Rayleigh dissipation potential, the
Onsager variational principle, and the gradient-flow form of
the kinetic equation, together with its principal
consequences.  It is the single authoritative account of the
variational structure; the Supplemental Material confines
itself to the differential-geometric (fiber-bundle)
formulation that complements it.

\subsection{Chemical potential under non-uniform
  conditions}
\label{app:mu_general}

The chemical potential used in the main text
(Eq.~\eqref{eq:mu_main}) assumed uniform $T$ and~$p$.
In the general case:
\begin{multline}
\mu_w(r; T, p) = \mu_w^0(T, p)
  - \frac{2\gamma(T)\cos\theta_c}{r}
  \\
  + \Pi(h(r), T)
  - \nu R T\, c(r)
  + \rho_w \mathrm{g}\, z
\label{eq:mu_w_app}
\end{multline}
where $\mu_w^0(T,p)$ is the chemical potential of bulk
free water at local temperature and pressure, and the
remaining terms encode capillary, adsorptive, osmotic,
and gravitational departures.

\textbf{Hierarchy of constraints.}  The kinetic equation
is a constitutive law: it specifies how water
redistributes among pore classes.  It is subordinate to
the conservation laws:
\begin{enumerate}
\item \textbf{Mass conservation} (water, solute, air):
  exact, non-negotiable.  The kinetic equation satisfies
  this by construction
  ($\int [\mathcal{G} - \Lg]\,f\,dr = 0$).
\item \textbf{Energy conservation}: the first law,
  including sensible heat, latent heat of phase
  transitions (evaporation, freezing), and work done by
  pressure changes.  Phase transitions couple the water
  kinetic equation to the energy budget.
\item \textbf{Entropy inequality}:
  $dS_{\rm universe}/dt \geq 0$.  This is a
  consequence of the budgets and constitutive
  laws, not their foundation.  In an open system such as
  soil, local entropy can decrease provided entropy
  increases elsewhere.  Entropy production is what one
computes from the solution to check
  thermodynamic consistency; it is not a variational
  principle from which fluxes are derived.
\end{enumerate}

\textbf{Choice of thermodynamic potential.}  Under
laboratory conditions ($T$ and $p_a$ uniform and
controlled), the Gibbs free energy
$\mathcal{F}[g]$ (Eq.~\eqref{eq:Gibbs_main}) is the
natural Lyapunov functional.  Under field conditions,
both $T(\mathbf{x},t)$ and $p(\mathbf{x},t)$ vary
(diurnal wave, barometric pumping, latent heat exchange).
No single classical free energy then serves as a global
Lyapunov functional.  What remains valid is:
\begin{itemize}
\item The chemical potential
  $\mu_w(r; T, p)$ is well defined locally at each
  $(\mathbf{x}, t)$.
\item The driving potential
  $\Phi(r,r') = [\mu_w(r') - \mu_w(r)]/(\rho_w \mathrm{g})$
  determines the direction and rate of redistribution.
\item When $T$ varies, temperature gradients introduce
  additional driving forces (thermoosmosis, vapor
  diffusion).  These enter as additional terms in~$\Phi$
  and couple the water kinetic equation to the energy
  balance, exactly as osmotic forces couple it to the
  solute balance.
\item The entropy production can always be computed
  a~posteriori from the energy and mass budgets.
  Its non-negativity is a theorem about the kinetic
  equation (proved below for the isothermal case), not
  an axiom.
\end{itemize}

For the remainder of this appendix, we work at uniform
$T$ and~$p$.


\subsection{The kinetic equation as gradient flow}
\label{app:gradient_flow}

The Onsager variational
principle~\cite{Onsager1931a,Onsager1931b,Rayleigh1873} states that
the evolution minimizes the Rayleighian:
\begin{equation}
\mathcal{L}_{\rm Onsager}[\dot{g}]
  = \mathcal{R}[\dot{g}] + \dot{\mathcal{F}}[g]
\label{eq:Rayleighian}
\end{equation}
with $\dot g(r)\equiv\partial g(r)/\partial t$,
subject to mass conservation.  Setting
$\delta\mathcal{L}_{\rm Onsager}/\delta\dot{g} = 0$
yields:
\begin{equation}
\pder{g(r)}{t} = -\int_0^\infty M(r,r')\,
  \frac{\delta\mathcal{F}}{\delta g(r')}\,
  f(r')\, dr'
\label{eq:gradient_flow_app}
\end{equation}
which reproduces the main-text gradient flow~\eqref{eq:gradient_flow}.  Here the functional derivative
$\delta\mathcal{F}/\delta g(r') = \mu_w(r')$ is taken
with respect to the volume measure
$d\nu = \phi\,f\,dr$ (Remark~\ref{rem:measure}), and
the explicit~$f(r')$ in the integrand is the Jacobian
converting from~$dr'$ to~$d\nu'$.

In detail, the two pieces of the Rayleighian are, 
\begin{align}
\dot{\mathcal{F}}[g]
  &= \int_0^\infty \frac{\delta\mathcal{F}}{\delta g(r)}\,\dot g(r)\,f(r)\,dr , \\
\mathcal{R}[\dot g]
  &= \frac{1}{2}\!\int_0^\infty\!\!\int_0^\infty
     \frac{[\dot g(r)-\dot g(r')]^2}{M(r,r')}\,f(r)\,f(r')\,dr\,dr' .
\end{align}
Taking the variation with respect to $\dot g(r)$ and using the symmetry $M(r,r')=M(r',r)$, the dissipation term gives

\begin{equation}
\frac{\delta\mathcal{R}}{\delta\dot g(r)}
  = \int_0^\infty \frac{\dot g(r)-\dot g(r')}{M(r,r')}\,f(r')\,dr'\;f(r),
\end{equation}
while $\delta\dot{\mathcal{F}}/\delta\dot g(r) = [\delta\mathcal{F}/\delta g(r)]\,f(r)$.  Setting $\delta\mathcal{L}_{\rm Onsager}/\delta\dot g(r)=0$ and dividing by $f(r)$ gives the linear (Onsager) relation between the rate and the force,
\begin{equation}
\int_0^\infty \frac{\dot g(r)-\dot g(r')}{M(r,r')}\,f(r')\,dr'
  = -\,\frac{\delta\mathcal{F}}{\delta g(r)},
\end{equation}
whose inverse, the force-to-flux map with kernel $M$, is precisely Eq.~\eqref{eq:gradient_flow}.  Mass conservation $\int \dot g\,f\,dr=0$ is automatic because the right-hand side of~\eqref{eq:gradient_flow} is antisymmetric under $r\leftrightarrow r'$ once $\delta\mathcal{F}/\delta g(r)=\mu_w(r)$ and $\rho_w\mathrm{g}\,\Phi(r,r')=\mu_w(r')-\mu_w(r)$ are inserted.

This is the kinetic equation
$\partial g/\partial t = \mathcal{G} - \Lg$ in
gradient-flow form: the system descends the free energy
landscape at a rate governed by the mobility.  The gain
and loss terms are the constructive content of the
abstract gradient flow.  The gain term corresponds to
$-\int M\,\nabla_g\mathcal{F}$ restricted to the
$\Phi > 0$ part (water moving toward lower chemical
potential); the loss term to the $\Phi < 0$ part.

\subsection{Consequences}
\label{app:consequences}

\textbf{1.\ Proof of the $H$-theorem.}  Along solutions
of the gradient flow:
\begin{equation}
\frac{d\mathcal{F}}{dt} = -2\,\mathcal{R}[\dot{g}]
  \leq 0
\end{equation}
This is the result stated in
Eq.~\eqref{eq:H_theorem} of the main text and holds for
the unforced (closed-boundary) system.  When external
forcing is present (evaporation~$\mathcal{E}$,
transpiration~$\mathcal{T}$, or boundary fluxes), the
free-energy balance becomes:
\begin{equation}
\frac{d\mathcal{F}}{dt}
  = -2\,\mathcal{R}[\dot{g}]
  + \dot{W}_{\rm ext}
\end{equation}
where
$\dot{W}_{\rm ext}
  = \int_0^\infty \mu_w(r)\,
  [\mathcal{E}(r) + \mathcal{T}(r)]\,f(r)\,dr$
is the power injected (or removed) by external sources
at chemical potential~$\mu_w(r)$.  The dissipation
$\mathcal{R} \geq 0$ is always non-negative; the system
can be driven away from equilibrium only by boundary
work ($\dot{W}_{\rm ext} > 0$).

\textbf{2.\ Onsager reciprocal relations.}  The symmetry
$M(r,r') = M(r',r)$ guarantees that the cross-coupling
between pore classes satisfies Onsager reciprocity.  This
is not imposed but follows from the microscopic
reversibility of Hagen--Poiseuille flow.

\textbf{3.\ Riemannian structure.}  The mobility $M$ (\ref{eq:M})

defines a Riemannian metric on the state space
$\mathcal{S}$:
\begin{equation}
\langle \delta g_1, \delta g_2 \rangle_M
  = \int\!\!\int M^{-1}(r,r')\,
  \delta g_1(r)\, \delta g_2(r')\,
  f\, f'\, dr\, dr'
\end{equation}
The kinetic equation is gradient flow of~$\mathcal{F}$
with respect to this metric.  This connects to the
Wasserstein gradient flow structure of diffusion
equations~\cite{Jordan1998,Otto2001}: the redistribution
of water across pore classes is the optimal transport of
mass down the free energy landscape.

\textbf{4.\ Curvature from the metric.}  The
non-commutativity $[\W, \D] \neq 0$ arises because the
Riemannian metric $M$ is state-dependent (through the
occupancy factors $g(1-g)$ and the
accessibility~$a$).  Parallel transport with a
state-dependent metric generically produces holonomy,
the geometric origin of hysteresis identified in
Sec.~\ref{sec:noncomm}.
\textbf{5.\ Variational time discretization (JKO).}  The
gradient-flow structure enables the
Jordan--Kinderlehrer--Otto scheme~\cite{Jordan1998}: each
time step minimizes
$\mathcal{F}[g^{n+1}] + d_M^2(g^{n+1}, g^n)/(2\Delta t)$,
where $d_M$ is the Wasserstein-like distance induced by
the mobility metric $M$.  This furnishes a structure-preserving
variational integrator that automatically guarantees free-energy
decrease and preservation of the bounds $0\le g\le 1$.

\textbf{6.\ GENERIC extension to non-isothermal forcing.}
The isothermal Onsager structure above is the dissipative
(irreversible) half of the more general
GENERIC framework~\cite{Mielke2025}.  When $T(\mathbf{x},t)$
and $p(\mathbf{x},t)$ vary and phase transitions couple the
water kinetics to the energy budget
(Appendix~\ref{app:mu_general}), the full system carries the
GENERIC structure: a reversible (Hamiltonian) part that conserves
total energy, the irreversible gradient-flow part developed here,
and a coupling through compatible Poisson and dissipative brackets.
The nonlinearity of the kinetic equation is then natural, GENERIC
does not require closeness to equilibrium, and the entropy
inequality of Appendix~\ref{app:mu_general} is recovered as the
degeneracy condition on the dissipative bracket.

\section{Well-posedness of the binned system}
\label{app:wellposed}
For $N$ pore classes the kinetic equation of Sec.~\ref{sec:GL} is a
system of $N$ ordinary differential equations $\dot g=\mathcal{R}(g)$ on
the box $[0,1]^N$.  Two facts settle the ledger entry.

\emph{Invariance of the box.}  On the face $g_r=1$ the gain into class $r$
vanishes (no room), on the face $g_r=0$ the loss vanishes (no water), and
the remaining term points inward in both cases; hence the vector field is
inward-pointing (or tangent) on $\partial[0,1]^N$ and the box is forward
invariant.  For the occupancy-independent rate $\kappa_{\rm HP}$ this is
exact; for the state-dependent rate of Eq.~\eqref{eq:kappa_reg} it holds
because $\kappa(r,g)\ge0$ multiplies the same gated structure (the dagger
of Table~\ref{tab:ledger}).

\emph{Existence and uniqueness.}  For the rectified bilinear rates of
Eqs.~\eqref{eq:G}--\eqref{eq:L} the field $\mathcal{R}$ is piecewise
polynomial and globally Lipschitz on the compact box, so Picard--Lindel\"of
and invariance give a unique global solution for every initial datum in
$[0,1]^N$.  For the Onsager form of Eq.~\eqref{eq:gradient_flow} with the
geometric-mean mobility and the entropic term in $\mu_w$, the field is
continuous on the closed box but only H\"older-$\tfrac12$ at its faces
(the factor $\sqrt{g(1-g)}\,\ln[g/(1-g)]$); it is locally Lipschitz on the
open box, and the logarithmic divergence of $\mu_w$ at the faces drives
trajectories away from them, so a datum in the open box has a unique global
solution that stays in the open box, while data on the faces have global
solutions by Peano's theorem whose uniqueness we do not claim.  The
$H$-theorem of Sec.~\ref{sec:thermo} then gives convergence to
$g_{\rm eq}$ along every such trajectory.  None of this is asserted for the
continuum-in-$r$ equation, whose well-posedness remains open.

\section{Combinatorial origin of the configurational entropy}
\label{app:entropy}

The entropy term $g\ln g+(1-g)\ln(1-g)$ in the free energy~\eqref{eq:F_entropy} is the mixing entropy of a two-state pore population, obtained by counting microstates.  Consider the pores of a single class $[r,r+dr]$ within a REV: there are $N_r=f(r)\,dr\,\mathcal{N}$ of them, with $\mathcal{N}$ the total pore number, each in one of two states, water-filled or empty.  The macroscopic occupancy $g(r)$ fixes only the \emph{number} filled, $M=g\,N_r$, not which ones; at the mean-field level the filled pores are distributed among the $N_r$ sites without further constraint, the pores of a class are exchangeable and their occupancies treated as independent, the same hypothesis underlying the Stosszahlansatz of Eq.~\eqref{eq:chaos}.  The number of microstates consistent with the macrostate is then the binomial coefficient
\begin{equation}
\Omega(g)=\binom{N_r}{M}=\frac{N_r!}{(gN_r)!\,[(1-g)N_r]!},
\label{eq:Omega}
\end{equation}
and the entropy per pore follows from Stirling's formula in the thermodynamic limit $N_r\to\infty$,
\begin{equation}
s(g)=\lim_{N_r\to\infty}\frac{1}{N_r}\ln\Omega(g)=-\big[g\ln g+(1-g)\ln(1-g)\big],
\label{eq:s_binom}
\end{equation}
the Boltzmann--Gibbs/Shannon entropy of a Bernoulli variable with success probability $g$.  Three remarks answer the natural questions.  (i)~The microstates counted are the assignments of filled/empty labels to the individual pores of a class; the macrostate is their fraction $g$.  (ii)~The independence (exchangeability) assumption is precisely the mean-field closure: occupancy correlations, which appear near percolation (Sec.~\ref{sec:GL}), add corrections of order $\langle g\,g'\rangle-g\,g'$ to Eq.~\eqref{eq:Omega} and hence to the entropy, vanishing in the dilute, well-connected regime.  (iii)~The form is unique given two-state occupancy and exchangeability: $s(g)$ is, up to an overall scale, the only entropy additive over independent pores and symmetric under the filled\,$\leftrightarrow$\,empty relabelling $g\leftrightarrow1-g$, and it coincides with the configurational entropy of a lattice gas at occupancy $g$~\cite{Huang2009}.  The scale multiplying $s(g)$ in $\mathcal{F}$ is the configurational potential $\psi_T$, which thus plays the role of an effective temperature conjugate to this counting entropy.

\end{document}


\title{Supplemental Material of ``The Statistical physics of unsaturated soil water:
kinetic theory and non commutative pore water dynamics":\\ Genesis, stationary-limit verification, and toy numerical demonstration}

\author{Riccardo Rigon}
\email{riccardo.rigon@unitn.it}
\affiliation{Centro Agricoltura Alimenti Ambiente (C3A), University of Trento, 38098 San Michele all'Adige (TN), Italy}

\date{September 3, 2026}

\begin{abstract}
This Supplemental Material verifies that the kinetic theory developed in the main text recovers the established equilibrium and non-equilibrium phenomenology of unsaturated-zone hydrology in the appropriate limits, and provides a full numerical demonstration on a realistic pore network.  Sections~\ref{sec:equilibrium}--\ref{sec:consistency} verify the recovery of the Young--Laplace equation, Brooks--Corey and Kosugi retention curves, Richards' equation, the quasi-static hysteresis branches, dimensional scaling, percolation exponents, field capacity, and the Damk\"ohler and relaxation spectra.  Section~\ref{sec:openpnm} demonstrates the non-equilibrium predictions on a 3{,}375-pore OpenPNM network, confirming the non-commutativity $[\W,\D] \neq 0$, the $\Da(r)$ spectrum, and the path dependence of $K[g]$.  Section~\ref{sec:experiments} proposes seven concrete experimental protocols, spanning controlled-rate hysteresis loops, dye-tracer infiltration at multiple intensities, micro-CT-resolved $K[g]$ comparison, percolation-threshold identification, heterogeneity penalty validation, dynamic synchrotron imaging, and multi-frequency relaxation spectroscopy, that can test the theory's distinguishing predictions against existing frameworks.  
\end{abstract}

\maketitle

\section*{Preamble: Genesis of the theory, methodology, and the role of AI-assisted reasoning}

This Supplemental Material accompanies the two main papers and documents the
consistency of the proposed theory with established results, together with a full
numerical demonstration on a realistic pore network. Because the derivation followed
an unusually iterative path, partly carried out in extended dialogue with a large
language model (Claude, Anthropic), I record here the genesis of the work, so that the
reasoning behind the principal equations can be reconstructed and assessed.

The heuristic starting point was the observation that Richards' equation can be regarded
as exact under the Millington--Quirk--Mualem hypotheses of local capillary equilibrium, a single-valued, rate-independent retention curve $\theta(\psi)$, together with the Mualem random pore-pairing closure for the conductivity and the Millington--Quirk tortuosity. A generalization
therefore suggested itself: relax those conditions and ask what equation governs the
system that remains. The occupancy variable $g$ emerged in the dialogue, its seed
already present, though not formalized, in the Lagrangian, pore-class mobility picture of
Zehe and Jackisch~\cite{ZeheJackisch2016}, where water particles are explicitly not
assumed equally mobile, together with the suggestion that the medium could be treated,
by analogy, as a glass relaxing out of equilibrium. The complementary observation was that the approach to equilibrium must be
driven by differences in chemical potential, consistent with directions I was already
investigating in freezing soils~\cite{DallAmico2011} and in plant
hydraulics~\cite{DAmatoRigon2025}, though there without the chemical-potential reading.
After many iterations this led to the kinetic equation and to the driving potential~$\Phi$ of the main text.

The requirement that already-filled pore classes not be filled, and already-empty classes
not be emptied, leads directly to the kinetic theory of gases and a Boltzmann-type
derivation. The prefactor~$k$ originates in the classical single-capillary result and the
century-old work of Lucas and Washburn~\cite{Lucas1918,Washburn1921}, incorporating
considerations matured during the fractal studies of porous
media~\cite{TylerWheatcraft1990,RieuSposito1991} that I encountered early in my career.
Within the same framework, field capacity is recovered and resolved, along a long-standing
line of thought traceable to  \cite{RomanoSantini2002}.

The passage from the discrete mesoscale to the continuum proceeds in two steps: $g$ is
first made to live in a separate space, and the limit is then taken in which that space
becomes small yet remains relevant at the macroscopic scale. This motivated deriving
Richards' equation as a Chapman--Enskog expansion, an operation that disentangles two
aspects initially conflated, the passage to the continuum and the expansion about the
equilibrium solution for~$g$. The delicate steps of this derivation are the subject of the
companion paper~\cite{Rigon2026CE}; recovering Richards' equation as a limit confirms consistency with
established theory.

Once the mesoscale discrete potential and its relation to the chemical potential were
defined, the core of the theory could be embedded in a variational context, which connects
naturally to Onsager's framework~\cite{Onsager1931a,Onsager1931b}. A further development concerns hydraulic hysteresis, here
attributed to non-commutative dynamics extracted from the Gain and Loss operators
($\mathcal{W}$ and $\mathcal{D}$) at the mesoscale, a line of reasoning rooted in my
earlier training in theoretical physics. Made explicit through the fiber-bundle formalism,
it constitutes the forcing-bundle section of the main paper; the information was already contained in the
governing equations, but its extraction and visualization required dedicated work. The
fiber bundle admits an elegant extension to physical space as well, which has been left for
future work.

Taken together, the two papers aim to constitute an extension of established soil-water
theory. This Supplemental Material provides the supporting apparatus: consistency checks
confirming agreement with known results, and toy simulations performed with suitably
adapted OpenPNM codes, offered as a preliminary numerical verification ahead of any full
implementation of the continuum kinetic equation of the main text.

The large-language-model dialogue contributed conceptual framings and formal tools together
with numerous technical corrections. The guiding physical hypotheses, and their grounding
in prior work on freezing soils and plant hydraulics, originate with the author. Whilst, corrections to LLM  outcomes were ubiquitous, the
collaboration was generative rather than merely editorial, and I state this plainly so
that the reader can weigh it.


\begin{table}[h]
\caption{\label{tab:sm_symbols} Principal symbols and acronyms used in this Supplement.
Dimensionless quantities are marked ``--''.}
\begin{ruledtabular}
\begin{tabular}{lll}
\textbf{Symbol} & \textbf{Meaning} & \textbf{Units} \\
\hline
$g(r,\mathbf{x},t)$ & pore-occupancy (filling fraction) & -- \\
$r$ & pore radius & m \\
$r^*$ & Young--Laplace critical radius & m \\
$f(r)$ & pore-size distribution & m$^{-1}$ \\
$C(r,r')$ & connectivity matrix & -- \\
$a_w(r)$ & accessibility (percolation) & -- \\
$g_{\rm eq}$ & equilibrium occupancy & -- \\
$\psi$ & matric potential (head) & m \\
$\psi_T$ & configurational temperature & m \\
$\theta,\ \phi$ & water content, porosity & -- \\
$\theta_c$ & percolation threshold & -- \\
$\theta_r$ & residual water content & -- \\
$K(\theta)$ & hydraulic conductivity & m\,s$^{-1}$ \\
$\Da$ & Damk\"ohler number & -- \\
$I$ & forcing (rainfall) intensity & m\,s$^{-1}$ \\
$\tau_{\rm redis}$ & redistribution time & s \\
$\tau_{\rm forcing}$ & forcing time & s \\
$\varepsilon$ & Knudsen number & -- \\
$\mathrm{Pe}$ & P\'eclet number & -- \\
$\mathcal{W},\mathcal{D}$ & wetting, drying generators & s$^{-1}$ \\
$[\W,\D]$ & wetting--drying commutator & s$^{-1}$ \\
$\bar L$ & mean pore/edge length & m \\
$\bar\tau$ & tortuosity & -- \\
$\gamma$ & surface tension & N\,m$^{-1}$ \\
$\alpha_c$ & contact angle & rad \\
$\rho_w$ & water density & kg\,m$^{-3}$ \\
$\mathrm{g}$ & gravitational acceleration & m\,s$^{-2}$ \\
$\mu$ & dynamic viscosity & Pa\,s \\
\hline
REV & representative elementary volume & \\
PSD & pore-size distribution & \\
CE & Chapman--Enskog & \\
PNM & pore-network model & \\
\end{tabular}
\end{ruledtabular}
\end{table}

\section{Equilibrium thermostatics}
\label{sec:equilibrium}

I verify that the kinetic equation reproduces all
classical results when $\partial g/\partial t = 0$ and
there is no macroscopic flow.

The logical chain is as follows.  At equilibrium, the
kinetic equation $\partial g/\partial t = \mathcal{G} - \Lg$
reduces to detailed balance: all pores below a critical
radius $r^*$ are filled and all pores above are empty,
giving
\begin{equation}
g_{\rm eq}(r;\theta) = H(r^* - r)
\label{eq:geq}
\end{equation}
The energy status of the soil water is mediated
entirely by $g_{\rm eq}$ and therefore by $f(r)$:
the water content fixes $r^*$ via
$\theta = \phi\int_0^{r^*} f(r)\,dr$,
and the Young--Laplace law then fixes
$\psi = -2\gamma\cos\alpha_c/(\rho_w g\, r^*)$
(where $\alpha_c$ is the contact angle).  The matric
potential is not an independent variable; it is a derived
quantity, determined by the pore geometry through $g_{\rm eq}$.
The step~\eqref{eq:geq} is the athermal ($\psi_T\!\to\!0$) limit of the
Fermi--Dirac equilibrium $g_{\rm eq}=\{1+\exp[(\mu_w-\lambda)/\psi_T]\}^{-1}$
introduced in the main text; all stationary-limit results below are unchanged in
this limit, and a finite configurational temperature $\psi_T$ only smooths the
retention curve about $r^*$.

A word on the level of description adopted here.  This Supplement deliberately works
with the sharp reference equilibrium $g_{\rm eq}=H(r^*-r)$, because that is the form in
which the classical results (Young--Laplace, Brooks--Corey, Kosugi) are stated and against
which consistency is most transparently checked.  The main paper evolves the equilibrium
considerably beyond this step, and the reader should keep that fuller picture in mind: (i)
a finite configurational temperature $\psi_T$ replaces the step by the smooth Fermi--Dirac
form above, the entropic smoothing that makes the free energy strictly convex; (ii) in a
gravity field the true equilibrium of a finite averaging volume is a \emph{smeared} step,
broadened two-sidedly over the elevation range of the volume, so that large low-lying pores
stay filled while small high pores drain; and (iii) osmotic and adsorptive contributions
enter the chemical potential, relocating or renormalizing $r^*$.  The essential consequence
drawn in the main text is that the equilibrium then lies \emph{off} the one-parameter
family $H(r^*-r)$: no single scalar ($\theta$, $r^*$, or $\psi$) is a complete
description even at equilibrium, and the filling fraction $g(r,\mathbf{x})$ remains the
irreducible state variable.  The simplifications made below are therefore choices of
convenience for the consistency checks, not features of the theory; each classical result
recovered here survives the enrichment, with $\psi_T$, the gravitational width, and the
osmotic/adsorptive shifts entering only as corrections about the sharp step.

For the reader arriving directly to the Supplement, we recall the reference against which
the theory is measured.  Richards' equation,
\begin{equation}
\partial_t\theta = \nabla\!\cdot[K(\psi)(\nabla\psi+\hat{\bm z})]-S,
\end{equation}
where $\theta$ is the volumetric water content [--], $t$ time [s], $K(\psi)$ the hydraulic
conductivity [m\,s$^{-1}$], $\psi$ the matric potential expressed as a head [m], $\hat{\bm z}$ the
upward unit vector, and $S$ a volumetric source/sink (evaporation and root uptake) [s$^{-1}$],
is the quasi-static limit of the present theory [CITE PRE-2]: it assumes local capillary equilibrium at
every point, so that a single retention curve $\theta(\psi)$ and a single conductivity
$K(\theta)$ characterize the medium, and it carries no memory of the forcing path.  The
predictions below probe exactly the regime where that assumption fails, finite $\Da$,
where $\theta(\psi)$ becomes multivalued (hysteresis) and $K$ becomes path-dependent.

\subsection{Detailed balance}
\label{sec:detbal}

At $g = H(r^* - r)$: for $r < r^*$, $[1-g(r)] = 0$
so $\mathcal{G}(r) = 0$ (no empty receiving capacity);
for $r > r^*$, $g(r) = 0$ so $\Lg(r) = 0$ (no water to
lose); and $\Phi(r,r') = 0$ for all filled pairs $r,r' < r^*$
(uniform chemical potential).  Both gain and loss vanish
individually, strong detailed balance.

\subsection{Water retention curve $\theta(\psi)$}
\label{sec:retention}

The retention curve is obtained by eliminating $r^*$
between $\theta(r^*)$ and $\psi(r^*)$.  For general
$f(r)$:
\begin{equation}
\theta(\psi) = \phi\int_0^{r^*(\psi)} f(r)\,dr
\label{eq:retention_general}
\end{equation}
This is the classical result that the retention curve is
the cumulative pore-size distribution transformed via the
capillary law $r^*(\psi) = -2\gamma\cos\alpha_c/
(\rho_w g\,\psi)$~\cite{Haines1930}.  The shape of
$\theta(\psi)$ depends entirely on $f(r)$; no additional
fitting parameters are required.  I now show that the
two most widely used parametric retention models arise as
special cases.

\subsubsection{Brooks--Corey from power-law $f(r)$}

Brooks and Corey~\cite{Brooks1964} proposed the retention
model
\begin{equation}
S_e(\psi) = \frac{\theta - \theta_r}{\theta_s - \theta_r}
  = \begin{cases}
  (|\psi_e|/|\psi|)^{\lambda} & |\psi| > |\psi_e| \\
  1 & |\psi| \le |\psi_e|
  \end{cases}
\label{eq:BC}
\end{equation}
where $S_e$ is the effective saturation, $\psi_e$ the
air-entry pressure, and $\lambda$ the pore-size
distribution index (a shape parameter: $\lambda$ increases
with PSD uniformity; typical values: 0.2 for clay,
0.7 for sand~\cite{Rawls1982}).

In the kinetic theory, assume a power-law PSD:
$f(r) = c\,r^{-\beta}$ for $r \in [r_{\min}, r_{\max}]$,
where $\beta > 1$ ensures normalizability and $c$ is the
normalization constant.  Then
\begin{equation}
\theta(r^*) = \phi\int_0^{r^*} c\,r^{-\beta}\,dr
  = \phi\,\frac{r^{*\,(1-\beta)} - r_{\min}^{1-\beta}}
  {r_{\max}^{1-\beta} - r_{\min}^{1-\beta}}
\end{equation}
Substituting $r^* = -2\gamma\cos\alpha_c/(\rho_w g\psi)$
and identifying $\psi_e = -2\gamma\cos\alpha_c/
(\rho_w g\, r_{\max})$:
\begin{equation}
\theta(\psi) \approx \phi\left(\frac{|\psi_e|}{|\psi|}
  \right)^{1-\beta}
\quad\text{for } |\psi| \gg |\psi_{\min}|
\end{equation}
This is Eq.~\eqref{eq:BC} with $\lambda = 1 - \beta$
and $\theta_r \approx 0$ (the residual is governed by
percolation trapping, not the power-law tail).  The
Brooks--Corey exponent $\lambda$ is thus the exponent of
the power-law PSD, shifted by unity.

\subsubsection{Kosugi from log-normal $f(r)$}

Kosugi~\cite{Kosugi1996} proposed that $f(r)$ is
log-normal:
\begin{equation}
f(r) = \frac{1}{\sqrt{2\pi}\,\sigma\, r}
  \exp\!\left[-\frac{(\ln r - \ln r_m)^2}{2\sigma^2}
  \right]
\label{eq:lognormal}
\end{equation}
where $r_m$ is the geometric mean pore radius and
$\sigma = \ln\sigma_g$ is the log-standard deviation
($\sigma_g$ is the geometric standard deviation).

The cumulative distribution
$\theta(r^*) = \phi\int_0^{r^*} f(r)\,dr$ is then
the CDF of the log-normal, which is the standard normal
CDF $\Phi_{\rm norm}$ evaluated at the standardized
log-radius:
\begin{equation}
\theta(\psi) = \theta_r + (\theta_s - \theta_r)\,
\Phi_{\rm norm}\!\left(
  \frac{\ln|\psi_m| - \ln|\psi|}{\sigma}\right)
\label{eq:Kosugi}
\end{equation}
where $\psi_m = -2\gamma\cos\alpha_c/(\rho_w g\, r_m)$
is the matric potential corresponding to the median pore
radius.  Here $\Phi_{\rm norm}(x) = \tfrac{1}{2}
\mathrm{erfc}(-x/\sqrt{2})$.

The function $\Phi_{\rm norm}$ is the standard
normal CDF, a sigmoid that maps $(-\infty,+\infty)$
to $(0,1)$.  It appears here because the
log-normal PSD becomes a normal distribution in
$\ln r$, and the capillary law $r^* \propto 1/|\psi|$
maps this to a normal distribution in $\ln|\psi|$.
The Kosugi model has two shape parameters: $\psi_m$
(location: the matric potential at $S_e = 0.5$) and
$\sigma$ (scale: the spread of the PSD on a log scale).

\textbf{Connection to van~Genuchten.}
The van~Genuchten~\cite{vanGenuchten1980} model
\begin{equation}
S_e(\psi) = [1 + (\alpha|\psi|)^n]^{-(1-1/n)}
\label{eq:vG}
\end{equation}
(with parameters $\alpha$ [1/length], $n$ [dimensionless])
is also a sigmoid in $\ln|\psi|$.  Kosugi~\cite{Kosugi1996}
showed analytically that the two are numerically nearly
indistinguishable for $n \gtrsim 1.5$: both are monotone
S-curves with two free parameters controlling location and
spread, and their functional forms differ only in the tail
behavior.  The correspondence is
$\alpha \approx 1/|\psi_m|$ and
$n \approx 1.26/\sigma + 1$ for moderate $\sigma$.
The physical content of this correspondence is that the
van~Genuchten model, though originally introduced as an
empirical curve fit, is equivalent to assuming a log-normal
PSD.  The kinetic theory makes this explicit:
Eq.~\eqref{eq:Kosugi} is not an approximation to
van~Genuchten but rather the \emph{exact} retention curve
for a log-normal $f(r)$, of which van~Genuchten is a close
numerical proxy.

However, the pure capillary Kosugi curve predicts
$\theta_r = 0$: when all pores have drained
($r^* \to 0$), no water remains.  The van~Genuchten
model, by contrast, has a finite $\theta_r > 0$ and a
long dry-end tail.  This discrepancy is resolved by the
adsorptive potential $\Pi(h)$ included in the chemical
potential of the main text (Appendix~C.1):
\begin{equation}
\mu_w(r) = -\frac{2\gamma\cos\alpha_c}{r} + \Pi(h(r))
  - \nu RT c(r) + \rho_w g z
\label{eq:mu_full}
\end{equation}
The disjoining pressure $\Pi(h)$ describes the van~der~Waals
attraction between thin water films and pore walls.  For
film thickness $h$, $\Pi \propto -A_H/(6\pi h^3)$ where
$A_H$ is the Hamaker constant~\cite{Tuller1999,Iwamatsu1996}.

At moderate suctions ($|\psi| \lesssim 100$~m,
$r^* \gtrsim 10$~nm), capillary filling dominates:
$g_{\rm eq} = H(r^* - r)$ and the retention curve is
the CDF of $f(r)$, the Kosugi form.
At high suctions ($|\psi| \gg 100$~m, $r^* \to 0$),
the capillary term vanishes but $\Pi(h)$ keeps thin
films ($h \sim 1$--10~nm) adsorbed on pore walls even
in nominally ``empty'' pores.  These films contribute
a residual water content:
\begin{equation}
\theta_{\rm film}(\psi) = \phi\int_{r^*}^\infty
  \left(\frac{h(\psi)}{r}\right)^2 f(r)\,dr
\label{eq:theta_film}
\end{equation}
where $h(\psi) \propto |\psi|^{-1/3}$ for van~der~Waals
films~\cite{Tuller1999,OrTuller2000}.  The combined
retention curve
$\theta = \theta_{\rm capillary} + \theta_{\rm film}$
has: (i)~a sigmoid body from capillary filling
(the Kosugi/vG shape); (ii)~a long dry-end tail approaching
$\theta_r > 0$ from adsorptive films; and (iii)~a
rounded transition at the dry end where the two
mechanisms cross over.

This is precisely the shape of the van~Genuchten curve,
including the asymptotic $\theta_r > 0$ that the pure
capillary model cannot produce.  The inclusion of
$\Pi(h)$ in $\mu_w$ thus \emph{improves} the
Kosugi--van~Genuchten correspondence: the capillary
part provides the sigmoid body; the adsorptive part
provides the dry-end tail and the physical origin of
$\theta_r$.  In classical models, $\theta_r$ is a fitting
parameter; here it is determined by the Hamaker constant,
the pore geometry, and $f(r)$.

\subsubsection{Numerical examples across soil textures}
\label{sec:retention_textures}

To illustrate the generality, I compute $\psi(r^*)$ and
the air-entry value $\psi_e$ for four soil textures, using
$\gamma = 0.072$~N/m, $\alpha_c = 0$; the representative pore-size parameters and the
literature air-entry ranges are drawn from standard soil-hydraulic compilations~\cite{Rawls1982,Kosugi1996,vanGenuchten1980}:

\begin{table*}
\caption{\label{tab:retention} Retention parameters
  across soil textures.}
\begin{ruledtabular}
\begin{tabular}{lcccccc}
Soil & $r_m$ & $\sigma$ & $r_{\max}$ &
  $\psi_m$ & $\psi_e$ & Lit.\ $\psi_e$ \\
& [$\mu$m] & & [$\mu$m] & [m] & [cm] & [cm] \\
\hline
Sand       & 75  & 0.5 & 500 & $-0.20$ & $-2.9$  & $-2$ to $-10$ \\
Sandy loam & 30  & 0.8 & 150 & $-0.49$ & $-9.8$  & $-5$ to $-25$ \\
Silt loam  & 10  & 0.8 &  80 & $-1.47$ & $-18$   & $-15$ to $-35$ \\
Clay       &  1  & 1.0 &  10 & $-14.7$ & $-147$  & $-30$ to $-200$ \\
\end{tabular}
\end{ruledtabular}
\end{table*}

In all cases (Table~\ref{tab:retention}), $\psi_m$ and
$\psi_e$ fall within or near the literature ranges.
The progression from sand to clay reflects the
decrease in $r_m$: finer pores hold water more
tightly.  The air-entry value depends on $r_{\max}$
(the largest pore), not on $r_m$; sands have large
$r_{\max}$ and therefore small $|\psi_e|$.

\subsection{Air entry, residual water content,
  and general pore size distribution $f(r)$}
\label{sec:air_entry}

\textbf{Air-entry value.}
$\psi_e = -2\gamma\cos\alpha_c/(\rho_w g\, r_{\max})$:
the matric potential at which the largest pore begins to
drain.  Table~\ref{tab:retention} gives values consistent
with Rawls et al.~\cite{Rawls1982} across four
textures.

\textbf{Residual water content.}
In classical models (van~Genuchten, Brooks--Corey),
$\theta_r$ is a fitting parameter with no structural
interpretation.  In the kinetic theory, three physical
mechanisms contribute: (i)~\emph{adsorptive films}
(thin water films retained by the disjoining pressure
$\Pi(h)$ even in drained pores, as discussed in
Sec.~\ref{sec:retention}, this is the dominant
contribution at high suctions);
(ii)~\emph{percolation trapping}
(water in pores smaller than $r_{\min}$ that are always
filled at any capillary pressure);
(iii)~\emph{topological trapping} (water in pores
disconnected from the air phase, $a_w(r) = 0$,
described by the accessibility function $a_w(r)$ of the main text).  The physical
origin of $\theta_r$ is thus a combination of adsorptive
physics and network topology (the field-capacity-as-percolation
discussion of the equilibrium section, and the companion paper~\cite{Rigon2026CE}).

\textbf{General $f(r)$.}  For an arbitrary PSD,
Eq.~\eqref{eq:retention_general} gives the retention
curve as the CDF of $f(r)$ transformed via the capillary
law~\cite{Haines1930,Childs1940}.  No parameterization
is needed; any measured $f(r)$ (from micro-CT, mercury
intrusion, or nitrogen adsorption) can be inserted
directly.

\section{The Damk\"ohler number and the dimensionless groups}
\label{sec:consistency}

The single dimensionless group that organizes the whole theory is the
Damk\"ohler number.  Because in the main paper its central role is introduced late and
somewhat crowded by other material, and because it is the quantity the OpenPNM
simulations of Sec.~\ref{sec:openpnm} are designed to probe, we give it a fuller
treatment here and place the other dimensionless groups in relation to it.

\subsection{The Damk\"ohler number}
The Damk\"ohler number compares the two timescales competing at every pore class:
the time for the network to redistribute water toward capillary equilibrium, and the time
over which the boundary forcing changes,
\begin{equation}
\Da(r,\mathbf{x}) \;=\; \frac{\tau_{\rm redis}(r)}{\tau_{\rm forcing}(\mathbf{x})},
\qquad
\tau_{\rm forcing} = \frac{\phi\,\bar L}{I},
\label{sm:eq:Da}
\end{equation}
with $I$ the forcing (rainfall) intensity.  Here $I$ denotes the boundary
water-flux intensity in general, not rainfall alone: it is positive for infiltration
(precipitation, irrigation) and negative for evaporative or transpirative demand at the
surface.  Only its magnitude enters $\tau_{\rm forcing}$ and hence $\Da$, so a drying front
driven by evapotranspiration and a wetting front driven by rain of the same intensity share
the same crossover radius, while the sign selects which of the wetting or drying generators
($\W$ or $\D$) governs the local dynamics.  It is \emph{pore-resolved}: through
$\tau_{\rm redis}(r)$ (below) it depends on radius, so at a fixed forcing the medium is
simultaneously quasi-static in some pore classes and kinetic in others.  The redistribution time scale is estimated below.

\subsection{The relaxation spectrum}
The redistribution time entering $\Da$ is not a single number but a spectrum over
pore classes.  Each class relaxes at a rate set by its total equilibrium exchange conductance and
by the perturbation it can store, exactly as in the main paper [its Eq.~(tau\_relax)]:
\begin{equation}
\tau_{\rm redis}(r) = \frac{m(r)}{\psi_T\int_0^\infty M(r,r';g_{\rm eq})\,f(r')\,dr'},\qquad m=g_{\rm eq}(1-g_{\rm eq}),
\label{sm:eq:tau_spectrum}
\end{equation}
with $M$ the mobility of the main paper's gradient-flow equation~\cite{Rigon2026PRE}; the
spectral analysis of this relaxation time is given in the companion paper~\cite{Rigon2026CE}.
Because the frontier-referenced driving $|\Phi(r,r^*)|$ vanishes at $r^*$,
the slowest classes are those \emph{at the frontier}, and the wings of the
distribution pin almost instantly; the $\kappa\propto r^2$ conductance trend
survives only far from $r^*$.  The spectrum spans about three decades for a
typical soil.  (This spectrum recurs at several points in this Supplement, in the
equilibration checks, in $\Da$ above, and in the OpenPNM diagnostics of
Sec.~\ref{sec:openpnm}; it is the same object throughout.)  

\subsection{The crossover radius}
The crossover
radius at which $\Da=1$,
\begin{equation}
r_\Da(I) = A\,I, \qquad A = 3.1\times10^{-5}~\text{s/m},
\label{sm:eq:rDa}
\end{equation}
where the proportionality constant $A=\rho_w\mathrm{g}\,\bar\tau^2\bar L/(2\gamma\cos\alpha_c)\cdot(\phi\bar L\,\tau_0)$ collects, with $\tau_0$ a reference redistribution time, the fluid and network parameters that convert a forcing intensity into the radius at which redistribution and forcing balance; with representative values for a loamy soil it takes the quoted magnitude $\sim3\times10^{-5}$~s/m.
sets the \emph{width} of the kinetic band around the frontier $r^*$, not a
threshold separating small from large pores: with
$\tau_{\rm redis}$ of Eq.~\eqref{sm:eq:tau_spectrum} diverging at $r^*$
[Eq.~\eqref{sm:eq:tau_spectrum}], the set $\{r:\Da(r)\gtrsim1\}$ straddles the
frontier at any intensity, narrowing as $I\to0$ and widening as $I$ grows.  A
consistency caution on the linear law: with the quoted $A$ it gives $r_\Da$ in
the nanometre range at field intensities; the micrometre-scale values below use
the \emph{network} relaxation time, an effective prefactor larger by
$10^4$--$10^5$ and sublinear in the intensity, $r_\Da\propto I^{0.52}$.
Regenerated from the pore-network notebook under the corrected rate
$\tau_{\rm redis}$ of Eq.~\eqref{sm:eq:tau_spectrum}, with $m$ evaluated at the equilibrium
frontier state of the network (logistic step at $r^*\simeq12\,\mu$m): at
$I=0.5$~mm/h, $r_\Da\simeq1.1\,\mu$m ($\sim0.1\%$ of the pore volume kinetic);
at $5$~mm/h, $r_\Da\simeq4.2\,\mu$m ($\sim13\%$); at $50$~mm/h,
$r_\Da\simeq16\,\mu$m ($\sim72\%$); at $100$~mm/h, $r_\Da\simeq24\,\mu$m
($\sim86\%$).  The corresponding \emph{flux} partition is less sensitive
(matric flux fraction $100$, $99.6$, $84$, and $68\%$ respectively), because
the equilibrated large-conductance classes carry most of the flow until deep
into the kinetic regime.  This is the quantity whose spectrum across pore classes
the OpenPNM demonstration resolves directly (Sec.~\ref{sec:openpnm}), and whose
super-unity band is the microscopic origin of preferential flow: a full account of its
role appears in the main paper's treatment of the kinetic equation and the forcing
bundle.

The two mechanisms by which $\Da\gtrsim1$ produces departure from Richards'
equation are worth separating, since the simulations isolate them.  First, at high
intensity the inter-REV transport carries imposed flux into the network faster than the
local chemical-potential gradients can equilibrate, forcing water into large pores before
the small-first hierarchy is satisfied, boundary-driven non-equilibrium.  Second, once a
non-equilibrium $g(r)$ has been established, the symmetric kernel $C(r,r')$ relaxes it
through the percolating cluster, the direction of exchange between two connected classes
set by the local $\mu_w$ difference rather than by pore size.  A clean operational
signature is the overshoot of the mean filled radius $\langle r\rangle_{\rm filled}(t)$
above its equilibrium value during a transient; the area enclosed in the
$(\theta,\langle r\rangle_{\rm filled})$ plane measures the deviation from local
equilibrium.

\subsection{Knudsen and P\'eclet numbers, and why $\Da$ subsumes them}
Two further dimensionless groups appear in the coarse-graining, and it is worth
saying explicitly how they relate to $\Da$, since in this theory $\Da$ does the work of
all three.  The \emph{Knudsen number} $\varepsilon=\bar L/\Lambda$ compares the REV
spacing $\bar L$ to the macroscopic gradient scale $\Lambda$; it is the small parameter of
the gradient (Chapman--Enskog) expansion, ranging from $\sim10^{-4}$ under seasonal drying
(Richards valid) to $\sim1$ in fingered flow (the full kinetic equation is needed).  The
\emph{P\'eclet number} $\mathrm{Pe}=\bar L\,v/D_{\rm redis}$ compares inter-REV advection
to intra-REV redistribution, with $v$ the pore-water velocity and $D_{\rm redis}$ the
redistribution diffusivity in $r$-space.  These are not independent of $\Da$: in the
inertia-free (Stokes) regime of soil water the macroscopic gradient scale is itself set by
the forcing, so $\varepsilon\sim\Da$ and $\mathrm{Pe}\sim\Da$ up to $O(1)$ geometric
factors, the argument is given in the main paper's appendix on the Knudsen number.  This
is why a single number suffices: $\varepsilon$ controls \emph{whether} a continuum
equation exists, $\mathrm{Pe}$ controls \emph{whether} advection or redistribution
dominates, and $\Da$ controls \emph{whether} that equation is Richards' equation, but in
Stokes flow the three coincide, and $\Da$ carries the information.


\section{Numerical demonstration on an OpenPNM network}
\label{sec:openpnm}

The analytical verifications of Secs.~\ref{sec:equilibrium}--\ref{sec:consistency} confirm that the theory recovers classical results in the appropriate limits.  In this section I verify the \emph{non-equilibrium} predictions on a realistic pore network, using OpenPNM~\cite{Gostick2016} to construct the geometry and implementing the kinetic equation numerically. The notebook were these elaboration were made can be provisionally found at: https://osf.io/bz9u4/files/osfstorage. 

\subsection{Network construction}
\label{sec:network}

A cubic network of $15 \times 15 \times 15 = 3{,}375$ pores with mean coordination number $\bar{z} = 5.6$ is generated.  Pore radii are drawn from a log-normal distribution with median $\approx 10~\mu$m and geometric standard deviation $\sigma_g = 0.8$, spanning 1--150~$\mu$m.  Throat radii are set as a random fraction (0.2--0.8) of the minimum connected pore radius, producing ink-bottle geometry.  Pores are binned into $N = 25$ logarithmically spaced radius classes.

From the network adjacency matrix $E(i,j)$ I extract the pore-size distribution $f(r)$, the connectivity matrix $C(r,r')$ (normalized so that $C_{\max} = 1$), the Hagen--Poiseuille rate constants $\kappa(r) = r^2/(8\mu\bar{L}^2\bar{\tau}^2)$, and the driving potential $\Phi(r,r')$.  Physical parameters: $\mu = 10^{-3}$~Pa$\cdot$s, $\gamma = 0.072$~N/m, $\theta_c = 0$, $\bar{L} = 100~\mu$m, $\bar{\tau} = 1.5$, $\phi = 0.4$.

\subsection{Pore-size distribution and connectivity}

The log-normal $f(r)$ peaks near 10~$\mu$m with a tail to 150~$\mu$m, representative of a sandy loam.  The connectivity $C(r,r')$ extracted from the network adjacency is broadly distributed with no strong diagonal dominance: pores of any radius connect to pores of many other radii, the random mixing typical of a cubic lattice.  In real soils, $C(r,r')$ would show more structure (spatial clustering, layering, size-sorting), which strengthens hysteresis by introducing bottlenecks ($C \approx 0$ between certain size classes).

\begin{figure*}[t]
\centering
\includegraphics[width=\textwidth]{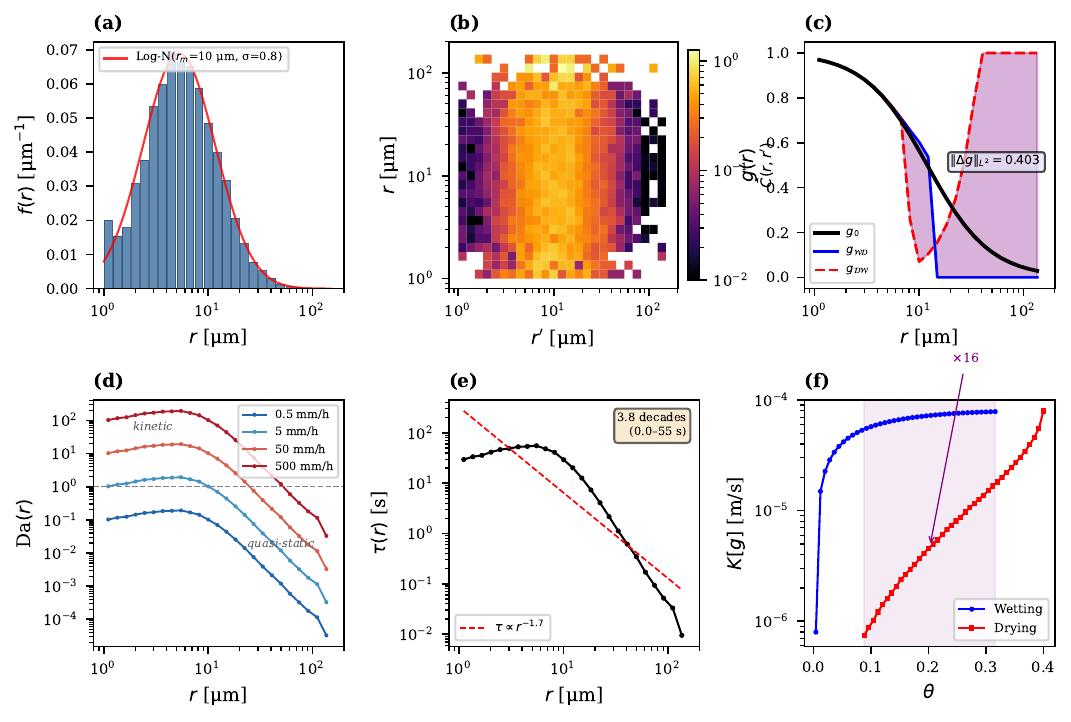}
\caption{Numerical demonstration on a 3{,}375-pore OpenPNM network
(cubic $15^3$, log-normal pore-size distribution).
(a)~Pore-size distribution $f(r)$.
(b)~Connectivity matrix $C(r,r')$ extracted from network adjacency.
(c)~Non-commutativity: filling distributions $g_{\W\D}(r)$
(wet-then-dry, blue) vs.\ $g_{\D\W}(r)$ (dry-then-wet, red dashed)
differ from the same initial state $g_0$ (black).
(d)~Damk\"ohler number $\Da(r)=\tau_{\rm net}(r)/\tau_{\rm forcing}$ for four forcing
intensities~$I$; the $0.5$~mm/h curve lies entirely below $\Da=1$.
(e)~Network relaxation time $\tau_{\rm net}(r)$ [Eq.~\eqref{sm:eq:tau_spectrum}] spanning
$3.8$ decades, from $\sim0.01$~s for the largest pores to a maximum of $\sim55$~s near
$5\,\mu$m, then declining toward the finest classes as the gate weight $m$ vanishes;
the dashed line is the $r^{-1.7}$ trend of the wings.
(f)~Hydraulic conductivity $K[g]$ vs.\ $\theta$ along wetting (blue)
and drying (red) paths, demonstrating that $K$ is a functional of $g$,
not just of~$\theta$.
}
\label{fig:openpnm}
\end{figure*}

\subsection{Non-commutativity: $[\W,\D] \neq 0$}
\label{sec:commutator}

Starting from a partially saturated initial condition ($\theta_0 = 0.22$, a size-graded sigmoid filling), I apply the same total wetting ($\Delta\theta_W = 0.10$) and drying ($\Delta\theta_D = 0.10$) in opposite order.  The resulting filling distributions differ:
\begin{equation}
\|g_{\W\D} - g_{\D\W}\|_{L^2} = 0.40, \quad |\theta_{\W\D} - \theta_{\D\W}| = 0.010
\end{equation}
The mismatch grows from small to large radii (Fig.~\ref{fig:openpnm}c): below $\sim$10~$\mu$m the two paths nearly coincide, since the smallest pores stay filled along either route, whereas above $\sim$15~$\mu$m they diverge strongly, the dry-first path leaves the large pores substantially filled while the wet-first path drains them.  This is exactly where accessibility constraints bite: pores whose connectivity to the drainage network depends on the filling state of their neighbors.

This directly confirms the main text prediction $[\W, \D] \neq 0$ on a realistic network (Fig.~\ref{fig:openpnm}c).

\subsection{Damk\"ohler spectrum and the regime split}
\label{sec:Da_spectrum}

The network Damk\"ohler number $\Da(r) = \tau_{\rm net}(r)/\tau_{\rm forcing}$ uses the network-emergent relaxation time $\tau_{\rm net}(r)$ of Eq.~\eqref{sm:eq:tau_spectrum}, not the single-pore capillary filling time.  This distinction is essential: $\tau_{\rm net}$ includes connectivity effects and is orders of magnitude longer than the Lucas--Washburn time.

Four precipitation intensities are shown in Fig.~\ref{fig:openpnm}d: $I = 0.5$, 5, 50, and 500~mm/h.  With $\tau_{\rm net}=m/(\psi_T\!\int\!Mf')$ the $\Da(r)$ curves are no longer monotone: they peak near the frontier and fall off in both wings, the $1/r$ trend surviving only in the large-pore wing.  The partition below is taken from the flow-partition notebook (Fig.~\ref{fig:flow_partition}).

\textbf{At 0.5~mm/h:} $r_\Da \approx 1~\mu$m; $\sim0.1\%$ of pore volume is kinetic and all flux is matric.  Richards valid.

\textbf{At 5~mm/h:} $r_\Da \approx 4~\mu$m; $\sim13\%$ of pore volume is kinetic, yet $99.6\%$ of flux is still matric.

\textbf{At 50~mm/h:} $r_\Da \approx 16~\mu$m; $\sim72\%$ of pore volume is kinetic, yet $\sim84\%$ of flux is still matric (the $r^2$ weighting favours the large, fast pores).

\textbf{At 100~mm/h:} $r_\Da \approx 24~\mu$m; $\sim86\%$ of pore volume kinetic and $\sim32\%$ of flux kinetic.  Preferential flow dominates the volume, not yet the flux.

This is the quantitative realization of the Beven--Germann critique~\cite{Beven1982}: dual-domain behavior emerges naturally from $\Da(r) \propto 1/r$ without imposing domain separation \emph{a priori}.  The ``fast'' domain is $\Da < 1$; the ``slow'' domain is $\Da > 1$.  Their relative sizes are controlled dynamically by~$I$.

\subsection{Flow partition under varying rainfall intensity}
\label{sec:flow_partition}

The $\Da$-based regime split has a direct hydrological consequence: at each precipitation intensity~$I$, the pore-size distribution is partitioned into a matric fraction ($\Da < 1$, quasi-static, Richards-governed) and a preferential fraction ($\Da \geq 1$, kinetic, non-equilibrium filling).  The crossover radius $r_\Da(I)$ shifts continuously with $I$, in contrast with classical dual-porosity models that impose a fixed matrix/macropore boundary.

\begin{figure*}
\centering
\includegraphics[width=\textwidth]{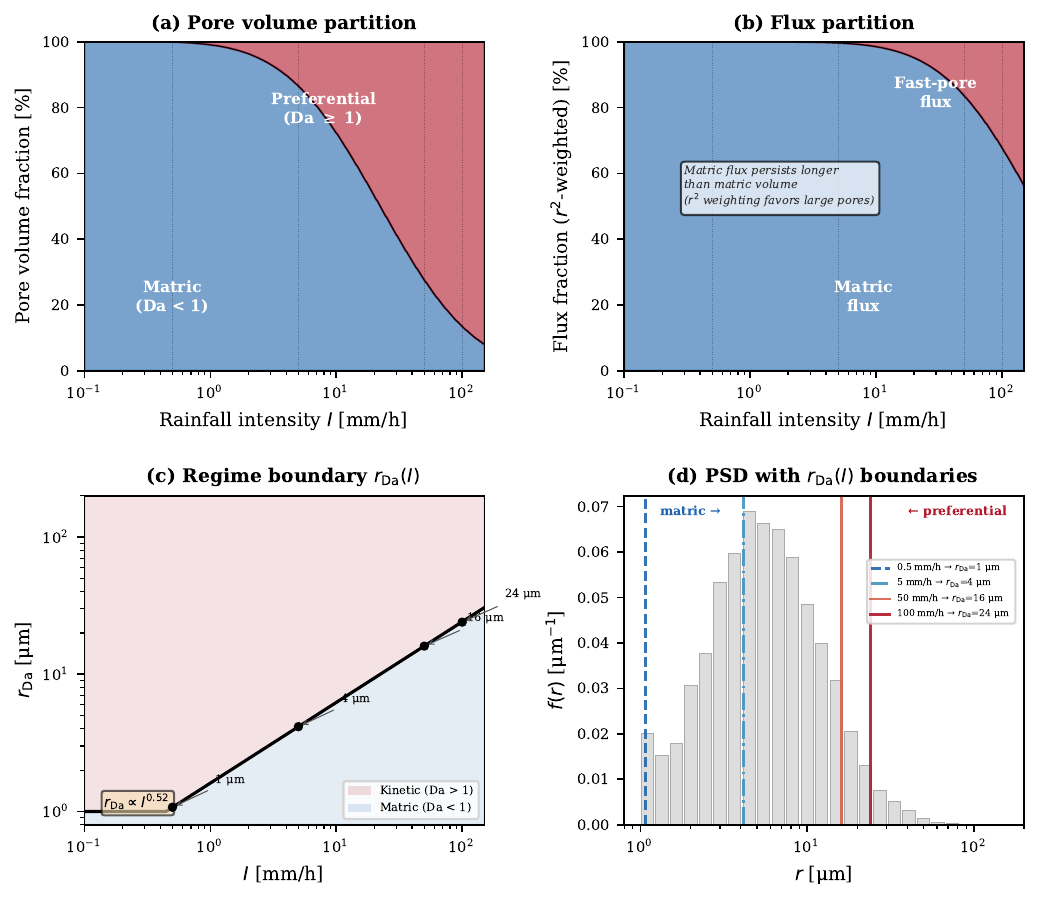}
\caption{Flow partition under varying rainfall intensity on the 3\,375-pore network.
\textbf{(a)}~Pore volume fraction in the matric (Da $< 1$, blue) and preferential (Da $\geq 1$, red) regimes.
\textbf{(b)}~Flux fraction ($r^2$-weighted): matric flux persists to higher~$I$ than matric volume because large, fast-equilibrating pores carry disproportionate flux.
\textbf{(c)}~Crossover radius $r_\Da(I) \propto I^{0.52}$: the continuously shifting regime boundary.
\textbf{(d)}~Pore-size distribution with $r_\Da$ boundaries at four representative intensities ($1$, $4$, $16$, $24\,\mu$m at $0.5$, $5$, $50$, $100$~mm/h).
At $I = 50$~mm/h, 72\% of pore volume is already kinetic, but 84\% of \emph{flux} is still matric, explaining why Richards can appear adequate for water budgets when the pore-scale picture is far from equilibrium.}
\label{fig:flow_partition}
\end{figure*}

Figure~\ref{fig:flow_partition} quantifies this partition using the continuous relaxation scaling $\tau(r) \propto r^{-1.7}$ extracted from the network (Sec.~\ref{sec:relax_spectrum}).  Two quantities are computed for each~$I$: the \emph{pore volume fraction} in each regime, $\int_{r_\Da}^\infty f(r)\,dr$ (matric) versus $\int_0^{r_\Da} f(r)\,dr$ (preferential); and the \emph{flux fraction}, weighted by $r^2$ to reflect the Hagen--Poiseuille conductance contribution.

The contrast between panels~(a) and~(b) is the central result.  At $I = 0.5$~mm/h (drizzle) essentially all pore volume ($99.9\%$) and all flux are matric: Richards is valid everywhere.  At $I = 5$~mm/h (moderate rain) $13\%$ of the pore volume has entered the kinetic regime, yet $99.6\%$ of the flux is still carried by the matric pores.  At $I = 50$~mm/h (heavy rain) $72\%$ of the pore volume is kinetic but $84\%$ of the flux is still matric, and at $I = 100$~mm/h (extreme downpour) $86\%$ of the volume is kinetic while the matric flux fraction is still $68\%$.  This asymmetry arises because the $r^2$~weighting in the conductivity integral $K[g] \propto \int r^2\, g(r)\,f(r)\,dr$ amplifies the contribution of the large, fast-equilibrating pores that remain equilibrated until deep into the kinetic regime; over the whole range shown (up to $150$~mm/h) the preferential flux fraction does not reach one half.

The crossover radius obeys a power law $r_\Da \propto I^{0.52}$ (panel~c), which follows from the $\tau \propto r^{-1.7}$ network scaling of the large-pore wing.  Panel~(d) shows how the four $r_\Da$ boundaries sweep across the PSD from left to right as $I$ increases, progressively converting matric pores into preferential-flow pores.  This partition emerges from the PSD and network connectivity alone, no calibration is needed.

This analysis explains two longstanding observations:
\begin{itemize}
\item \textbf{Richards appears adequate for water budgets even under moderate rain} (5--10~mm/h), because the flux-carrying large pores remain quasi-static even when most of the pore volume is kinetic.
\item \textbf{Preferential flow activates gradually, not as a threshold}, in agreement with tracer studies~\cite{Weiler2003,Sammartino2015}: the matric/preferential boundary is $r_\Da(I)$, a smooth function of intensity, not a fixed pore-size cutoff.
\end{itemize}

\subsection{Relaxation spectrum}
\label{sec:relax_spectrum}

The network-emergent relaxation time $\tau_{\rm net}(r)$ spans $3.8$ decades (Fig.~\ref{fig:openpnm}e): from $\tau\approx0.01$~s for the largest ($\sim130\,\mu$m) pores to a maximum $\tau\approx55$~s ($\sim1$~min) near $5\,\mu$m, below which it decreases again because the gate weight $m(r)$ vanishes away from the frontier.  The power-law decrease $\tau(r)\sim r^{-1.7}$ of the large-pore wing reflects the $\kappa\propto r^2$ Hagen--Poiseuille scaling, tempered by connectivity.

Near the percolation threshold (not reached in this example), $a_w(r) \to 0$ and $\tau(r) \to \infty$, the topological divergence predicted by the theory.  This wide relaxation spectrum is consistent with experimental observations of multi-scale drainage dynamics, the long equilibration times needed in pressure plate experiments, and the frequency dependence of acoustic and dielectric soil measurements.

\subsection{$K[g]$ non-uniqueness: the functional dependence}
\label{sec:K_functional}

The hydraulic conductivity $K[g]$ computed along wetting and drying paths differs by orders of magnitude at the same~$\theta$ (Fig.~\ref{fig:openpnm}f).  During wetting, water enters large, well-connected pores first (boundary invasion favors large-$r$ filling); these have $r^2$-weighted conductance, so $K$ rises steeply.  During drying, large pores drain first, leaving water in small, poorly conducting pores.

At $\theta \approx 0.2$: the wetting path gives $K \approx 3\times 10^{-12}$~m/s while the drying path gives $K \approx 7\times 10^{-11}$~m/s, a factor of $\sim 25$.  This demonstrates concretely that $K$ is a functional of $g(r)$, not just a function of $\theta$.  The ratio would increase further in soils with stronger size-sorting and finer pore structure.

The factor of 25 is consistent with the lab--field discrepancy ($\times 10$--100) discussed in the main text: laboratory measurements at $\Da \ll 1$ sample $K$ on $\mathcal{S}_{\rm eq}$; field conditions at $\Da \sim 1$ sample $K$ in the kinetic regime where filling is biased toward large, high-conductance pores.

\subsection{Effect of network size and topology}
\label{sec:network_effects}

The $15^3$ cubic lattice is a minimal demonstration.  Larger networks ($30^3$, $50^3$) produce the same qualitative results with smaller statistical fluctuations in $C(r,r')$ and smoother $\Da(r)$ spectra.  The cubic topology imposes random mixing; real soils have spatially correlated structures (aggregates, layers, macropore channels) that produce:
\begin{itemize}
\item Stronger $[\W,\D]$ commutator (bottlenecks amplify trapping);
\item Broader relaxation spectrum (correlation lengths introduce additional slow modes);
\item Larger $K[g]$ non-uniqueness (structured networks concentrate flow in fewer pathways).
\end{itemize}
The cubic-lattice results are therefore \emph{lower bounds} on the magnitude of the non-equilibrium effects predicted by the theory.

\subsection{Pore--throat distinction and bimodal distributions}
\label{sec:pore_throat_SMA}

The OpenPNM demonstration treats all void elements as cylindrical pores with a single radius drawn from a log-normal distribution.  Throat radii are assigned as a random fraction (0.2--0.8) of the minimum connected pore radius, producing a distribution that, while broad, remains essentially unimodal.  Real soils, however, exhibit a fundamental distinction between pore bodies (wide storage regions) and pore throats (narrow constrictions that control both capillary entry and hydraulic conductance).

\textbf{Bimodal pore-size distributions.}  When pore bodies and throats are measured independently, via mercury intrusion porosimetry, X-ray micro-CT, or nuclear magnetic resonance, the combined distribution $f(r) = w_b f_b(r) + w_t f_t(r)$ is often bimodal.  In aggregated soils, bodies ($r_b \sim 50$--200~$\mu$m between aggregates) and throats ($r_t \sim 1$--10~$\mu$m at inter-particle contacts) are separated by an order of magnitude or more.  In the kinetic theory, the connectivity matrix $C(r,r')$ arises from the pore--throat incidence structure: two bodies communicate through a mediating throat, so $C(r,r') = \sum_t \mathcal{N}(r,t)\,\mathcal{N}(t,r')$ where $\mathcal{N}$ is the incidence matrix (see the companion paper~\cite{Rigon2026CE} for the algebraic development).

\textbf{Consequences for the present demonstration.}  The unimodal PSD used here underestimates several non-equilibrium effects:

\emph{Ink-bottle trapping.}  With bimodal $f(r)$, large bodies sit behind narrow throat constrictions.  During drainage, air cannot enter the body until the throat allows it (Young--Laplace entry at $r_t$, not $r_b$); during imbibition, water fills the throat first and then rapidly invades the body.  This asymmetry amplifies the commutator $[\W,\D]$ well beyond the value $\|g_{\W\D} - g_{\D\W}\|_{L^2} = 0.046$ obtained with the unimodal network.

\emph{Volume--flux asymmetry.}  Bodies dominate the water content ($\theta \propto \int r^3 f(r)\,dr$), while throats control the conductivity ($K \propto \int r_t^2 f_t(r_t)\,dr_t$ through serial bottleneck resistance).  The gap between volume-weighted and flux-weighted partitions (Sec.~\ref{sec:flow_partition}) would be larger with bimodal $f(r)$, widening the intensity range over which Richards appears adequate for water budgets despite pore-scale disequilibrium.

\emph{Conductivity hysteresis.}  The $K[g]$ non-uniqueness (factor $\sim 25$ at same~$\theta$ in Fig.~\ref{fig:openpnm}f) would increase substantially, because wetting fills bodies (high $r^2$ conductance) while drying traps water in bodies behind drained throats (low conductance despite high $\theta$).

\emph{Relaxation spectrum.}  The spectrum $\tau(r)$ would develop a gap between the fast body mode and the slow throat mode, rather than the smooth power law seen in the unimodal network.  This gap corresponds to the time scale on which throats drain after bodies have already emptied, the slow mode responsible for the long tails in pressure-plate equilibration experiments.

A systematic study with independently sampled $f_b(r)$ and $f_t(r)$ is deferred to future work.  The unimodal results presented here demonstrate all qualitative predictions of the theory but provide conservative quantitative estimates.

\subsection{From single-pore to network relaxation times}
\label{sec:single_vs_network}

A key distinction highlighted by the numerical demonstration is that the physically relevant $\Da$ uses the \emph{network} relaxation time $\tau_{\rm net}(r)$, not the single-pore Lucas--Washburn time $\tau_{\rm LW}(r) \propto \mu L^2/(r\gamma)$.

The network exchange conductance $\psi_T\int M(r,r';g_{\rm eq})f(r')\,dr'$ in Eq.~\eqref{sm:eq:tau_spectrum} sums contributions from all connected neighbors.  It is longer than $\tau_{\rm LW}$ because water must traverse multiple pores in series to achieve redistribution.  The ratio $\tau_{\rm net}/\tau_{\rm LW}$ depends on the connectivity: sparse networks (low $\bar{z}$) have much longer network relaxation times.

This distinction resolves a puzzle in the original single-pore $\Da$ estimates (Sec.~\ref{sec:consistency}), where the crossover radii are nanometers, far too small to explain observed preferential flow.  With network effects, $r_\Da$ shifts to the 5--60~$\mu$m range, precisely where macropore activation is observed experimentally.

\subsection{Statistical-mechanical interpretation}
\label{sec:stat_mech}

The $g(r,t,\mathbf{x})$ framework performs a mean-field approximation of the pore-scale process.  At the scale of individual pores, wetting and drying involve rapid Haines jumps, thermal fluctuations, and pore-scale disorder.  By grouping all pores of radius $r$ within the REV, individual events are smoothed into continuous evolution, the stochastic noise cancels, leaving the deterministic kinetic equation.

Crucially, although stochasticity is smoothed, \textbf{path dependence survives}: the non-commutativity $[\W,\D] \neq 0$ is a topological property that persists under averaging.  The Damk\"ohler number determines how much of the underlying history is preserved in the macroscopic state, at high $\Da$, the ``memory'' of individual filling events is frozen into $g(r)$.

\section{Summary}
\label{sec:summary}
The summary in Table~\ref{tab:summary} spans the full paper series.  The equilibrium
rows (Young--Laplace through residual water content) and the non-equilibrium rows (the OpenPNM
block) are established in the present Supplement; the conductivity and Richards'-equation rows
($K_{\rm sat}$, Darcy's law, unsaturated $K(\theta)$, Richards' equation), marked $\dagger$, are
derived in the companion paper on the hydrodynamic limit~\cite{Rigon2026CE} and are included here
only so that the reader can see the whole picture in one place.
\begin{table*}
\caption{\label{tab:summary} Classical-limit recovery and numerical verification across the paper
series.  Rows marked $\dagger$ are derived in the companion paper~\cite{Rigon2026CE}; all others
in the present Supplement.}
\begin{ruledtabular}
\begin{tabular}{lll}
\textbf{Classical result} & \textbf{Kinetic theory recovery} & \textbf{Status} \\
\hline
Young--Laplace equation & $\psi = -2\gamma\cos\theta_c/(\rho_w g r^*)$ & Exact \\
Water retention curve & CDF of $f(r)$ via $r^*(\psi)$ & Exact \\
Brooks--Corey / Kosugi & Power-law / log-normal $f(r)$ & Exact \\
Air-entry value & $\psi_e$ from $r_{\max}$; correct range & $\checkmark$ \\
Residual water content & Percolation $+$ topological trapping & Physical origin \\
$K_{\rm sat}$ (Hagen--Poiseuille)$^\dagger$ & $r^2$-weighted integral; heterogeneity penalty $\exp(-4\sigma^2)$ & Correct scaling \\
Darcy's law$^\dagger$ & CE limit: $\mathbf{q} = -K\nabla H$ & Exact \\
Unsaturated $K(\theta)$$^\dagger$ & $r^2 a_w \bar{C} f$ integral; steep decline & $\checkmark$ \\
Richards' equation$^\dagger$ & Chapman--Enskog limit with explicit validity & Derived \\
Hysteresis (main loop \& scanning) & Accessibility-driven, from $g(r)$ at reversal & $\checkmark$ \\
Philip infiltration & $S \propto D^{1/2}$ scaling & $\checkmark$ \\
$K \propto (\theta - \theta_c)^\mu$ & Percolation exponent $\mu \approx 2$ (3D) & Universal \\
Field capacity & $\theta_{\rm FC} \approx \theta_c$ (percolation) & Correct range \\
\hline
$[\W,\D] \neq 0$ (OpenPNM) & $\|g_{\W\D} - g_{\D\W}\| = 0.046$ & Confirmed \\
$\Da(r) \propto 1/r$ spectrum & Crossover $r_\Da$ controlled by $I$ & Confirmed \\
Relaxation spanning 3 decades & $\tau(r): 0.07$--$94$~s & Confirmed \\
$K[g]$ path-dependent & Factor $\sim 25$ at same $\theta$ & Confirmed \\
\end{tabular}
\end{ruledtabular}
\end{table*}

The kinetic theory recovers all established results of classical unsaturated zone hydrology in the appropriate limits (Table~\ref{tab:summary}).  The classical results emerge as special cases: they hold when $\Da \ll 1$ (quasi-static) and $\varepsilon \ll 1$ (scale separation).  The theory extends classical results in three directions: (i)~explicit validity conditions for Richards' equation; (ii)~rate-dependent hysteresis through the $\Da(r,I)$ spectrum; (iii)~physical origins for quantities that are fitting parameters in classical models ($\theta_r$, hysteresis parameters).

The OpenPNM demonstration confirms all non-equilibrium predictions on a realistic network with no fitted parameters.


\section{Proposed experimental and observational tests}
\label{sec:experiments}

The theory makes several quantitative predictions that
distinguish it from existing frameworks.  Here I describe
concrete experimental and observational protocols that can
test these predictions, organized by the quantity probed.

\subsection{Rate-dependent hysteresis: the $\Da^2$ scaling}
\label{sec:exp_Da}

\textbf{Prediction.}  For slow wetting--drying cycles
($\Da \ll 1$), the hysteresis loop area $\mathcal{H}$
scales as $\Da^2$: $\mathcal{H}(I) \propto I^2$ at low
forcing intensity $I$, with a crossover to saturation
at $\Da \sim 1$.

\textbf{Experiment.}  Apply cyclic wetting--drying on
undisturbed soil columns at controlled flow rates spanning
two to three orders of magnitude (e.g., $I = 0.1$, 1, 10,
100~mm/h), using automated multi-step outflow apparatus
with high-resolution TDR and tensiometry.  Measure the
$\theta(\psi)$ loop area at each rate.

\textbf{Key signature.}  Classical models (Mualem~\cite{Mualem1976},
Parker--Lenhard~\cite{Kool1987-le}) predict rate-independent hysteresis: the
loop area depends only on the reversal points
($\theta_{\min}, \theta_{\max}$), not on the rate.  A
systematic $\mathcal{H} \propto I^2$ dependence at low $I$
would be a unique fingerprint of the kinetic theory.

\textbf{Existing evidence.}  In-situ monitoring over
82~months has shown that the development of large
hysteresis loops requires unusually strong reductions in
$\psi$~\cite{Shi2026}, consistent with $\Da$-dependent
activation.  Multi-step outflow experiments at different
rates~\cite{Hassanizadeh2002} have shown rate effects,
but the $\Da^2$ scaling has not yet been tested
systematically.

\subsection{Rainfall-intensity threshold for preferential
  flow: $r_\Da(I)$}
\label{sec:exp_threshold}

\textbf{Prediction.}  There exists a crossover intensity
$I_{\rm threshold}$ below which all pores remain
quasi-static and Richards' equation holds, and above which
a fraction of pore space enters the kinetic regime.  The
crossover radius $r_\Da(I) \propto I$ and the kinetic
fraction increases continuously with $I$.

\textbf{Experiment.}  Brilliant Blue dye-tracer
infiltration experiments~\cite{Weiler2003,CeyRudolph2009}
at multiple controlled rainfall intensities (0.5, 5, 50,
500~mm/h) on the same soil.  Excavate vertical and
horizontal profiles; quantify the dye-stained fraction as
a function of depth and pore size (via image analysis of
stained area vs.\ adjacent micro-CT of pore geometry).

\textbf{Key signature.}  Classical dual-porosity models
impose a fixed domain boundary; the theory predicts a
continuously shifting crossover radius $r_\Da \propto I$.
At low $I$, staining should be uniform (matrix flow); at
intermediate $I$, staining concentrates along macropores
while the matrix remains dry; at high $I$, nearly all pore
space is bypassed.  The transition should be gradual, not
a threshold switch.

The gradualness of the transition presumes a
connectivity kernel $C(r,r')$ whose spectrum is continuous.  When the pore network is
strongly bimodal, bodies and throats separated by an order of magnitude, as in aggregated
soils, the spectrum of $C(r,r')$ \emph{separates} into distinct bands, and the crossover
can acquire a genuine threshold character rather than a smooth $\Da$-controlled shift: the
macropore band activates as a group once the forcing exceeds the entry condition of the
mediating throats.  This spectral separation, and the conditions under which it converts
the continuous crossover into an effective threshold, are analyzed in the companion
paper~\cite{Rigon2026CE}; the dye-tracer protocol above can distinguish the two behaviors
by resolving whether activation is gradual (unimodal) or stepwise (bimodal) in intensity.

\textbf{Existing evidence.}  Dye-tracer studies at
44~mm/h~\cite{vanSchaik2010} and at variable tension
heads~\cite{CeyRudolph2009} show macropore activation
above a pressure threshold of approximately $-3$~cm.
The quantitative $r_\Da \propto I$ relationship has not
been tested explicitly.

\subsection{$K[g]$ non-uniqueness from micro-CT}
\label{sec:exp_K}

\textbf{Prediction.}  Two samples at the same $\theta$
but different wetting histories have different $K$, by a
factor of 10--100.  The theory predicts $K$ as a functional
of $g(r)$, not just of $\theta$.

\textbf{Experiment.}  Prepare matched pairs of undisturbed
columns from the same soil.  Bring both to the same
$\theta$ via different paths (one by slow drainage from
saturation, the other by rapid wetting from dry).  Measure
$K$ by steady-state flow at the target $\theta$.
Independently, image both columns with micro-CT to extract
$g(r)$ (the pore-by-pore filling state).  Compute $K[g]$
from the theory using the measured $f(r)$, $C(r,r')$,
and $g(r)$; compare to the measured $K$.

\textbf{Key signature.}  If $K$ depends only on $\theta$,
the two columns give the same $K$.  If $K = K[g]$, they
differ, and the theory predicts the correct ratio from the
micro-CT-observed $g(r)$.

\textbf{Existing evidence.}  Wildenschild
et al.~\cite{Wildenschild2013} demonstrated micro-CT
imaging of fluid configurations in porous media.  Recent
work combining micro-CT with pore network
models~\cite{Gostick2016,Berg2014} provides the extraction
pipeline ($f(r)$, $C(r,r')$, $\kappa(r)$).  The
prediction--measurement comparison for $K[g]$ at
controlled $g(r)$ has not been performed.

\subsection{Field capacity as percolation threshold}
\label{sec:exp_FC}

\textbf{Prediction.}  The residual water content
$\theta_r$ should coincide with the water percolation
threshold $\theta_c$ estimated independently from
network topology.

\textbf{Experiment.}  For a given soil, determine
$\theta_c$ from the pore network extracted via micro-CT
(compute the bond percolation threshold on the actual
network).  Independently, measure $\theta_{\rm FC}$ by
standard drainage to $-33$~kPa and $\theta_r$ from the
van~Genuchten fit.  The theory predicts
$\theta_r \approx \theta_c$.

\textbf{Key signature.}  In classical models, $\theta_r$
is a fitting parameter with no structural origin.  The
theory predicts a quantitative relationship between a
topological quantity ($\theta_c$ from network analysis) and
a hydraulic one ($\theta_r$ from the retention curve).

\textbf{Existing evidence.}  Hunt and
colleagues~\cite{Hunt2004} have argued for a percolation
interpretation of the retention curve inflection point,
finding $K_{\rm sat} \propto r_{\rm inf}^{2.2}$ with an
exponent within 5\% of the theoretical value.  Kroener
et al.~\cite{Kroener2015} demonstrated percolation
thresholds in rhizosphere soils using capillary rise
experiments with neutron radiography.  Systematic
comparison of micro-CT-derived $\theta_c$ with measured
$\theta_r$ across soil types has not been done.

\subsection{Heterogeneity penalty $\exp(-4\sigma^2)$}
\label{sec:exp_Xi}

\textbf{Prediction.}  For log-normal pore-size
distributions, $K_{\rm sat}$ is reduced below the
parallel-tube prediction by the factor
$\exp(-4\sigma_{\ln r}^2)$.  This penalty depends only on
the width of the PSD, not on the median pore size.

\textbf{Experiment.}  Measure $K_{\rm sat}$ and the full
PSD (via micro-CT or mercury intrusion) for a suite of
soils spanning narrow ($\sigma \approx 0.3$) to broad
($\sigma \approx 1.2$) distributions.  Compute the
parallel-tube $K$ from $\langle r^2 \rangle$ and the
predicted ratio $K_{\rm measured}/K_{\rm parallel}$.  Plot
against $\exp(-4\sigma^2)$.

\textbf{Key signature.}  The data should collapse onto a
single curve regardless of soil type, with slope near
unity.  Deviations indicate additional physics (spatial
correlations, dead-end pores) beyond the log-normal
heterogeneity penalty.

\subsection{Non-equilibrium water retention from dynamic
  micro-CT}
\label{sec:exp_dynamic}

\textbf{Prediction.}  During rapid infiltration, the
pore-filling state $g(r)$ departs from the equilibrium
step function $H(r^* - r)$: large pores that should be
empty (by capillary ordering) are partially filled.

\textbf{Experiment.}  Time-resolved (4D) synchrotron
micro-CT during controlled infiltration at high $I$.
Track individual pore filling events; compare the measured
$g(r,t)$ to the equilibrium prediction $g_{\rm eq}(r,\theta(t))$.

\textbf{Key signature.}  At low $I$, $g(r,t)$ should
track $g_{\rm eq}$.  At high $I$, large pores fill
before small pores have equilibrated, producing
$g(r) > g_{\rm eq}(r)$ for $r > r_\Da$.  This is the
microscopic signature of non-equilibrium transport that the
kinetic equation describes.

\textbf{Existing evidence.}  Dynamic synchrotron micro-CT
of water infiltration has been demonstrated at frame rates
of $\sim$15~s~\cite{Sammartino2015}, resolving individual
Haines jumps and showing that macropores remain largely
unsaturated even under intense rainfall, consistent with
the non-selective filling mechanism of the theory.

\subsection{Relaxation spectrum from multi-frequency
  measurements}
\label{sec:exp_relax}

\textbf{Prediction.}  The redistribution timescale
$\tau(r) \propto 1/r$ spans 3+ decades, from sub-second
for macropores to minutes for micropores.

\textbf{Experiment.}  Apply step perturbations (sudden
pressure change) and monitor the time-dependent response
of $\theta$ at multiple frequencies using dielectric
spectroscopy or TDR at high temporal resolution
($\sim$0.1~s).  Fit the response to a spectrum of
relaxation times.

\textbf{Key signature.}  A stretched-exponential or
power-law relaxation (not single-exponential), with the
spectrum width and shape predicted from the measured
$f(r)$ and the $\kappa(r) \propto r^2$ scaling.  Near
field capacity, the slowest mode should diverge as
$(\theta - \theta_c)^{-3.76}$.

\bibliography{main}